\documentclass[aip, rmp, reprint]{revtex4-1}
\usepackage{LXpackagesdefs}
\graphicspath{{figs/}}
\newcolumntype{C}{>{$}c<{$}}
\AtBeginDocument{
\heavyrulewidth=.08em
\lightrulewidth=.05em
\cmidrulewidth=.03em
\belowrulesep=.65ex
\belowbottomsep=0pt
\aboverulesep=.4ex
\abovetopsep=0pt
\cmidrulesep=\doublerulesep
\cmidrulekern=.5em
\defaultaddspace=.5em
}
\usepackage{CJKutf8}

\renewcommand{\RevisedText}[1]{{#1}}

\begin{document}


\title{Turbulent drag reduction by polymer additives: fundamentals and recent advances}



\begin{CJK*}{UTF8}{gbsn}
\CJKtilde 
\CJKindent 
\author{Li Xi\RevisedText{~(奚力)}}
\email[corresponding author, E-mail: ]{xili@mcmaster.ca}
\homepage[]{http://www.xiresearch.org}

\affiliation{Department of Chemical Engineering, McMaster Universtiy, Hamilton, Ontario L8S 4L7, Canada}

\date{\today}



\begin{abstract} 
A small amount of polymer additives can cause substantial reduction in the energy dissipation and friction loss of turbulent flow. The problem of polymer-induced drag reduction has attracted continuous attention over the seven decades since its discovery.
However, changes in research paradigm and perspectives have triggered a wave of new advancements in the past decade.
This review attempts to bring 
\RevisedText{researchers of all levels, from beginners to experts,}
to the forefront of
\RevisedText{this area}.
It starts with a comprehensive coverage of fundamental knowledge
\RevisedText{and}
classical findings and theories. It then highlights several recent developments that are bringing
\RevisedText{fresh}
insights into long-standing \RevisedText{problems}. Open questions and ongoing debates are also discussed.
\end{abstract}

\maketitle 
\end{CJK*}


\RevisedText{\tableofcontents}

\section{Introduction}\label{sec:intro}
In 1883, Osborne Reynolds reported his historical experiments on the transition \RevisedText{to} turbulence in pipe flow~\citep{Reynolds_PTRSL1883}: as fluid velocity increases, its trajectories in the flow change from a steady linear pattern (laminar flow) to a dynamic and sinuous one \RevisedText{(turbulence)}.
Dimensional analysis leads to the conclusion that for the same \RevisedText{type of} flow setup, the transition, as well as the whole flow behaviors, would depend solely on one non-dimensional parameter\RevisedText{, later known as the Reynolds number}
\begin{gather}
	\mathrm{Re}=\frac{\rho Ul}{\eta}
	\label{eq:Re}
\end{gather}
where $\rho$ and $\eta$ are the fluid density and viscosity
\RevisedText{(to be specific, dynamic viscosity as against the kinematic viscosity $\eta/\rho$), respectively, and} $U$ and $l$ are the characteristic velocity and length scales of the flow.
The laminar-turbulent (L-T) transition is marked by sharp changes in flow statistics. Most notably, the friction factor rises abruptly\RevisedText{. I}n the turbulent regime, friction loss is significantly higher than that of a laminar flow when compared at the same $\mathrm{Re}$.
Techniques for friction drag reduction (DR) is thus of significant practical interest\RevisedText{~\citep{Choi_Kim_JFM1994,Lumley_Blossey_ARFM1997}}.

Polymer additives have long been known as highly potent drag-reducing agents. In 1948, \citet{Toms_P1INTCRHEOL1948} reported that dissolving a minute amount ($O(10)\;\si{wppm}$\RevisedText{---}parts per million by weight) of poly(methyl methacrylate) (PMMA) into monochlorobenzene can substantially reduce the friction drag\RevisedText{,} compared with that of the pure solvent\RevisedText{,} in high-$\mathrm{Re}$ pipe flow.
Similar observations were subsequently \RevisedText{made} in a wide variety of polymer-solvent pairs and, under certain circumstances, the percentage drop in friction drag can be as high as $80\%$~\citep{Lumley_ARFM1969,Virk_AIChEJ1975}.
The ubiquity of DR across different chemical species shows the effect to be purely mechanical, caused by the coupling between polymer dynamics and turbulent flow motions rather than any specific chemical interaction.
The most effective drag-reducing polymer molecules are linear long chains with flexible backbones~\citep{Lumley_ARFM1969,Virk_AIChEJ1975,White_Mungal_EXPFL2004,Voulgaropoulos_Markides_AIChEJ2019}, although rigid polymers are also known to cause DR~\citep{Paschkewitz_Dimitropoulos_POF2005,Benzi_Ching_PRE2008}.

The phenomenon of polymer-induced turbulent DR has been continuously studied for seven decades since its discovery and findings were extensively reviewed in the literature.
The classical review by \citet{Virk_AIChEJ1975} \RevisedText{is a comprehensive} and well-organized account of major experimental observations \RevisedText{up to that time}, especially, flow statistics and their parameter dependence.
Later availability of new research tools, in particular, direct numerical simulation (DNS)~\citep{Sureshkumar_Beris_POF1997} and particle image velocimetry (PIV)~\citep{Warholic_Hanratty_EXPFL2001,White_Mungal_EXPFL2004}, allows the access to detailed flow and polymer stress fields, which has led to \RevisedText{significant new} discoveries in the past two decades. Many of \RevisedText{those} advances were covered in more recent reviews by \citet{Graham_RheologyReviews2004} and by \citet{White_Mungal_ARFM2008}.

Despite a long history of research, this area has witnessed a wave of recent advances that pushed the boundaries of our knowledge. These developments, which mostly occurred over the past ten years, were largely triggered by the shift of focus from ensemble flow statistics of turbulence to its dynamical heterogeneity, intermittency, and 
\RevisedText{transitions between different}
flow states.
Significant breakthroughs have been made in areas of long-standing difficulty, which is the primary motivation for this review.
These newest developments will be covered in \cref{sec:recent}.
Established phenomenology and classical theories will still be reviewed, in \cref{sec:fundamental}, to present a self-contained overview.
Note that separation between the two sections is not strictly chronological.
A number of very recent and interesting contributions are covered in \cref{sec:fundamental} for their better conceptual alignment with the more established framework of study.

Finally, as \RevisedText{is} the case in any other review, the current work is inevitably limited by the author's scope of knowledge and expertise. \RevisedText{A}lthough every effort has been made to stay neutral \RevisedText{and objective when} describing previous findings and observations, interpretation and discussion naturally reflects my own opinion.
The reader is \RevisedText{advised} to consult a number of earlier reviews for \RevisedText{more balanced viewpoints}~\citep{Virk_AIChEJ1975,Lumley_ARFM1969,Lumley_JPSMacroRev1973,Graham_RheologyReviews2004,White_Mungal_ARFM2008,Graham_POF2014,Benzi_Ching_ARCMP2018}.


\section{Fundamentals}\label{sec:fundamental}
This section starts with an introduction to the basic concepts\RevisedText{,} terminology\RevisedText{, and theoretical foundation} in polymer DR (\cref{sec:basic}). I attempt to take a pedagogical approach and write for the broadest readership possible,
\RevisedText{which I believe is necessary for}
the interdisciplinarity of this area\RevisedText{. It is indeed} an unusual marriage between two otherwise distant fields of physics, i.e., flow turbulence and polymer dynamics.
\RevisedText{Consequently}, it is rather rare for a beginner to have prior exposure to both fields.
After that, \cref{sec:phenomenology} will summarize major transitions between flow regimes and phenomenological observations in each regime. Classical understanding and theories will be reviewed and discussed in \cref{sec:mechanism}.


\subsection{Basic concepts and \RevisedText{background knowledge}}\label{sec:basic}
\subsubsection{Flow geometry and setup}
\begin{figure}
	\centering				
	\includegraphics[width=\linewidth, trim=0mm 0mm 0mm 0mm, clip]{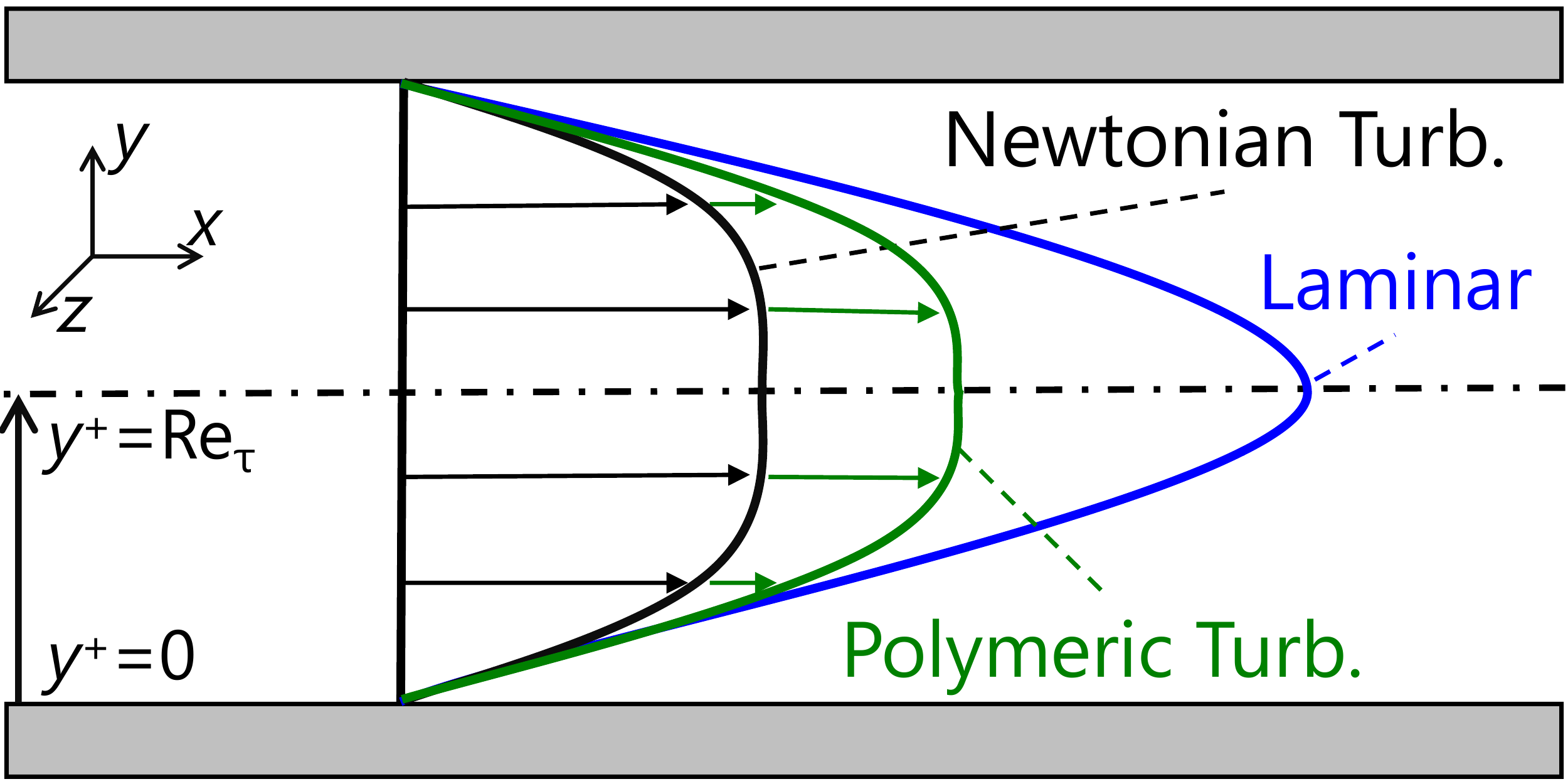}			
	\caption{Drag reduction in turbulent channel flow under constant pressure gradient (thus same wall shear stress).
	}
	\label{fig:channel}
\end{figure}

The most studied flow geometries are pipe and channel (\cref{fig:channel}) flows where the flow is driven by a pressure difference between the inlet and outlet.
The flow can be set up either with fixed average pressure gradient (\RevisedText{where} flow rate is allowed to vary) or with fixed bulk average velocity \RevisedText{$U_\text{avg}$} (\RevisedText{where} pressure drop is allowed to vary\RevisedText{).
(Hereinafter,} bulk average \RevisedText{refers to} average over the volume of the flow domain\RevisedText{.)}
The latter seems to be more common in experiments while the former is more often seen in simulations, although with exceptions.
When the bulk velocity is the controlled variable, it becomes the natural choice of the characteristic velocity $U$ in the $\mathrm{Re}$ definition (\cref{eq:Re}).
When the average pressure gradient is \RevisedText{held} constant, time-averaged velocity is availably only \textit{ex post facto}. Else, a velocity scale can be deduced from the applied pressure gradient, e.g., using laminar flow \RevisedText{velocity generated from} the same pressure gradient.
For the characteristic length $l$, the pipe diameter or half (occasionally, full) channel gap height is the common choice.

Other \RevisedText{flow setups} often used include plane Couette (flow between two parallel plates driven by their relative velocity \RevisedText{instead of} pressure gradient)~\citep{Page_Zaki_JFM2014}, boundary layer (flow generated near the solid surface when a uniform \RevisedText{bulk} flow is moving \RevisedText{over} a stationary plate)~\citep{Tamano_Graham_JFM2011}, and duct (pressure-driven flow in a straight conduit with uniform square or rectangular cross-sections) flows~\citep{Shahmardi_Rosti_JFM2019}.

\subsubsection{Definition of drag reduction}\label{sec:drdef}
Friction loss is measured by the Fanning friction factor $C_\text{f}$, defined \RevisedText{as~\citep{Bird_Stewart_2002}
\begin{gather}
	C_\text{f}\equiv\frac{\tau_\text{w}}{\frac{1}{2}\rho U_\text{avg}^2}
	\label{eq:Cf}
\end{gather}
where $\tau_\text{w}$ is the average wall shear stress (subscript ``w'' indicates quantities at the wall(s)) and $(1/2)\rho U_\text{avg}^2$ is the average fluid kinetic energy per volume.
At steady states,}
\begin{gather}
	\Delta P A_\text{cross-section}= \RevisedText{\tau_\text{w}} A_\text{wall}
	\label{eq:forcebal}
\end{gather}
\RevisedText{is a balance of the flow's driving force (left-hand side---LHS), provided by the average pressure drop}
\begin{gather}
	\Delta P\equiv P_\text{inlet}-P_\text{outlet},
\end{gather}
\RevisedText{and resistance due to wall friction (right hand side---RHS). A}reas of the flow cross section and the walls,
\RevisedText{$A_\text{cross-section}$ and $A_\text{wall}$, can be calculated from the flow geometry, which, combined with \cref{eq:Cf,eq:forcebal}, leads to the specific forms of the definition} in pipe and channel flows:
\begin{gather}
	C_\text{f, pipe}\equiv\frac{1}{4}\left(\frac{D}{L}\right)\left(\frac{\Delta P}{\frac{1}{2}\rho U^2}\right)
	\label{eq:Cfpipe}
\end{gather}
($D$ and $L$ are the pipe diameter and length) and
\begin{gather}
	C_\text{f, channel}\equiv\left(\frac{H}{L}\right)\left(\frac{\Delta P}{\frac{1}{2}\rho U^2}\right)
\end{gather}
($H$ and $L$ are the channel half height and length).
Note that $\Delta P/L$ is the average pressure gradient.

The level of drag reduction is quantified by
\begin{gather}
	\mathrm{DR\%}\equiv\frac{C_\text{f}-C_\text{f,\RevisedText{N}}}{C_\text{f}}\times100\%
\end{gather}
where $C_\text{f,\RevisedText{N}}$ is the friction factor of a reference flow \RevisedText{of} a benchmark Newtonian fluid (``\RevisedText{N}'' stands for ``\RevisedText{Newtonian}'').
The definition is straightforward when the solution is dilute enough that its shear viscosity \RevisedText{is still constant and is indistinguishable from} that of the pure solvent. Turbulent flow of the solvent\RevisedText{,} which presumably is a Newtonian fluid\RevisedText{, in} the same flow geometry and \RevisedText{with the} same pressure drop or flow rate\RevisedText{,} is a natural choice of the reference flow.
When compared at the same flow rate (same \RevisedText{$U_\text{avg}$}), DR is reflected in the decrease of pressure drop\RevisedText{. L}ikewise, when compared at the same $\Delta P$, DR is reflected in the increase of flow rate (\cref{fig:channel}).

At higher \RevisedText{polymer} concentrations,
\RevisedText{the solution viscosity is no longer constant and decreases with shear rate $\dot\gamma$, which is known as the shear-thinning effect.
Its value at the zero-shear limit
\begin{gather}
	\eta_0\equiv\lim_{\dot\gamma\to0}\eta\left(\dot\gamma\right)
	\label{eq:eta0}
\end{gather}
is usually considerably higher than the solvent viscosity.
In such cases, i}%
t is common to define the benchmark fluid as a hypothetical Newtonian liquid having the same viscosity as the polymer solution \RevisedText{shear viscosity $\eta(\dot\gamma_\text{w})$} corresponding to the \RevisedText{average} shear rate measured at the walls \RevisedText{$\dot\gamma_\text{w}$} (see, e.g., \citet{Warholic_Hanratty_EXPFL1999, Ptasinski_Nieuwstadt_JFM2003}).
This is so that any reduction of \RevisedText{the} friction factor caused by shear thinning\RevisedText{,} i.e., \RevisedText{reduction of} viscous shear stress resulting from the lowering viscosity\RevisedText{,} is not \RevisedText{considered. Rather,} $\mathrm{DR\%}$ \RevisedText{should only include} the reduction of turbulent friction loss.
\Citet{Lumley_ARFM1969} insisted a more stringent criterion which uses the pure solvent\RevisedText{,} whose viscosity is lower than or at best equal to that of the polymer solution\RevisedText{,} as the benchmark fluid\RevisedText{.}

\subsubsection{Turbulent inner layer: scales and \RevisedText{flow} characteristics}\label{sec:inner}
Flow statistics of wall turbulence are often reported in nondimensionalized quantities using the so-called viscous scale\RevisedText{s} or turbulent inner scales~\citep{Pope_2000}. The practice is very common in DR literature but can be confusing to beginners.
\RevisedText{At its heart, it is rooted in the separation of scales between turbulence in the bulk flow and that in near-wall layers.}

For Newtonian flow, $\mathrm{Re}$ (\cref{eq:Re}) \RevisedText{is} the only \RevisedText{dimensionless} parameter in the governing equations if $U$ and $l$\RevisedText{, sometimes called the outer scales,} are used to nondimensionalize all
\RevisedText{flow}
quantities.
When the flow is turbulent, boundary layers develop near the walls, within which it is expected that the most relevant scales are not those of the bulk flow ($U$ and $l$), but scales derived from
\RevisedText{flow quantities at the wall, with wall shear stress}
\begin{gather}
	\tau_\text{w}=\RevisedText{\eta\dot\gamma_\text{w},}
\end{gather}
\RevisedText{where}
\begin{gather}
	\RevisedText{\dot\gamma_\text{w}\equiv}\left.\frac{dU_\text{m}}{dy}\right|_{y=y_\text{w}},
\end{gather}
\RevisedText{being the only choice available.
Note that in this paper, spatial axis labels are assigned according to the commonly accepted convention in wall turbulence (also marked in \cref{fig:channel}):%
\begin{CmpctItem}%
	\item[--]$x$: streamwise direction, aligned with the mean flow;
	\item[--]$y$: wall-normal direction, perpendicular to the wall;
	\item[--]$z$: spanwise direction, aligned with the vorticity of the laminar shear flow.
\end{CmpctItem}%
}%
\noindent
Here, $y_\text{w}$ is the position of the wall, $U_\text{m}$ is the mean velocity profile
\begin{gather}
	U_\text{m}(y)\equiv\langle v_x\RevisedText{(x,y,z,t)}\rangle
\end{gather}
\RevisedText{and}
$\langle\cdot\rangle$ indicates ensemble average, which in \RevisedText{wall turbulence} is average over homogeneous directions\RevisedText{,} $x$ and $z$\RevisedText{,} and time $t$.

\RevisedText{Combining $\tau_\text{w}$ with fluid properties $\rho$ and $\eta$, the only velocity and length scales to be derived, i.e., the turbulent inner scales,}
are the friction velocity
\begin{gather}
	u_\tau\equiv\sqrt\frac{\tau_\text{w}}{\rho}
	\label{eq:utau}
\end{gather}
and viscous length scale
\begin{gather}
	l_\text{v}\equiv\frac{\eta}{\rho u_\tau}.
	\label{eq:lv}
\end{gather}
\RevisedText{The latter is also} colloquially known as the ``wall unit''.
The friction Reynolds number
\begin{gather}
	\mathrm{Re}_\tau=\frac{\rho u_\tau l}{\eta}
	\label{eq:Retau}
\end{gather}
is defined based on friction velocity and the flow geometric length scale $l$ (instead of $l_\text{v}$ which would give \RevisedText{the same trivial} value of 1 \RevisedText{for all cases}).
Comparing \cref{eq:lv,eq:Retau}, it is clear that
\begin{gather}
	\mathrm{Re}_\tau=\frac{l}{l_\text{v}}
	\label{eq:Retaulv}
\end{gather}
-- i.e., the number of wall units in the flow domain size $l$ (\RevisedText{see} \cref{fig:channel}).
Combining \cref{eq:utau,eq:Retau}, we get $\mathrm{Re}_\tau=\sqrt{\rho\tau_\text{w}}l/\eta$\RevisedText{. F}or flow under the constant-$\Delta P$ constraint, $\tau_\text{w}$ is constant (\RevisedText{see \cref{eq:forcebal}}) \RevisedText{and thus} $\mathrm{Re}_\tau$ is constant and predetermined.

Quantities nondimensionalized with inner scales are marked with the superscript ``+'', e.g.,
\begin{gather}
	U_\text{m}^+\equiv\frac{U_\text{m}}{u_\tau}.
\end{gather}
When discussing near\RevisedText{-}wall turbulence using the inner scales, it is also customary to use the wall coordinate in the wall-normal direction
\begin{gather}
	y^+\equiv\frac{|y-y_\text{w}|}{l_\text{v}}
\end{gather}
which measures the distance from the wall in wall units (\cref{fig:channel}).

\begin{figure}
	\centering
 	\includegraphics[width=0.88\linewidth, trim=0mm 0mm 0mm 0mm, clip]{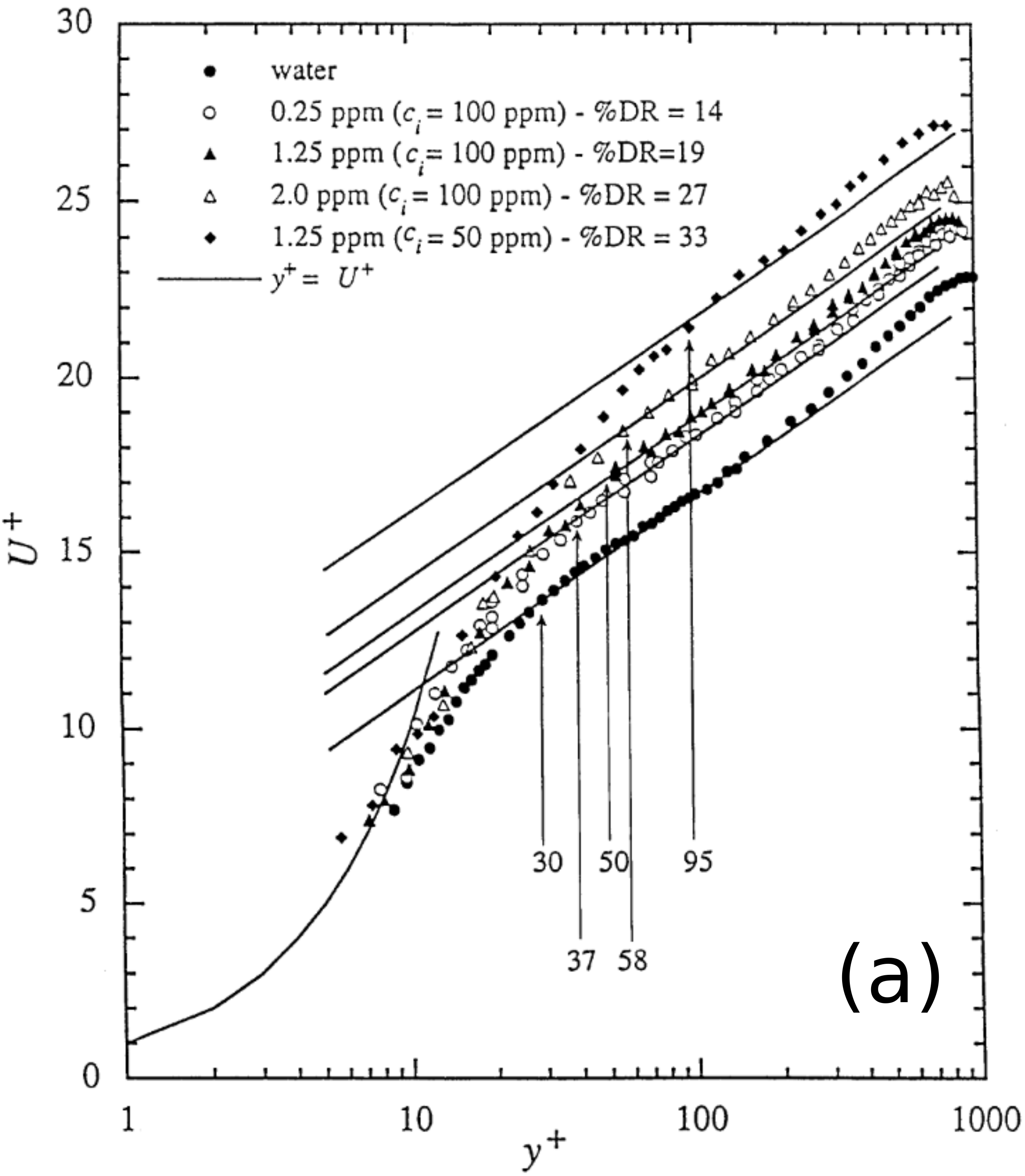}\\
	\includegraphics[width=0.88\linewidth, trim=0mm 0mm 0mm 0mm, clip]{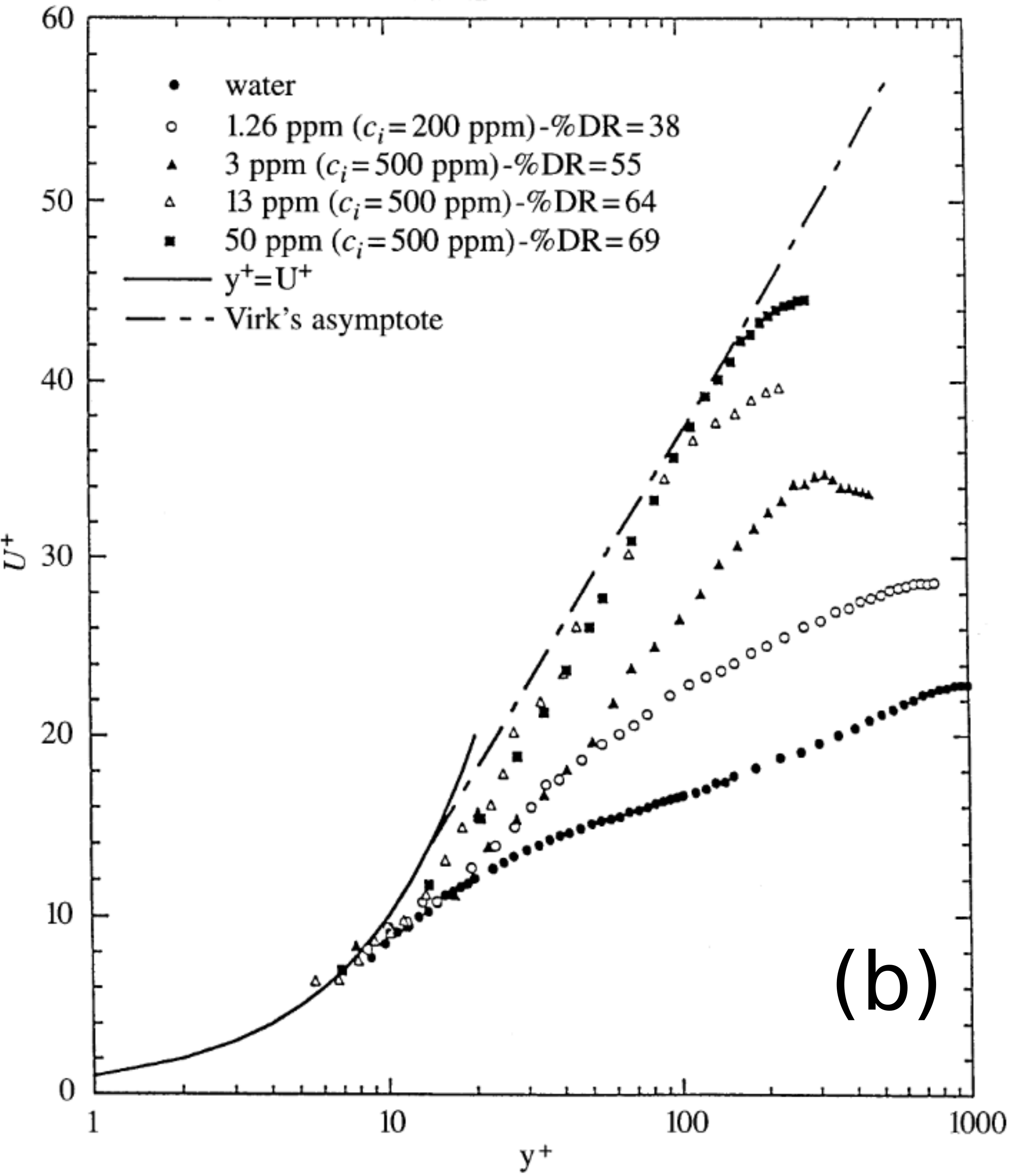}		
	\caption{Mean velocity profiles $U_\text{m}^+(y^+)$ from \RevisedText{channel flow experiments} ($\mathrm{Re}\approx\num{2e4}$ based on bulk velocity and water viscosity):
	(a) LDR (lowest straight solid line shows the von~K\'arm\'an law\RevisedText{---}\cref{eq:loglaw}; numbers annotate the lower-bound $y^+$ position for the logarithmic relation);
	(b) HDR and MDR (dot-dashed line shows the Virk MDR profile\RevisedText{---}\cref{eq:virk}).    
	The first ppm number of each case is the polymer concentration \RevisedText{in wppm measured} in the test section.
	(Reprinted by permission from Springer Nature: Springer, Experiments in Fluids,
	\RevisedText{27, 461,
	Influence of drag-reducing polymers on turbulence: effects of {R}eynolds number, concentration and mixing,
	\citeauthor*{Warholic_Hanratty_EXPFL1999},}
	Copyright (\citeyear{Warholic_Hanratty_EXPFL1999}).)
	}
	\label{fig:Um}
\end{figure}

\RevisedText{At sufficiently high $\mathrm{Re}$, many flow quantities scale with the inner scales up to $|y-y_\text{w}|~\sim O(0.1)l$. These regions are collectively called the turbulent inner layer and, correspondingly, the bulk of turbulence beyond the near-wall region is called the outer layer.}
The best known example of the success of the inner scaling is the von K\'arm\'an law of wall~\citep{Pope_2000} which shows that for different flow geometries and for vastly different $\mathrm{Re}$, the near-wall mean velocity profile of Newtonian turbulence follows a universal logarithmic relation when scaled in inner units
\begin{gather}
	U_\text{m}^+=A^+\ln y^+ +B^+\quad (y^+\gtrsim30)
	\label{eq:loglaw}
\end{gather}
where $A^+$ and $B^+$ are constants believed to be universal among Newtonian wall flows, with some small variations between different sources\RevisedText{.}
\citet{Pope_2000} found that literature values are within $5\%$ variation of $A^+=2.44$ and $B^+=5.2$\RevisedText{.}
DNS data from \citet{Kim_Moin_JFM1987} were best fit by $A^+=2.5$ and $B^+=5.5$.
The logarithmic profile is found to hold over most of the near\RevisedText{-}wall region from $y^+\approx 30$ up to $|y-y_\text{w}|\sim 0.3l$, which is often called the ``log-law layer''.
\RevisedText{In \cref{fig:Um}, a clear logarithmic dependence is observed in the water case at $y^+\gtrsim 30$.
The polymeric cases will be discussed in \cref{sec:intermediate}.}

In regions closest to the wall, viscous dissipation dominates and the flow is essentially a laminar shear flow
\begin{gather}
	U_\text{m}^+=y^+\quad (y^+\lesssim 5).
	\label{eq:vsublayer}
\end{gather}
\RevisedText{This region} is \RevisedText{called} the ``viscous sub-layer'' \RevisedText{and \cref{eq:vsublayer} is sketched in \cref{fig:Um} as a curved solid line.}
Between $y^+\approx 5$ and $y^+\approx 30$ is a transition region called the ``buffer layer''.
In \RevisedText{addition} to flow statistics, inner scales are also found to describe flow structures well: e.g., the average spacing between near-wall velocity streaks (again, for Newtonian flow) is found to be around $100$ wall units for various flow conditions~\citep{Smith_Metzler_JFM1983,Jimenez_Moin_JFM1991,Robinson_ARFM1991}.

\subsubsection{Experimental \RevisedText{systems and} parameters}
Experimental parameters include the flow geometric size (pipe diameter or channel height), flow rate, and polymer solution properties. The latter include the polymer and solvent species, polymer molecular weight, and concentration. 
Although a variety of polymer-solution pairs have been tested especially in earlier years, after it was established that DR depends only on the mechanics of polymer molecules, aqueous solutions of flexible hydrophilic polymers, such as poly(ethylene oxide) (PEO) and polyacrylamide (PAM), have become the most widely used systems for fundamental research~\citep{Lumley_ARFM1969,Virk_AIChEJ1975,Owolabi_Poole_JFM2017,Voulgaropoulos_Markides_AIChEJ2019}.
More rigid bio-based polymers, in particular polysaccharides such as scleroglucan and xanthan gum (XG), are also often studied~\citep{JapperJaafar_Poole_JNNFM2009,Jaafar_Poole_JApplSci,Pereira_Soares_JNNFM2013,Mohammadtabar_Ghaemi_POF2017}.
Typical concentrations are in the range of $O(10)-O(100)\;\si{wppm}$. For most drag-reducing polymers, $O(100)\;\si{wppm}$ is sufficient to observe substantial increase in the zero-shear viscosity and clear shear-thinning. A shear viscosity versus shear rate curve needs to be measured for the proper calculation of $\mathrm{DR}\%$\RevisedText{, which, as discussed in \cref{sec:drdef}, requires the $\eta(\dot\gamma_\text{w})$ value for each flow condition}.
Most of \RevisedText{those} molecules have very high molecular weights, $O(\num{e6})\;\si{Dalton}$\RevisedText{, corresponding to} over $O(\num{e4})$ repeating units\RevisedText{,} or higher.

\subsubsection{DNS: constitutive models \RevisedText{and physical basis}}
\begin{figure}
	\centering				
 	\includegraphics[width=\linewidth, trim=0mm 0mm 0mm 0mm, clip]{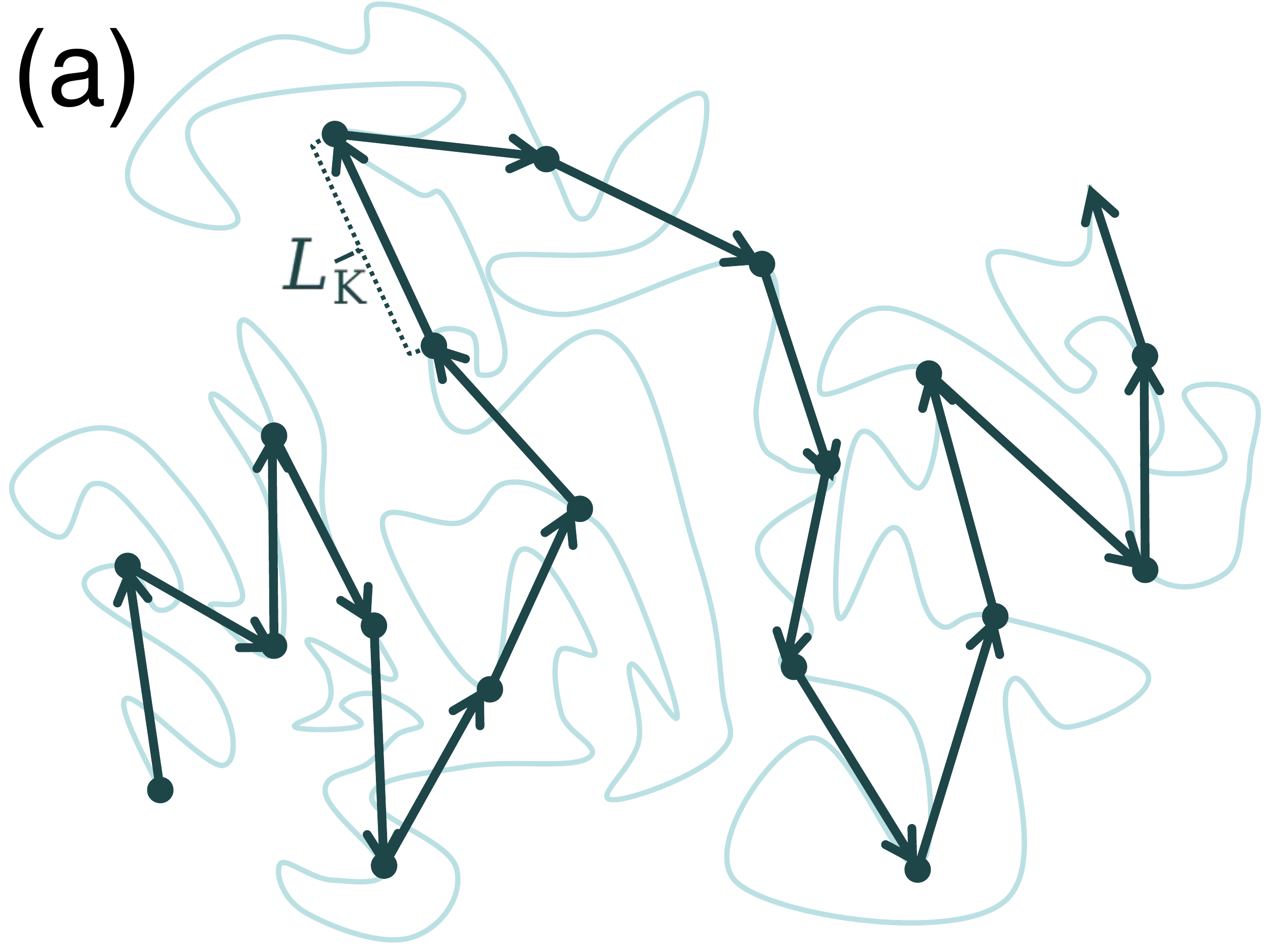}
 	\includegraphics[width=\linewidth, trim=0mm 0mm 0mm 0mm, clip]{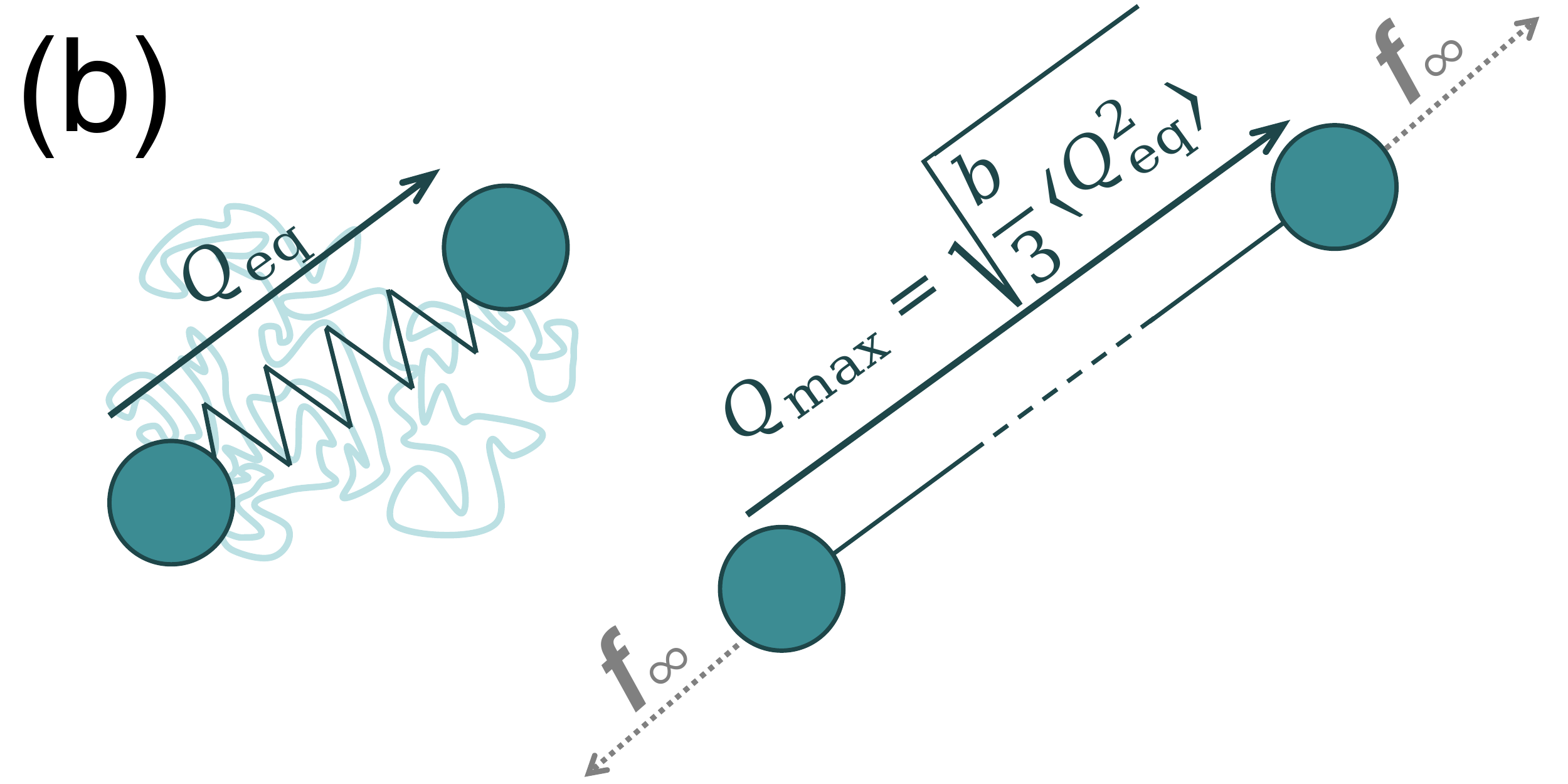}
	\caption{Models for flexible polymer chains:
	(a) an un-stretched polymer chain in a \texttheta-solvent or \RevisedText{at} ideal-chain condition (curve) and $N_\text{K}$-step random walk representation (freely-joined chain\RevisedText{---}FJC) with step size $L_\text{K}$; 
	(b) FENE dumbbell model in un-stretched (left) and fully stretched (right; with infinite force $\mbf f_\infty$) states.
	The FJC model tracks the motion of $N_\text{K}+1$ equispaced markers on the chain and neglects the detailed conformation between markers\RevisedText{. O}rientation of each Kuhn segment is uncorrelated with other ones.
	The FENE dumbbell model tracks the motion of two ends of the chain and neglects all internal conformation. The spring force models the entropic resistance of the chain \RevisedText{to} stretching. An accurate spring-force law can be derived from the free energy of a stretched FJC\RevisedText{, to which the FENE spring force (\cref{eq:fene}) is an approximation}~\citep{Rubinstein_Colby_2003,Graham_2018}.
	}
	\label{fig:chain}
\end{figure}

\paragraph{Introduction: \RevisedText{DNS and} constitutive equations}
Ever since its first successful implementation by \citet{Sureshkumar_Beris_POF1997}, DNS has become an indispensable tool in DR research which complements experiments with fully-detailed and time-resolved flow field data as well as direct access to polymer conformation and stress fields.
In DNS, the time-dependent Navier-Stokes (N-S) equation (momentum and mass balances) is numerically solved and the polymer stress contribution to momentum balance is modeled with a viscoelastic constitutive equation.
\Citet{Sureshkumar_Beris_POF1997} adopted the FENE-P constitutive model (\RevisedText{FENE-P stands for} \emph{f}initely-\emph{e}xtensible \emph{n}onlinear \emph{e}lastic dumbbell model with the \emph{P}eterlin approximation)~\citep{Bird_Curtis_1987}, whereas Oldroyd-B and Giesekus models have also been occasionally used in later studies~\citep{Dimitropoulos_Beris_JNNFM1998,Min_Choi_JFM2003a,Yu_Kawaguchi_JNNFM2004,Tamano_Itoh_POF2007}.

All three models treat polymer molecules as elastic dumbbells\RevisedText{---}two beads or force points connected by a spring force (\cref{fig:chain}(b)).
This effectively reduces the chain conformation into its end-to-end vector $\mbf Q$\RevisedText{,} which describes the orientation and extension
of the chain and ignores all internal degrees of freedom.
\RevisedText{(I}n this paper, boldface symbols indicate vectors or tensors\RevisedText{.)} 
Both Oldroyd-B and Giesekus models use the Hooke's law for the spring force\RevisedText{:} i.e., the force is proportional to the chain extension $|Q|$\RevisedText{.
(Hereinafter,}
$|\cdot|$ denotes the $L^2$-norm of a vector, i.e.,
\begin{gather}
	|Q|\equiv\sqrt{\mbf Q\cdot\mbf Q}
\end{gather}
\RevisedText{.) This}
is clearly unrealistic at the limit of large deformation since polymer extension is bounded by its contour length whereas a Hookean spring can be stretched infinitely.
In a FENE dumbbell, the spring force depends non-linearly on $|Q|$
\begin{gather}
	\mbf f^\text{FENE}=\frac{H_\text{s}\mbf Q}{1-\left(|Q|/Q_\text{max}\right)^2}
	\label{eq:fene}
\end{gather}
where $H_\text{s}$ is the spring force constant and $Q_\text{max}$ is the upper limit of $|Q|$\RevisedText{: $Q_\text{max}$ equals} the chain contour length.
It is easily verifiable that the force diverges as $|Q|\to Q_\text{max}$, which ensures the upper-boundedness of $|Q|$ (\cref{fig:chain}(b)).

\RevisedText{The} mechanical model behind FENE-P is clearly physically sounder and most closely represents the drag-reducing fluids of concern: i.e., dilute solutions of flexible polymers. Indeed, without \RevisedText{the} finite extensibility \RevisedText{constraint}, Oldroyd-B fails to predict shear-thinning and erroneously predicts infinite extensional viscosity at finite extension rate.
Giesekus model \RevisedText{incorporates} interactions between Hookean dumbbells\RevisedText{, making it a natural fit for}
more concentrated solutions.
\RevisedText{As a result}, FENE-P has been by far the most widely adopted model in the DNS of viscoelastic turbulence. \RevisedText{Meanwhile, l}imited comparisons between constitutive models \RevisedText{found} in the literature did not show any significant difference in \RevisedText{their} physical results~\citep{Dimitropoulos_Beris_JNNFM1998,Min_Choi_JFM2003a}.
\RevisedText{T}he following discussion will \RevisedText{thus} focus on the FENE-P model\RevisedText{.}

\paragraph{\RevisedText{FENE-P: basic concepts and assumptions}}\label{par:fenepbg}
\RevisedText{%
Here, we briefly go over a few basic concepts in polymer physics in relation to dilute solutions~\citep{Rubinstein_Colby_2003}, which will be mentioned repeatedly in this review. It is followed by a very brief introduction to the physical basis of the FENE-P model and its associated approximations~\citep{Graham_2018,Bird_Curtis_1987}.
}%

\subparagraph{\RevisedText{Ideal chain and solvent conditions}}
	\RevisedText{
	An ideal chain model neglects non-bonded interactions, such as van~der~Waals (vdW) and electrostatic interactions, between chain segments~\citep{Rubinstein_Colby_2003}.
	Obviously, real chain segments at least experience close-range steric repulsion and longer range vdW attraction.
	The latter is tunable through solvent effects.
	For a given polymer-solvent pair, the temperature at which the repulsion and attraction balance each other is called the \texttheta-temperature $T_\theta$. Polymer chains under this \texttheta-solvent condition effectively behave like ideal chains. 
	At $T>T_\theta$, known as the good-solvent condition, the polymer-solvent interaction is more favorable, allowing polymer chains to swell and become more exposed to solvent molecules.
	This can be described as an \emph{effective} reduction in the inter-segmental attraction.
	Likewise, at $T<T_\theta$, polymer chains contract, which is called the poor-solvent condition.
}

\subparagraph{\RevisedText{Dilute solution}}
\RevisedText{A polymer solution is considered dilute when individual polymer chains are so far apart from one another that their interactions can be neglected.
At equilibrium, polymer chains take a random coil conformation (see \cref{fig:chain}(a)) which is statistically most probable---i.e., having highest entropy.
At the infinitely dilute limit, each coil is effectively isolated from the rest. With increasing polymer concentration $C_\text{p}$, the average distance between coils shortens. 
Upon a critical concentration level $C_\text{p}^*$, this distance becomes shorter than the coil diameter and different coils start to penetrate into the space occupied by one another.
This so-called overlapping concentration is traditionally regarded as the upper bound of the dilute regime~\citep{Rubinstein_Colby_2003}.
Recent experiments showed that, in both shear and extensional flows, the polymer relaxation time $\lambda_H$---i.e., the time scale for an extended polymer chain to retract to its equilibrium coil conformation---starts to increase with $C_\text{p}$ well before $C_\text{p}^*$ is reached~\citep{Clasen_McKinley_JRh2006,DelGiudice_Shen_JRh2017}.
For polystyrene in both \texttheta- and good-solvent conditions, concentration-dependence was found at $C_\text{p}$ as low as $O(\num{e-2})C_\text{p}^*$.
For a molecular weight of $\approx\SI{7e6}{Da}$ studied in \citet{DelGiudice_Shen_JRh2017}, $C_\text{p}^*$ was estimated at $\approx\SI{5e3}{wppm}$ in a \texttheta-solvent and even lower in a good solvent.
This indicates that under flow conditions, considerable inter-chain interactions may start to exist at $C_\text{p}\lesssim\SI{100}{wppm}$, which covers most DR systems.
One plausible explanation is that, as polymer molecules are extended by the flow, each chain would span a larger volume than an equilibrium coil~\citep{DelGiudice_Shen_JRh2017}.%
}

\subparagraph{\RevisedText{Physical basis and assumptions of FENE-P}}
FENE-P can be derived for polymer chains from a molecular mechanics approach\RevisedText{, by considering the free energy of stretching polymer chains in dilute solutions,} with a few assumptions and approximations\RevisedText{~\citep{Rubinstein_Colby_2003,Bird_Curtis_1987,Graham_2018}}\RevisedText{:}
\begin{CmpctItem}
	\item the free energy of chain extension \RevisedText{derived assuming ideal-chain conformation}	leads to a force law $\mbf f(\mbf Q)$ \RevisedText{in} the form of an inverse Langevin function;
	\item the \RevisedText{inverse Langevin} force law \RevisedText{can be approximated by} a mathematically simpler empirical form of \cref{eq:fene}; and
	\item a mathematical Peterlin approximation \RevisedText{is introduced to allow} the closure of the constitutive model.
\end{CmpctItem}
\RevisedText{As such},  FENE-P is an idealized and approximate model for drag-reducing polymer solutions. Validity of and errors from these simplifications depend on the specific polymer type and solvent conditions.
\RevisedText{%
There has not been much effort among DNS researchers in establishing the connection between FENE-P and realistic drag-reducing polymer solutions, which is one of the areas open for future research (see discussion in \cref{sec:summary}).
Existing DNS studies, in a way, can thus be viewed as describing the DR behaviors of a generic type of ``FENE-P polymer'' solutions.
}%

\paragraph{Polymer conformation tensor}
Numerical solutions from DNS contain the time-resolved three-dimensional (3D) turbulent velocity $\mbf v(\mbf r, t)$ and pressure $p(\mbf r, t)$ fields ($\mbf r$ denotes 3D spatial coordinates and $t$ is time).
Constitutive models such as FENE-P \RevisedText{solve} for the non-dimensionalized polymer conformation tensor
\begin{gather}
	\mbf\alpha(\mbf r, t)\equiv\frac{H_\text{s}}{k_\text{B}T}\langle\mbf Q\mbf Q\rangle
	\label{eq:alpha}
\end{gather}
field, from which the polymer stress contribution $\mbf\tau_\text{p}$ can be calculated. (Here, $k_\text{B}$ is the Boltzmann constant, $T$ is absolute temperature, and $\langle\cdot\rangle$, again, denotes ensemble average\RevisedText{---}in the case of \cref{eq:alpha}, it is the average over all individual polymer molecules at the local flow region of $(\mbf r, t)$.
\RevisedText{S}ymbol $\mbf C$ is sometimes used instead of $\mbf\alpha$ \RevisedText{in the literature}.)
Note that the $xx$-, $yy$-, and $zz$-components of the $\langle\mbf Q\mbf Q\rangle$ tensor are $\langle Q_x^2\rangle$, $\langle Q_y^2\rangle$, and $\langle Q_z^2\rangle$, respectively\RevisedText{. Thus, t}%
he trace of the $\mbf\alpha$ tensor
\begin{gather}
	\mathrm{tr}(\alpha)\equiv\sum_i\alpha_{ii}=\frac{H_\text{s}}{k_\text{B}T}\langle|Q|^2\rangle
	\label{eq:tralpha}
\end{gather}
is the non-dimensional mean square end-to-end distance of polymer chains \RevisedText{and a common measure of chain extension}.

\paragraph{\RevisedText{DNS p}arameters \RevisedText{based on} the FENE-P constitutive model}\label{par:feneppara}
Dimensional analysis of the governing equations leads to four non-dimensional parameters (compared with only one in the case of Newtonian fluids) which fully \RevisedText{define} the system.
\subparagraph{Reynolds number $\mathrm{Re}$} same as \cref{eq:Re}, except that $\rho$ and $\eta$ are \RevisedText{now} the density and \RevisedText{zero-shear} viscosity \RevisedText{$\eta_0$ (\cref{eq:eta0})} of the polymer solution.
\subparagraph{Viscosity ratio $\beta$} defined as the ratio of the solvent viscosity to that of the solution
	\begin{gather}
		\beta\equiv\frac{\eta_\text{s}}{\RevisedText{\eta_0}}=\frac{\eta_\text{s}}{\eta_\text{s}+\eta_\text{p}}
	\end{gather}
	(%
	subscripts ``s'' and ``p'' represent solvent and polymer contributions, respectively). Of course, $\beta=1$ is the pure Newtonian fluid limit and $\beta$ decreases with increasing polymer concentration.
	Typical DNS studies use $\beta\geq 0.9$~\citep{Sureshkumar_Beris_POF1997,Xi_Graham_JFM2010}.
\subparagraph{Weissenberg number $\mathrm{Wi}$} defined as the product of polymer relaxation time $\lambda_H$
and a characteristic shear rate of the flow $\dot\gamma$
	\begin{gather}
		\mathrm{Wi}\equiv\lambda_H\dot\gamma
		\label{eq:Wi}
	\end{gather}
	in which, for elastic dumbbell models\RevisedText{~\citep{Bird_Curtis_1987}},
	\begin{gather}
		\lambda_H=\frac{\zeta}{4H_\text{s}}
		\label{eq:lambdaH}
	\end{gather}
	and $\zeta$ is the friction coefficient of a single bead in the dumbbell: i.e.,
	\RevisedText{each bead experiences a viscous drag force from the surrounding solvent}
	\begin{gather}
		\mbf f_\text{drag}=-\zeta\left(\mbf v_\text{b}-\mbf v(\mbf r_\text{b})\right)
	\end{gather}
	\RevisedText{that is proportional (and opposite) to its relative velocity---defined as the difference between bead velocity $\mbf v_\text{b}$ and}
	the solvent velocity at the bead location \RevisedText{$\mbf v(\mbf r_\text{b})$.}
	Note that $\dot\gamma^{-1}$ has the unit of time: \cref{eq:Wi} can be interpreted as
	\begin{gather}
	\begin{split}
		\mathrm{Wi}=&\frac{\lambda_H}{\dot\gamma^{-1}}\\
			=&\frac{\text{time scale of polymer chain relaxation}}{\text{time scale of chain deformation by the flow}}\RevisedText{.}
	\end{split}
	\end{gather}
	\RevisedText{H}igher $\mathrm{Wi}$ indicates slower relaxation compared with the flow deformation scale, resulting \RevisedText{in} stronger ``memory effect'' in the fluid and thus stronger elasticity. A purely viscous fluid would respond \RevisedText{instantly} to flow deformation and thus has $\mathrm{Wi}=0$.
	In DNS, the mean shear rate measured at the wall $\dot\gamma_\text{w}$ is the most \RevisedText{common} choice for the characteristic shear rate.
	With the advancement of numerical methods, $\mathrm{Wi}$ up to $O(100)$ \RevisedText{has} been reported in latest studies~\citep{Li_Khomami_PRE2015,Sid_Terrapon_PRFluids2018}.
\subparagraph{Finite extensibility parameter $b$} sometimes also denoted as $L^2$,
		\begin{gather}
			b\equiv\frac{H_\text{s}}{k_\text{B}T}Q_\text{max}^2
			\label{eq:b}
		\end{gather}
		enforces the finite extensibility of the FENE dumbbells\RevisedText{---}comparing \RevisedText{\cref{eq:b,eq:tralpha}}, it is clear that
		\begin{gather}
			\mathrm{tr}(\alpha)\leq b.
		\end{gather}
		On the other hand, since the equilibrium solution of \RevisedText{the} FENE-P equation is 
		\begin{gather}
			\mbf\alpha_\text{eq} = \left(\frac{b}{b+5}\right)\mbf\delta
		\end{gather}
		 where $\mbf\delta$ is the identity tensor\RevisedText{---}i.e.,
		\begin{gather}
			\delta_{ij}=
			\begin{cases}
				1	&(i=j)\\
				0	&(i\neq j)\\
			\end{cases},
		\end{gather}
		the mean square end-to-end distance at equilibrium is
		\begin{gather}
			\langle Q_\text{eq}^2\rangle
				=\frac{k_\text{B}T}{H_\text{s}}\mathrm{tr}(\alpha_\text{eq})
				=\frac{k_\text{B}T}{H_\text{s}}\frac{3b}{b+5}
				\approx\frac{3k_\text{B}T}{H_\text{s}}
			\label{eq:Qeqsq}
		\end{gather}
		-- the last ``$\approx$'' relation is because for flexible polymers $b\gg O(1)$.
		Comparing \cref{eq:Qeqsq} and \cref{eq:b}, one can write
		\begin{gather}
			\frac{b}{3}\approx\frac{Q_\text{max}^2}{\langle Q_\text{eq}^2\rangle}
			\label{eq:binterpret}
		\end{gather}
		-- i.e., $\sqrt{b/3}$ is the ratio between the polymer extension when fully stretched and that at equilibrium.
		Earlier \RevisedText{DNS} studies have used $b$ as low as $100$ but $b=O(10^3)-O(10^4)$ are commonly seen in the later literature~\citep{Sureshkumar_Beris_POF1997,Xi_Graham_JFM2010,Li_Khomami_PRE2015,Lopez_Hof_JFM2019}.

\paragraph{Connection with experimental parameters}\label{par:feneexp}
Let us now examine the relationship between model and experimental parameters for polymer solutions. (\RevisedText{F}low parameters are straightforward and thus not discussed.)
Note that in the definition of $\mathrm{Wi}$ (\cref{eq:Wi}), $\dot\gamma$ is a flow parameter and only $\lambda_H$ depends on the polymer solution used.
The following discussion \RevisedText{covers the} effects of polymer solution properties\RevisedText{---}polymer and solvent species, polymer concentration, and molecular weight\RevisedText{---}on $\beta$, $\lambda_H$, and $b$.

\subparagraph{Effects of polymer concentration}
\mbox{}\RevisedText{For a dilute solution in the strict sense (i.e., no inter-chain interactions), p}olymer concentration affects the governing equations only through the $\beta$ parameter.
Solving FENE-P for a simple shear flow, one can get the polymer contribution to viscosity at the $\dot\gamma\to0$ limit
\begin{gather}
	\eta_\text{p}=C_\text{p}\frac{\mathcal{N}_\text{Av}k_\text{B}T\lambda_H}{M_\text{w}}
		\left(\frac{b}{b+5}\right)
\end{gather}
($M_\text{w}$ is its molar mass and $\mathcal{N}_\text{Av}$ is the Avogadro constant)\RevisedText{. O}nce the polymer and solvent species as well as polymer $M_\text{w}$ are \RevisedText{determined} (i.e., $\lambda_H$, $b$, and $M_\text{w}$ are fixed)\RevisedText{, $\eta_p$ is proportional to polymer concentration $C_\text{p}$}.
Since
\begin{gather}
	1-\beta=\frac{\eta_\text{p}}{\eta_\text{s}+\eta_\text{p}},
\end{gather}
\RevisedText{and} for a very dilute solution ($C_\text{p}\to 0$ limit), $\eta_\text{p}\ll\eta_\text{s}$, \RevisedText{we get} $1-\beta\approx\eta_\text{p}/\eta_\text{s}\propto C_\text{p}$.
\RevisedText{Meanwhile, as discussed in section~\ref{par:fenepbg}, recent evidences suggested that at concentrations relevant to DR systems, inter-chain interactions may not be negligible at all.
In this case, concentration would also affect $\lambda_H$: higher $C_\text{p}$ \textrightarrow\mbox{} higher $\lambda_H$.}

\subparagraph{Effects of polymer molecular weight}
Effects of $M_\text{w}$ or, more accurately, chain length are illustrated with a scaling argument \RevisedText{for polymer conformation}~\citep{Rubinstein_Colby_2003,Graham_2018}.
\RevisedText{Without getting into the full complexity of solvent effects, the discussion here will be limited to the \texttheta-solvent (ideal chain) condition as a simple demonstration.
Effects of changing solvent conditions will be briefly discussed below but only at a qualitative level.
Also, the analysis here assumes true diluteness with no inter-chain interactions.}

As sketched in \cref{fig:chain}(a), if we \RevisedText{move along} the contour of a flexible chain over a sufficiently long distance or arc length (\RevisedText{long compared with} the persistence length of the chain), in the absence of inter-segment\RevisedText{al} interactions (ideal-chain \RevisedText{assumption}), \RevisedText{the} orientation of the local segment would be decorrelated from that of the starting point.
It is thus always possible to map the chain into a random walk or freely-jointed chain (FJC) model
\RevisedText{in which each step covers a sufficiently large number of repeating units that} directions \RevisedText{of} successive steps \RevisedText{are} uncorrelated\RevisedText{. The} step size $L_\text{K}$ \RevisedText{of this FJC is} commonly termed the ``Kuhn length''\RevisedText{.}
\RevisedText{For ideal chains, there is no energetic effect associated with changing chain conformation. Random coils are preferred} at equilibrium solely because \RevisedText{the conformational entropy would be lower with increased chain extension.}
\RevisedText{It can be shown that the ``}entropic force\RevisedText{''} pulling the chain ends \RevisedText{together is} equivalent to an elastic force with spring constant
\begin{gather}
	H_\text{s}=\frac{3k_\text{B}T}{N_\text{K}L_\text{K}^2},
	\label{eq:Hsscaling}
\end{gather}
which is the rationale for the elastic dumbbell model (\cref{fig:chain}(b)).

For a given polymer chemical species, the number of Kuhn steps $N_\text{K}\propto M_\text{w}$.
Note that $\lambda_H$ is proportional to the ratio between $\zeta$ and $H_\text{s}$ (\cref{eq:lambdaH}). Other than $H_\text{s}$, $\zeta$ also depends on $N_k$, which follows the Zimm scaling
\begin{gather}
	\zeta\sim\eta_\text{s}L_\text{K}N_\text{K}^\frac{1}{2}
	\label{eq:zetascaling}
\end{gather}
in a dilute solution with a \texttheta-solvent\RevisedText{.} (\RevisedText{The} ``$\sim$'' \RevisedText{sign} indicates that the two sides differ only by an $O(1)$ constant prefactor\RevisedText{.)}
Combining \cref{eq:Hsscaling,eq:zetascaling} with \cref{eq:lambdaH},
\begin{gather}
	\lambda_H\sim\frac{\eta_\text{s}}{k_\text{B}T}L_\text{K}^3N_\text{K}^\frac{3}{2}.
\end{gather}
For \RevisedText{the} $b$ parameter, since
\begin{gather}
	Q_\text{max}=N_\text{K}L_\text{K}
	\label{eq:QmaxRW}
\end{gather}
and, in a \texttheta-solvent,
\begin{gather}
	\langle Q_\text{eq}^2\rangle=N_\text{K}L_\text{K}^2,
	\label{eq:Qeq}
\end{gather}
from \cref{eq:binterpret}, we have
\begin{gather}
	b\sim N_\text{K}.
\end{gather}
Experimentally, parameters for the FENE-P model are obtained through fitting with rheological measurements and both scalings of $\lambda_{H}\propto M_\text{w}^{3/2}$ and $b\propto M_\text{w}$ have been observed \RevisedText{under certain conditions}~\citep{Anna_McKinley_JRh2001,Arratia_Gollub_NJPhys2009}.


\subparagraph{Effects of the solvent species} The solvent affects the FENE-P model in two aspects. The first is the solvent-polymer interaction which \RevisedText{influences} chain conformation.
For instance, \RevisedText{compared with the \texttheta-solvent condition discussed above,} a good solvent allows polymer molecules to expand at equilibrium, which, \RevisedText{according to the Flory theory~\citep{Rubinstein_Colby_2003}}, changes the scaling of \cref{eq:Qeq} to
\begin{gather}
	\langle Q_\text{eq, g.s.}^2\rangle\sim N_\text{K}^\frac{6}{5}L_\text{K}^2,
	\label{eq:Qeqgs}
\end{gather}
and thus the $Q_\text{max}^2/\langle Q_\text{eq}^2\rangle$ ratio must be reduced\RevisedText{---}so does $b$.
\RevisedText{Expansion of the polymer coil also changes} the elastic force and thus $\lambda_H$.
The second is solvent viscosity $\eta_\text{s}$ which directly controls the friction coefficient $\zeta$ (\cref{eq:zetascaling}) and thus $\lambda_H$.

\subparagraph{Effects of the polymer species} This is, of course, also included in the solvent-polymer interaction factor discussed above. In addition, changing chain mechanics directly affects the Kuhn length $L_\text{K}$.
Increasing chain rigidity means more backbone carbon atoms \RevisedText{must be represented by} each Kuhn segment (higher $L_\text{K}$). \RevisedText{W}hen compared at the same contour length $Q_\text{max}$ (\cref{eq:QmaxRW}), it means the total number of Kuhn segments $N_\text{K}$ must drop. The combined results are higher $\lambda_H$ and lower $b$.
This is, of course, assuming that the chain is still flexible enough to be modeled by FENE-P.

\paragraph{Artificial diffusion (AD) in DNS}\label{par:ad}
Constitutive models for viscoelastic polymer fluids, such as FENE-P, do not have a diffusion term. Such purely convective \RevisedText{partial differential equations} are prone to numerical instability at high $\mathrm{Wi}$.
\RevisedText{If}
pseudo-spectral methods
\RevisedText{are used for DNS}, the only viable option for \RevisedText{numerically stable solutions} is to artificially introduce a diffusion term\RevisedText{. I}n the case of FENE-P, the term 
\begin{gather}
	\frac{1}{\mathrm{Sc}\mathrm{Re}}\nabla^2\mbf\alpha
\end{gather}
is added to the RHS of the $\partial\mbf\alpha/\partial t$ dynamical equation\RevisedText{. The Schmidt number} $\mathrm{Sc}$ is defined as the ratio of \RevisedText{kinematic} viscosity $\eta/\rho$ to polymer stress diffusivity\RevisedText{---the} artificial diffusivity of the transport of \RevisedText{the} $\mbf\alpha$ tensor. Lower $\mathrm{Sc}$ corresponds to higher AD.

The practice of applying AD to viscoelastic DNS was first introduced by \citet{Sureshkumar_Beris_JNNFM1995} who made the case that the numerical solution would converge to the \RevisedText{accurate} solution with mesh refinement\RevisedText{,} if the magnitude of AD is kept small and scales with the mesh size\RevisedText{:} $1/\mathrm{Sc}\sim\delta x^2/\delta t$ as $\delta x\sim\delta t\to0$, where $\delta x$ and $\delta t$ are the characteristic mesh size and time step, respectively~\citep{Sureshkumar_Beris_POF1997,Dimitropoulos_Beris_JNNFM1998}.
\RevisedText{A} few finite-difference methods \RevisedText{developed later} can minimize and, in some cases, completely avoid \RevisedText{the need of} AD~\citep{Min_Choi_JNNFM2001,Yu_Kawaguchi_JNNFM2004,Dubief_Lele_FTC2005,Dallas_Vassilicos_PRE2010,Agarwal_Zaki_JFM2014,Zhu_Xi_JNNFM2019}.
Application of AD was prevalent in earlier years of DNS research and is still widespread among researchers.
For the most part, the results represent the physical system reasonably well if the choice of $\mathrm{Sc}$ is carefully tested\RevisedText{. I}ndeed, many meaningful insights were extracted from DNS studies using AD.
The reader is referred to \citet{Min_Choi_JNNFM2001} and \citet{Yu_Kawaguchi_JNNFM2004} (and to some extent, \citet{Dubief_Lele_FTC2005}) for detailed comparative studies on the effects of AD.

\RevisedText{T}he topic is brought up here because of the recent discovery of flow states where flow instabilities (not to be confused with numerical instabilities) are driven\RevisedText{, at least in part,} by polymer elasticity \RevisedText{(}which \RevisedText{will be the focus of} \cref{sec:edt}\RevisedText{)}.
Numerical schemes using AD are known to have difficulty with this particular class of flow states\RevisedText{,} because AD smears the \RevisedText{polymer stress field} at regions with steep stress gradients, which are crucial for those elastic flow instabilities~\citep{Gupta_Vincenzi_JFM2019}.
\RevisedText{For channel flow at $\mathrm{Re}_\tau=84.95$,}
\citet{Sid_Terrapon_PRFluids2018} reported that $\mathrm{Sc}<O(10)$ (typical pseudo-spectral methods require $\mathrm{Sc}\leq O(0.1)$ for stability) would completely \RevisedText{eradicate} those flow instabilities in the solution and even higher $\mathrm{Sc}$ affects its accuracy.
However,
\RevisedText{at a comparable $\mathrm{Re}_\tau$,}
\citet{Lopez_Hof_JFM2019} were able to capture those flow instabilities with $\mathrm{Sc}=0.5$ but only at a very high \RevisedText{$b=40000$}.
Since AD is not a physically meaningful quantity, in this review, it will only be mentioned when the \RevisedText{physical} interpretation of results is expected to be affected.

\subsubsection{Vortex identification}\label{sec:vortexid}
Vortex is an instrumental concept for \RevisedText{understanding} turbulent dynamics.
A vortex identification criterion would turn a fully-resolved 3D \RevisedText{instantaneous} velocity field $\mbf v(x,y,z)$ into a quantifiable and visualizable measure of vortex strength or intensity \RevisedText{as well as its spatial distribution}.
This topic is not directly relevant to DR, but it is necessary to understand many flow visualization images from DNS.

At first glance, additional vortex identification criteria seem redundant as one would intuitively resort to the vorticity field
\begin{gather}
	\omega\equiv\mbf\nabla\times\mbf v
\end{gather}
in which a streamwise ($x$-aligned) vortex will show as a region with large $\omega_x$ magnitude.
The necessity for quantitative criteria beyond vorticity becomes clear in the example of \RevisedText{a} simple shear flow: $v_x=\dot\gamma y$ and $v_y=v_z=0$, where $\omega_z=-\dot\gamma$ (proportional to shear rate) even though there is no vortex at all.
\RevisedText{A vortex identification criterion must effectively} differentiate swirling and shear flow motions.

The topic of vortex identification \RevisedText{has been} widely studied. Here\RevisedText{,} one of the most widely used method, the $Q$-criterion\RevisedText{~\citep{Hunt_Wray_CTR1988}}, \RevisedText{is used} as an illustrative example.
From the velocity field, the rate of strain
\begin{gather}
	\mbf S\equiv\frac{1}{2}\left(\mbf\nabla v+\mbf\nabla v^\text{T}\right)
\end{gather}
and vorticity tensors
\begin{gather}
	\mbf\Omega\equiv\frac{1}{2}\left(\mbf\nabla v-\mbf\nabla v^\text{T}\right)
\end{gather}
can be calculated and the scalar identifier is defined as
\begin{gather}
	Q\equiv\frac{1}{2}\left(\Vert\mbf\Omega\Vert^2-\Vert\mbf S\Vert^2\right)
	\label{eq:q}
\end{gather}
where $\Vert\cdot\Vert$ is the Frobenius tensor norm: e.g., $\Vert\mbf\Omega\Vert\equiv\sqrt{\sum_i\sum_j\Omega_{ij}^2}$.
In the simplest interpretation, \cref{eq:q} measures the difference between the magnitudes of fluid rotation $\Vert\mbf\Omega\Vert^2$ and strain $\Vert\mbf S\Vert^2$.
The sign of $Q$ reflects the local flow type\RevisedText{---}i.e.,
\begin{gather}
	Q
	\begin{dcases*}
		>0	&rotation	\\
		=0	&shear		\\
		<0	&extension
	\end{dcases*}
\end{gather}
-- and the magnitude measures \RevisedText{the} strength of \RevisedText{rotational or extensional} motion.
Vortices are defined as regions dominated by strong \RevisedText{rotation} with $Q>Q_\text{threshold}$.
In flow field visualization, it is common to use the isosurfaces of $Q=Q_\text{threshold}$ as a graphical representation of vortices.
The choice of the threshold value \RevisedText{$Q_\text{threshold}$} depends on the flow field and is somewhat arbitrary.
\RevisedText{An approach for its} systematic determination has been recently used \RevisedText{in} \citet{Zhu_Xi_JNNFM2018} and \citet{Zhu_Xi_JFM2019}\RevisedText{, which} was adapted from a similar approach for flow structure analysis by \citet{LozanoDuran_Jimenez_JFM2012}.

There are quite a few other options for vortex identification. Nearly all of them, like the $Q$-criterion, turn the $\mbf v$ field into a scalar field that indicates the flow type and strength.
For instance, the $\lambda_2$-criterion by \citet{Jeong_Hussain_JFM1995} calculates the $\lambda_2$ quantity from the velocity gradient tensor, in which $\lambda_2<0$ corresponds to rotational flow regions.
These criteria differ mathematically but in complex turbulent flow fields, the results are practically equivalent. The reader is referred to \citet{Chakraborty_Adrian_JFM2005} and \citet{Chen_Qi_POF2015} for \RevisedText{detailed} comparison.

\subsection{Phenomenology: different stages of DR}\label{sec:phenomenology}
\begin{figure}
	\centering				
 	\includegraphics[width=\linewidth, trim=0mm 0mm 0mm 0mm, clip]{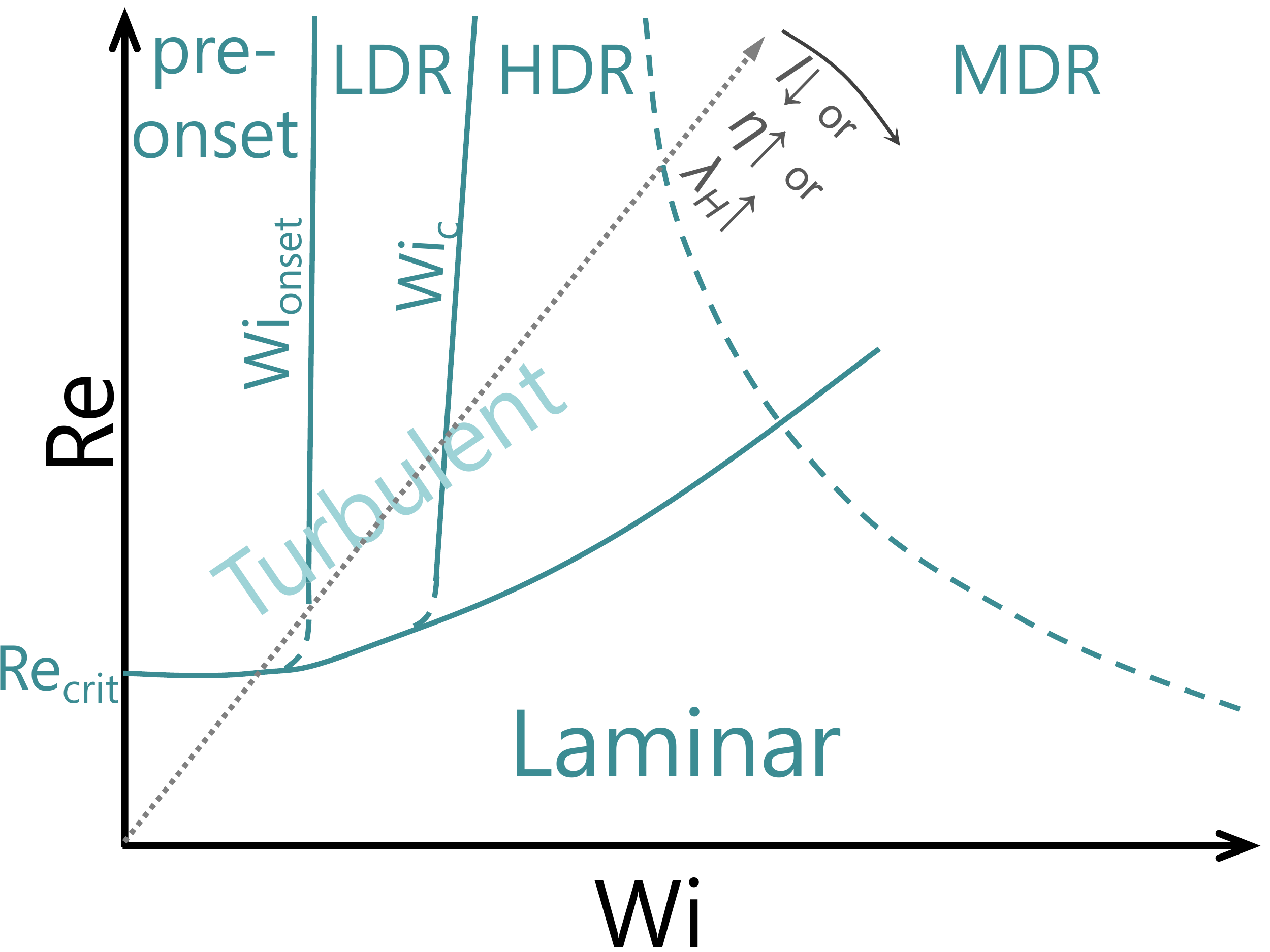}
	\caption{Different regimes of drag reduction behaviors in a $\mathrm{Re}$-$\mathrm{Wi}$ parameter space (i.e., fixed $\beta$ and $b$\RevisedText{---}same polymer solution and different flow conditions).
	Regime boundaries are sketched to reflect numerical and experimental findings of \citet{Zhu_Xi_JNNFM2018,Li_Graham_JFM2006} and \citet{Choueiri_Hof_PRL2018}. MDR boundary and those of other regimes near the L-T transition are less understood and shown in dashed lines.
	An experiment of increasing flow rate in a fixed flow apparatus with a given polymer solution will follow a line of constant $\mathrm{Re}/\mathrm{Wi}$ (dotted); decreasing $l$ or increasing $\eta$ or $\lambda_H$ decreases the slope of these experimental paths (see \cref{eq:expath}).}
	\label{fig:trans}
\end{figure}

The framework of DR phenomenology, in terms of
\RevisedText{the transitions between different flow regimes based on}
flow statistics, had been mostly established by the time \citet{Virk_AIChEJ1975} wrote his influential review\RevisedText{.}
\RevisedText{Meanwhile,} since the late 1990s,
\RevisedText{direct access to} turbulent flow structures and polymer conformation field \RevisedText{enabled} by PIV and DNS
\RevisedText{has fueled a wave of new observations, which greatly deepened our understanding of the dynamics within each regime.}
One key recent addition to the framework was the differentiation between low- and high-extent DR (LDR and HDR) first brought to light by the experiments of \citet{Warholic_Hanratty_JFM1999}. LDR and HDR were later shown to be two distinct flow stages driven by different DR mechanisms~\citep{Zhu_Xi_JNNFM2018} (see section~\ref{par:hdr} \RevisedText{below}).

An overview of different regimes of DR behaviors in the $\mathrm{Re}$-$\mathrm{Wi}$ parameter space is sketched in \cref{fig:trans}.
Numerical simulations typically explore the parameter space along horizontal lines\RevisedText{---}i.e., increasing $\mathrm{Wi}$ with fixed $\mathrm{Re}$, during which a series of transitions are typically observed, including the onset of DR, LDR-HDR transition, and convergence to maximum drag reduction (MDR).
In relation to MDR, LDR and HDR are collectively called intermediate DR\RevisedText{,} which was indeed \RevisedText{viewed as} one homogeneous stage \RevisedText{until} \citet{Warholic_Hanratty_JFM1999}\RevisedText{. W}e will call the regime before the onset the pre-onset stage.

Experiments performed in the same flow apparatus and with the same polymer solution would see both $\mathrm{Re}$ and $\mathrm{Wi}$ increasing with flow rate (\cref{eq:Re,eq:Wi}), but their ratio
\begin{gather}
	\frac{\mathrm{Re}}{\mathrm{Wi}}
		=\left(\frac{\rho}{\eta\lambda_H}\right)\left(\frac{Ul}{\dot\gamma}\right)
		\xlongequal[]{\dot\gamma=U/l}\frac{\rho l^2}{\eta\lambda_H}
	\label{eq:expath}
\end{gather}
would be constant.
\RevisedText{(Here,} $\dot\gamma=U/l$ \RevisedText{is taken} as the characteristic shear rate \RevisedText{without loss of generality: for the same flow type, any choice of $\dot\gamma\propto U/l$.)}
In \cref{fig:trans}, such data points would follow an inclined line with zero-intercept (whose slope depends on polymer solution properties and flow geometric size), which is termed an experimental path by \citet{Li_Graham_JFM2006} and \citet{Li_Graham_POF2007}.

\subsubsection{Onset of DR}
DR does not occur immediately upon introducing polymers. Rather, at sufficiently low fluid elasticity (low $\mathrm{Wi}$), flow statistics are indistinguishable from those of the Newtonian benchmark.
For given $\mathrm{Re}$, there is a critical \RevisedText{onset} $\mathrm{Wi}$\RevisedText{,} hereinafter denoted by $\mathrm{Wi}_\text{onset}$\RevisedText{,} above which DR becomes discernible.
Existence of this threshold is not surprising, considering that at the molecular level, polymer chains in a dilute solution are coiled at equilibrium and they unwind rather abruptly under increasing velocity gradient\RevisedText{---}the so-called ``coil-stretch'' (C-S) transition~\citep{DeGennes_JCP1974}.
In the case of FENE dumbbells in a uniaxial extensional flow, the C-S transition occurs when the Weissenberg number
\begin{gather}
	\mathrm{Wi}^\text{ext}\equiv\lambda_H\dot\epsilon
	\label{eq:Wiext}
\end{gather}
reaches $1/2$ \RevisedText{($\dot\epsilon$ is}
the extension rate\RevisedText{)}%
~\citep{Bird_Curtis_1987}.

The rheological consequence of the C-S transition is substantial, including, e.g., a drastic increase in \RevisedText{the} extensional viscosity \RevisedText{of the solution}. It is thus expected that for DR\RevisedText{,} $\mathrm{Wi}_\text{onset}=O(1)$.
\RevisedText{For $\mathrm{Wi}$ defined based on the wall shear rate $\dot\gamma_\text{w}$,}
the onset is observed in the range of $5\lesssim\mathrm{Wi}_\text{onset}\lesssim10$
\RevisedText{in DNS,}
with small variations between different $\mathrm{Re}$\RevisedText{~\citep{Min_Choi_JFM2003a,Housiadas_Beris_POF2003,Li_Khomami_PRE2015,Xi_Graham_JFM2010,Zhu_Xi_JNNFM2018}} and, possibly, different numerical settings\RevisedText{~\citep{Sureshkumar_Beris_POF1997,Dimitropoulos_Beris_JNNFM1998,Housiadas_Beris_POF2003}}.
The number is higher than the expected $O(1)$ magnitude because $\dot\gamma_\text{w}^{-1}$ is not the time scale directly associated with DR. Indeed, dynamics within the viscous sub-layer\RevisedText{, which $\dot\gamma_\text{w}$ directly measures,} is inconsequential as far as polymer-induced DR is concerned~\citep{Virk_Merrill_JFM1967,Donohue_Tiederman_JFM1972}.
Should we have a full grasp of the \RevisedText{complex} polymer-turbulence \RevisedText{dynamics}, a time scale of turbulent motion most relevant to DR\RevisedText{,} $\tau_\text{flow}$\RevisedText{,} would be identified and the corresponding $\mathrm{Wi}=\lambda_H/\tau_\text{flow}$ \RevisedText{must} be $O(1)$. (Rigorously speaking, this should be called Deborah number $\mathrm{De}$\RevisedText{---}see the $\mathrm{De}$ vs. $\mathrm{Wi}$ discussion of \citet{Poole_RhBull2012}.)
This $\lambda_H/\tau_\text{flow}$, although differs from the \RevisedText{mean-flow definition} $\mathrm{Wi}=\lambda_H\dot\gamma_\text{w}$ by one order of magnitude (\RevisedText{estimated by comparing} their onset magnitudes), is expected to \RevisedText{be proportional} to the latter.

At least in \RevisedText{a region immediately after} the onset, DR is primarily contributed by polymer effects in the buffer layer~\citep{Virk_AIChEJ1975,White_Mungal_ARFM2008,Graham_RheologyReviews2004,Zhu_Xi_JNNFM2018}, where turbulence is dominated by streamwise vortices~\citep{Robinson_ARFM1991}.
\citet{Li_Khomami_PRE2015} noted that the root-mean-square (RMS) streamwise vorticity fluctuations
\begin{gather}
	\omega_{x,\text{rms}}'\equiv \sqrt{\langle \omega_x^{\prime 2}\rangle}
\end{gather}
(apostrophe denotes the fluctuating component, i.e.
\begin{gather}
	\mbf\omega=\langle\mbf\omega\rangle+\mbf\omega'
\end{gather}
) in the buffer layer of Newtonian turbulence is $\omega_{x,\text{rms}}'\RevisedText{=O(0.1)}\dot\gamma_\text{w}$.
\RevisedText{Thus,} $\mathrm{Wi}$ defined \RevisedText{as $\lambda_H\omega_{x,\text{rms}}$} would be smaller than $\lambda_H\dot\gamma_\text{w}$ by a factor of $O(10)$. The onset \RevisedText{threshold} in the latter definition, as noted above, is $5\sim 10$, this leads to $(\lambda_H\omega_{x,\text{rms}}')_\text{onset}=O(1)$.
Of \RevisedText{course}, \RevisedText{$\omega_{x,\text{rms}}^{\prime -1}$} is only one of many plausible choices for $\tau_\text{flow}$.
How to choose this time scale and\RevisedText{,} \RevisedText{ultimately}, how to predict the onset, depend on one's interpretation of the DR mechanism, which itself is up for debate. More detailed discussion is deferred to \cref{sec:mechanism}.

\subsubsection{Intermediate DR: LDR vs. HDR}\label{sec:intermediate}
\paragraph{Mean flow}
After the onset, DR increases with fluid elasticity (usually by increasing polymer concentration or molecular weight in experiments and increasing $\mathrm{Wi}$ \RevisedText{or $b$} in DNS).
\Cref{fig:Um} shows typical mean velocity profiles for various levels of $\mathrm{DR}\%$ in channel flow experiments performed at constant flow rate~\citep{Warholic_Hanratty_EXPFL1999}.
\RevisedText{With the bulk velocity $U_\text{avg}$ fixed,} $\mathrm{Re}_\tau$ decreases with $\mathrm{DR}\%$ from $\approx 1000$ at the Newtonian (water) limit to $\lesssim 200$ at highest DR levels.
For the Newtonian (water) case, a pronounced log-law relation closely following the von~K\'arm\'an asymptote (\cref{eq:loglaw}) is found at $y^+\gtrsim 30$.
With increasing polymer concentration, the profile is elevated\RevisedText{. S}ince the profiles are scaled by the friction velocity (\cref{eq:utau}), higher $U_\text{m}^+$ indicates lower $\tau_\text{w}$ and thus higher $\mathrm{DR}\%$.
Up to $\mathrm{DR}\%\approx 30\%$ (\cref{fig:Um}(a)), increase in $U_\text{m}^+$ is caused by its higher slope in the buffer layer which \RevisedText{also} thickens with $\mathrm{DR}\%$\RevisedText{.} Note that the lower limit of the log law layer increases from $y^+\approx 30$ to $\approx 95$ as $\mathrm{DR}\%$ rises to $33\%$.
The logarithmic relation itself still follows the same slope ($A^+=A^+_\text{Newt.}=2.5$) but \RevisedText{with} higher intercept\RevisedText{s} ($B^+$) \RevisedText{as a direct result of} the \RevisedText{velocity gain} in the buffer layer.
At \RevisedText{$\mathrm{DR}\%>35\%$} (\cref{fig:Um}(b)), the profile lifts up across the channel and the slope is substantially higher in regions  where the von K\'arm\'an slope used to dominate (in Newtonian and LDR cases).
Based on this clear difference in the shape of $U_\text{m}(y^+)$, \RevisedText{\citet{Warholic_Hanratty_EXPFL1999} divided DR} into two regimes, LDR and HDR, roughly around the $DR\%\approx30-35\RevisedText{\%}$ line\RevisedText{.}
\RevisedText{This distinction}
was later confirmed in a large number of experimental and DNS studies of different flow geometries~\citep{Warholic_Hanratty_EXPFL2001,Min_Choi_JFM2003b,Ptasinski_Nieuwstadt_JFM2003,Li_Khomami_JNNFM2006,Xi_Graham_JFM2010,Dallas_Vassilicos_PRE2010,Elbing_Perlin_POF2013,Li_Khomami_PRE2015,Mohammadtabar_Ghaemi_POF2017,Zhu_Xi_JNNFM2018,Shaban_Ghaemi_POF2018}.

\paragraph{Fluctuations}\label{par:fluc}
Fluctuating velocities in transverse (perpendicular to the mean flow\RevisedText{---i.e., $y$ and $z$}) directions decrease with increasing $\mathrm{DR}\%$ and so does the Reynolds shear stress (RSS)~\citep{Sureshkumar_Beris_POF1997,Warholic_Hanratty_EXPFL1999,Ptasinski_Nieuwstadt_JFM2003,Li_Khomami_JNNFM2006,Xi_Graham_JFM2010,Thais_Mompean_IJHFF2013,Mohammadtabar_Ghaemi_POF2017,Zhu_Xi_JNNFM2018,Shaban_Ghaemi_POF2018}:
\begin{gather}
	\tau_{\text{R},xy}^+\equiv-\langle v_x^{\prime +}v_y^{\prime +}\rangle.
	\label{eq:rss}
\end{gather}
This is consistent with an intuitive general picture that, with DR, turbulent fluctuations are suppressed and turbulent motions are weakened, leading to less momentum redistribution to transverse directions and more efficient streamwise momentum transport.

Significance of $\tau_{\text{R},xy}$ is evident once we write out the transport equation for turbulent kinetic energy (TKE)~\citep{Pope_2000,Min_Choi_JFM2003b,Zhu_Xi_JNNFM2019},
\begin{gather}
	\frac{\partial k}{\partial t}+\mbf\nabla\cdot\mbf T^k = \mathcal{P}^k-\epsilon_\text{v}^k-\epsilon_\text{p}^k
	\label{eq:tkebal}
\end{gather}
where
\begin{gather}
	k\equiv\frac{1}{2}\sum_{i=x,y,z}\langle v_i^{\prime 2}\rangle
\end{gather}
is the TKE \RevisedText{and} $\mbf T^k$ is the total \RevisedText{TKE} flux\RevisedText{.}
The production rate \RevisedText{of TKE}
\begin{gather}
	\mathcal{P}^k=-\langle v_x'v_y'\rangle\frac{dU_\text{m}}{dy}
		=\tau_{\text{R},xy}\frac{dU_\text{m}}{dy}
	\label{eq:tkeP}
\end{gather}
is generally positive and correlates directly with RSS\RevisedText{---}reduction of $\tau_{\text{R},xy}$ thus suppresses turbulence generation\RevisedText{. The $\epsilon_\text{v}^k$ term} measures the rate at which TKE converts to heat via viscous dissipation. \RevisedText{It is always positive (}$-\epsilon_\text{v}^k$ is negative\RevisedText{--- i.e., net loss of TKE), which is a reflection} of the second law of thermodynamics\RevisedText{. Finally, $\epsilon_\text{p}^k$} is the rate \RevisedText{of} TKE conversion to the elastic energy stored in stretched polymer chains\RevisedText{,} which, in theory, can be either positive \RevisedText{(loss of TKE)} or negative \RevisedText{(gain of TKE)}.

\begin{figure}
	\centering
 	\includegraphics[width=\linewidth, trim=0 0 0 0, clip]{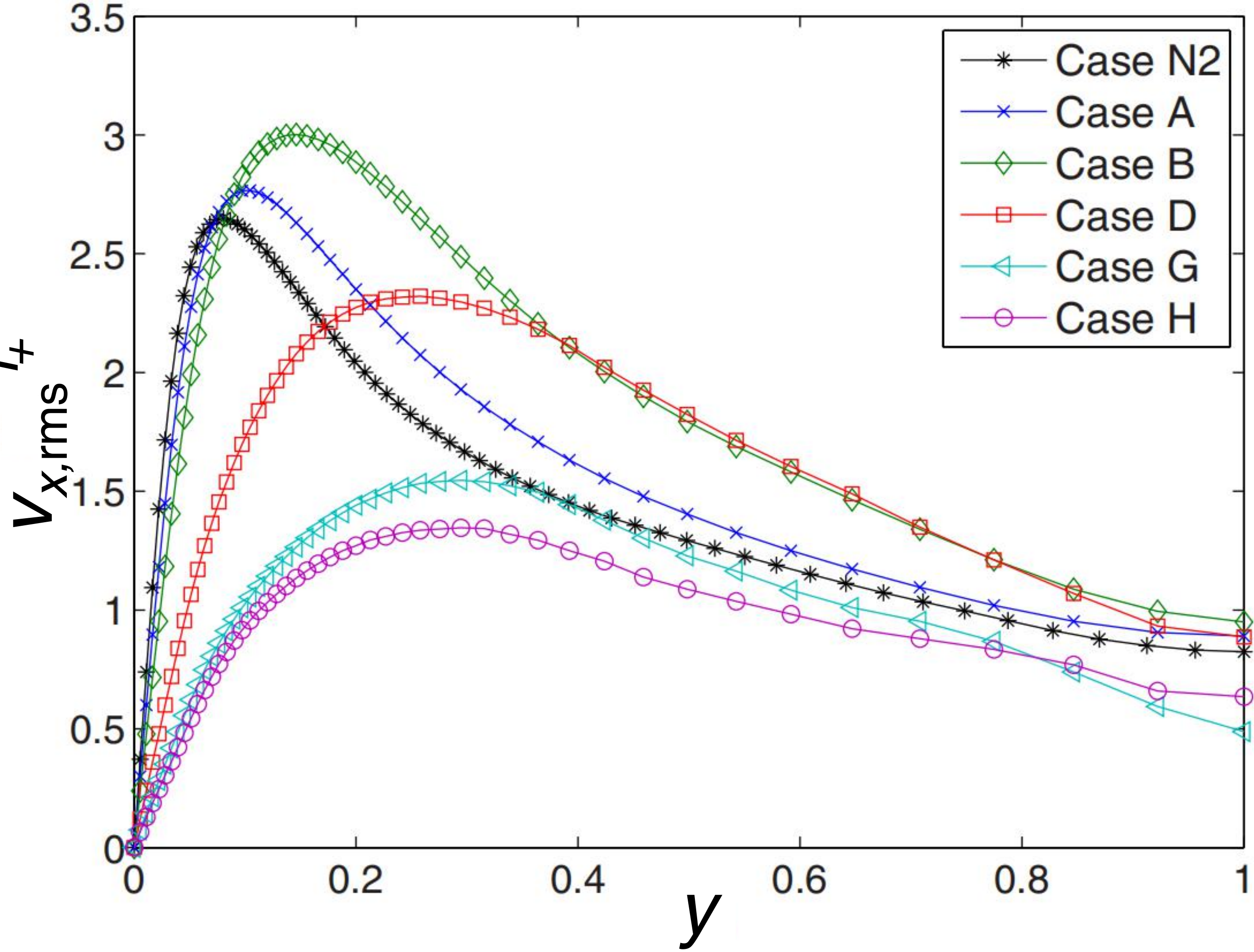}\\
	\caption{\RevisedText{%
	RMS streamwise velocity fluctuation $v_{x,\text{rms}}^{\prime +}$ profiles from DNS at $\mathrm{Re}=4250$ (based on bulk velocity), $\beta=0.9, b=14400$, and varying $\mathrm{Wi}$.
	Case N2 is Newtonian ($\mathrm{Re}_\tau=181$), A \& B are LDR, D \& G are HDR, and H is MDR ($\mathrm{Re}_\tau=107.8$).
	(Reprinted figure with permission from \citeauthor*{Dallas_Vassilicos_PRE2010}, Physical Review E, 82, 066303, \citeyear{Dallas_Vassilicos_PRE2010}. Copyright (\citeyear{Dallas_Vassilicos_PRE2010}) by the American Physical Society.)%
	}}
	\label{fig:vxrms}
\end{figure}

Observations about streamwise velocity fluctuations $v_{x,\text{rms}}'$
are more conflicted.
In the experimental $v_{x,\text{rms}}^{\prime +}(y^+)$ profiles of \citet{Warholic_Hanratty_EXPFL1999}, it was found that at LDR, with increasing $\mathrm{DR}\%$, $v_{x,\text{rms}}^{\prime +}$ decreases within the viscous sub-layer but increases moderately in buffer and log-law layers. The overall shape of the $v_{x,\text{rms}}^{\prime +}$ profile still resembles that of Newtonian flow, which has a sharp peak near the wall.
The peak position is at $y^+\approx 15$ for Newtonian turbulence and for LDR it gradually shifts away from the wall but still stays in the buffer layer.
At HDR, $v_{x,\text{rms}}^{\prime +}$ decreases with increasing $\mathrm{DR}\%$ across the entire flow domain. The overall profile takes a much flatter shape.
Similar decline of $v_{x,\text{rms}}^{\prime +}$ at HDR was found in DNS by \citet{Min_Choi_JFM2003b} and \citet{Dallas_Vassilicos_PRE2010} \RevisedText{(see \cref{fig:vxrms})}, but not in many other DNS studies~\citep{Ptasinski_Nieuwstadt_JFM2003,Housiadas_Beris_POF2003,Dubief_Lele_JFM2004,Li_Khomami_JNNFM2006,Xi_Graham_JFM2010,Thais_Mompean_IJHFF2013,Zhu_Xi_JNNFM2018}\RevisedText{---}rather, their $v_{x,\text{rms}}^{\prime +}$ continues with the same trend as LDR.
\citet{Dallas_Vassilicos_PRE2010} attributed the failure of the other studies in capturing the sharp drop of $v_{x,\text{rms}}^{\prime +}$ at HDR to the numerical artifact of using AD in the DNS (section~\ref{par:ad}).

Remarkably, two recent experimental reports from \RevisedText{the same} group (Ghaemi and co-workers)~\citep{Mohammadtabar_Ghaemi_POF2017,Shaban_Ghaemi_POF2018} using different polymers showed both types of behaviors.
For PAM, which is a flexible polymer, the behavior is consistent with that of \citet{Warholic_Hanratty_EXPFL1999} (and thus with \citet{Dallas_Vassilicos_PRE2010} and \citet{Min_Choi_JFM2003b}), but for XG, which is more rigid, the $v_{x,\text{rms}}^{\prime +}(y^+)$ profiles at HDR remain similar in shape as the Newtonian and LDR cases.
Similar $v_{x,\text{rms}}^{\prime +}(y^+)$ behaviors can also be seen in XG solutions measured by \citet{Escudier_Nickson_JNNFM2009}.
As discussed in section~\ref{par:ad}, one known spurious effect of AD is its suppression of flow instabilities (or, loosely speaking, ``turbulence'') that are driven, in part or in whole, by fluid elasticity~\citep{Sid_Terrapon_PRFluids2018,Gupta_Vincenzi_JFM2019}.
\RevisedText{(By contrast, turbulence, in the conventional sense,} is driven by fluid inertia and suppressed by polymers\RevisedText{---}see more detailed discussion in \cref{sec:edt}\RevisedText{.)}
In a way, DNS with AD can be viewed as a virtual experimental in which turbulent states that are elastic in nature are filtered. 
The lack of \RevisedText{non-monotonic} $v_{x,\text{rms}}^{\prime +}(y^+)$ behaviors \RevisedText{(as shown in \cref{fig:vxrms}) between LDR and HDR} in those \RevisedText{simulations}, as well as in rigid polymer experiments, suggests that the drastic decrease and flattening of $v_{x,\text{rms}}^{\prime +}(y^+)$ could be associated with those elastic turbulent states, which may appear at HDR.
Meanwhile, all other key features of HDR, including its characteristic behaviors of $U_\text{m}^+$ and $-\langle v_x^{\prime +}v_y^{\prime +}\rangle$ profiles as well as distinct flow structures \RevisedText{(shown below)}, are still captured, indicating that turbulence driven by elasticity is not a necessary condition for the transition to HDR.
Of course, this is so far only an educated guess\RevisedText{---}further research is needed for its validation.

Further complicating this issue, most DNS studies compared different $\mathrm{DR}\%$ cases at the same pressure drop (thus same $\tau_\text{w}$ and $\mathrm{Re}_\tau$), \RevisedText{whereas} experiments \RevisedText{such as \citet{Warholic_Hanratty_EXPFL1999}}, together with both aforementioned DNS cases reporting the $v_{x,\text{rms}}^{\prime +}(y^+)$ behaviors \RevisedText{of \cref{fig:vxrms}}~\citep{Min_Choi_JFM2003b,Dallas_Vassilicos_PRE2010}, compared cases at the same flow rate (thus decreasing $\mathrm{Re}_\tau$ with increasing $\mathrm{DR}\%$).
\RevisedText{This divide could also partially account for the differences in the observed $v_{x,\text{rms}}^{\prime +}$ magnitudes: e}ven for Newtonian flow, $v_{x,\text{rms}}^{\prime +}$ goes down with decreasing $\mathrm{Re}_\tau$~\citep{Moser_Kim_POF1999}.

\paragraph{Flow structures}\label{par:struct}
\begin{figure*}
	\centering
	\begin{minipage}[c][][c]{0.65\linewidth}
 	\includegraphics[width=\linewidth, trim=0 0 0 0, clip]{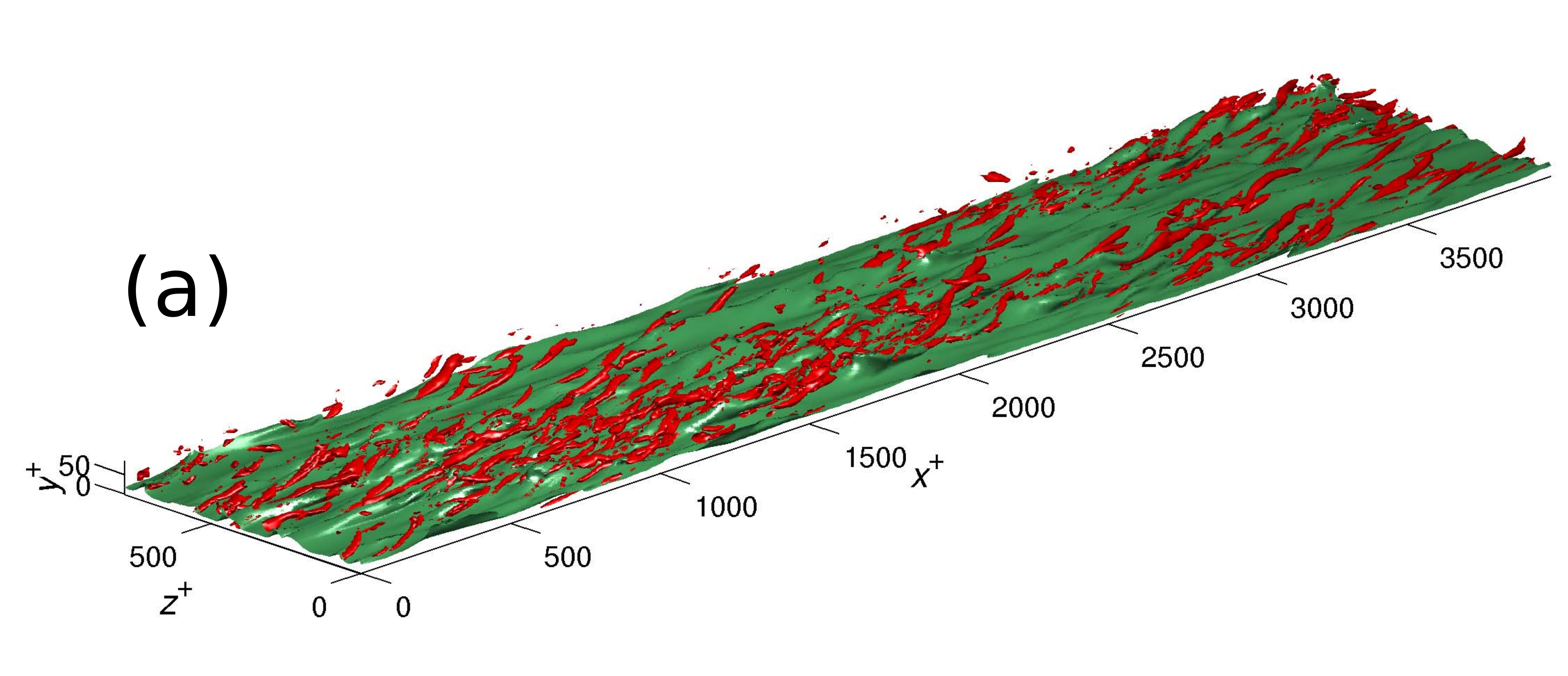}\\
 	\includegraphics[width=\linewidth, trim=0 0 0 0, clip]{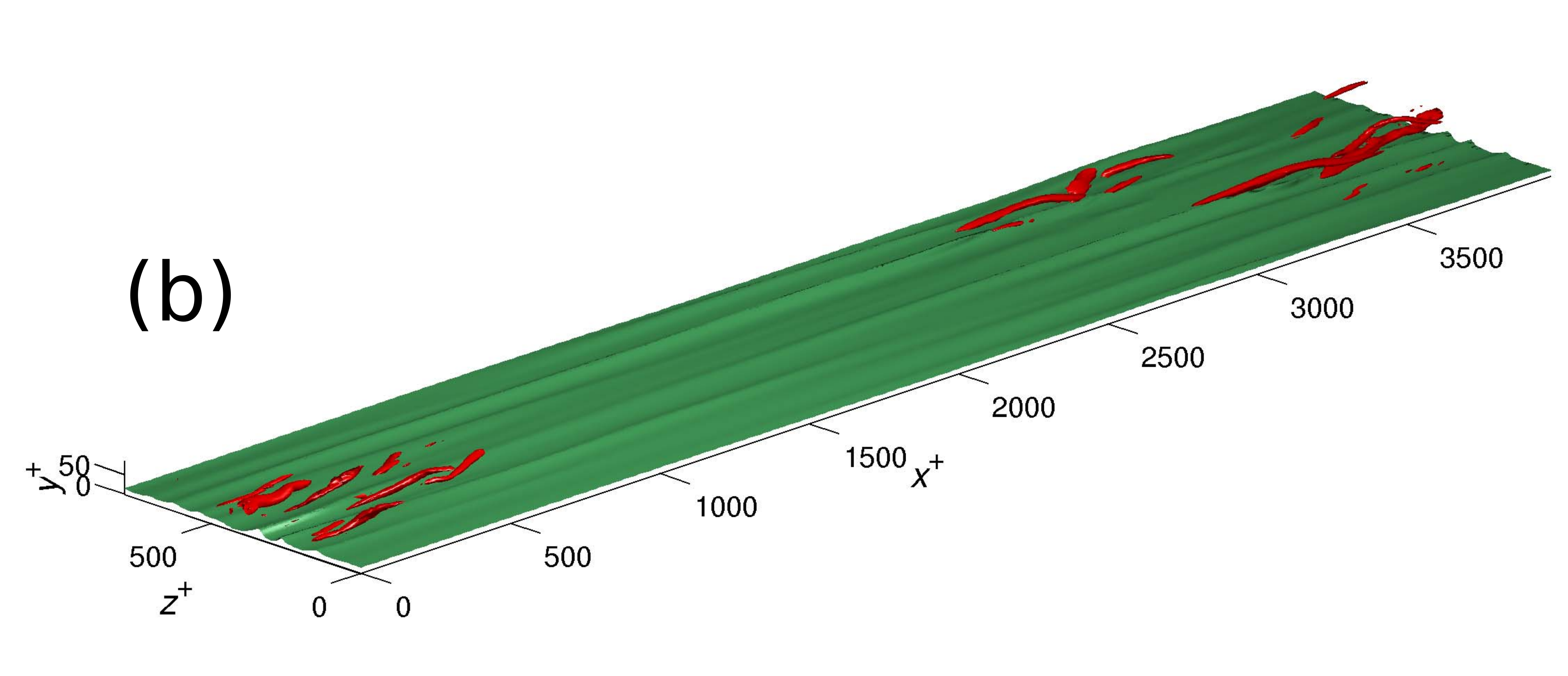}
	\end{minipage}%
	\begin{minipage}[c][][c]{0.35\linewidth}
	\caption{
	Flow field visualization from DNS for (a) Newtonian and (b) $\mathrm{Wi}=80$ and $\mathrm{DR}\%=57\%$ cases ($\mathrm{Re}_\tau=84.85$, $\beta=0.97$, and $b=5000$).
	Isosurfaces show constant $v_x$ (light/green) and $Q_\text{2D}$ (dark/red; $Q_\text{2D}$ is the 2D variant of $Q$ calculated in the $yz$-plane, which identifies $x$-aligned vortices\RevisedText{---}see \citet{Xi_PhD2009} for details). The same isosurface levels are used.
	Upward folds (or pleats) correspond to regions with ejection of slower near-wall fluids\RevisedText{---}i.e., low-speed streaks.
	(\RevisedText{Reproduced} with permission from \RevisedText{Xi, Ph.D. Dissertation, Univ.~Wisconsin-Madison, 2009}. Copyright (2009) by Li Xi.).
	}
	\label{fig:largebox}
	\end{minipage}
\end{figure*}
Near wall turbulence is populated and, in many ways, sustained by flow structures with well-recognizable patterns (\cref{fig:largebox}(a)). The existence and importance of \RevisedText{those} so-called ``coherent structures'' are well documented in the literature~\citep{Robinson_ARFM1991,Smith_Walker_PTRSLA1991,Jimenez_POF2013,Jimenez_JFM2018}.
The best-known conceptual model involves streamwise vortices and velocity streaks\RevisedText{. They are the characteristic structures in} the buffer layer \RevisedText{where} TKE production \RevisedText{is the highest}~\citep{Pope_2000,Moser_Kim_POF1999,Abe_Kawamura_JFluidsEng2001,Smits_McKeon_ARFM2011}.
\RevisedText{Those} vortices align in the direction of the mean flow. Neighboring vortices often rotate in opposite directions, which creates stripes of either slower fluid near the wall being washed up or faster fluid closer to the bulk flushed down\RevisedText{, forming low and high speed velocity streaks.}
(\RevisedText{Such events are also referred to as} ``ejections''  and ``sweeps''\RevisedText{,} respectively\RevisedText{,} in turbulence literature~\citep{Wallace_ARFM2016}\RevisedText{.})

\begin{figure}
	\centering
 	\includegraphics[width=\linewidth, trim=0 0 0 0, clip]{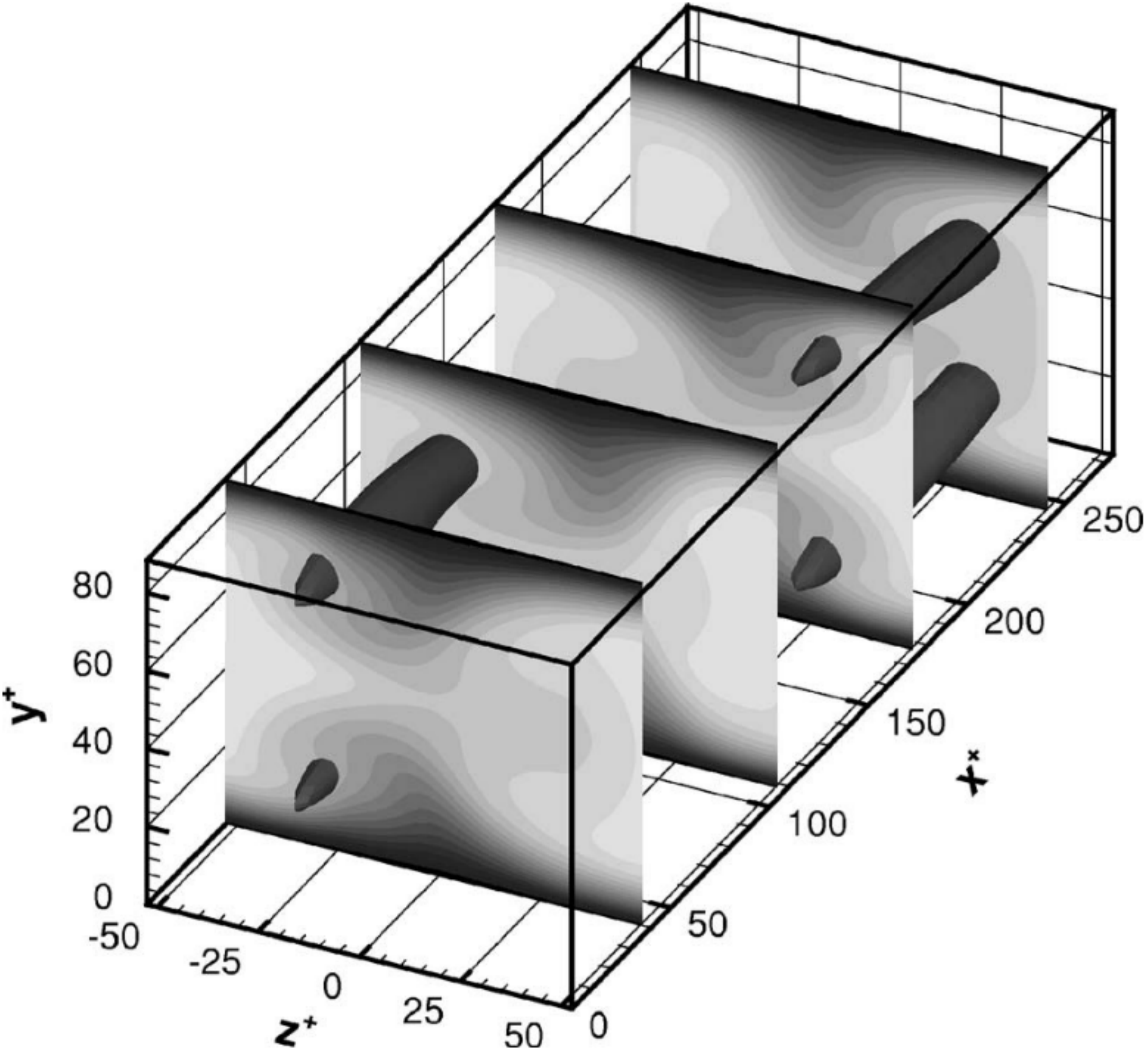}\\
	\caption{Exact coherent state (ECS) solution in a Newtonian channel flow at $\mathrm{Re}_\tau=44.2$. Flow fields at two sides of the center-plane are mirror images.
	For each side, the solution contains a pair of staggered counter-rotating streamwise vortices (dark tube-like isosurfaces of \RevisedText{$Q_\text{2D}$}).
	Contours on the slices show the distribution of streamwise velocity $v_x$ (light for high velocity).
	For the bottom side, the left and right vortices swirl counterclockwise and clockwise, respectively, which washes up the slower fluid near the wall to form a low-speed streak in between (an ``ejection'' event).
	(Reprinted from \RevisedText{\citeauthor*{Li_Graham_POF2007}, Physics of Fluids, 19, 083101, \citeyear{Li_Graham_POF2007}}, with the permission of AIP Publishing.)
	}
	\label{fig:ecs}
\end{figure}

Availability of invariant solutions to
the N-S equation enables the \textit{a priori} study of isolated coherent structures which are otherwise braided into the complexity of full scale turbulence in DNS~\citep{Nagata_JFM1990,Waleffe_PRL1998,Waleffe_JFM2001,Schneider_Gibson_PRL2010,Kawahara_ARFM2012}.
\RevisedText{They} are fully nonlinear solutions corresponding to the ``exact'' forms of specific coherent structures, also known as exact coherent states (ECSs).
The earliest-known form of ECS consists of a pair of staggered streamwise vortices separated by sinuous velocity streaks~\citep{Waleffe_JFM2001,Waleffe_JFM2001}, which corresponds directly to buffer layer structures.
(See \cref{fig:ecs}.
\RevisedText{Note that at $\mathrm{Re}_\tau=44.2$, the buffer-layer thickness is comparable to half channel height.})

Earlier attention on polymer DR effects \RevisedText{was} focused on the buffer layer\RevisedText{, which was thought to be the primary region for DR. It was later known that this presumption is only valid for LDR where increase in the $U_\text{m}^+(y^+)$ slope is limited to the buffer layer (\cref{fig:Um}).}
Indeed, until the \citet{Warholic_Hanratty_EXPFL1999} study, the whole intermediate DR stage was thought to follow the LDR-type behavior.
Based on this belief, \citet{Virk_JFM1971} proposed his well-known three-layer model, in which \RevisedText{DR} only occurs in the buffer layer and both the viscous sub-layer and log-law layer remain unaffected.
Polymers cause turbulence to be suppressed in the buffer layer, leading to \RevisedText{the enlargement of its thickness.}
Effects of polymers on buffer layer structures have thus been most extensively studied.

Earlier \RevisedText{dye} experiments were able to visualize velocity streak patterns. It was found that the average spacing between low-speed streaks increases with rising $\mathrm{DR}\%$, from 100 wall units in the Newtonian limit to over 200 wall units at high levels of DR~\citep{Donohue_Tiederman_JFM1972,Oldaker_Tiederman_POF1977}.
\RevisedText{In DNS,} the spanwise correlation length from velocity spatial autocorrelation functions was \RevisedText{also} found to increase with $\mathrm{DR}\%$%
~\citep{Sureshkumar_Beris_POF1997,DeAngelis_Piva_CompFl2002}.
Later application of PIV allowed more detailed view of velocity patterns, which showed that not only do the streaks grow wider, they, especially at HDR, also extend along the flow direction for much longer distances \RevisedText{without interruption} and their contours become smoother and not as rugged~\citep{Warholic_Hanratty_EXPFL2001,White_Mungal_EXPFL2004}.
This was again widely confirmed in DNS~\citep{Yu_Kawaguchi_JNNFM2004,Housiadas_Beris_POF2005,Li_Khomami_JNNFM2006,Xi_PhD2009,Zhu_Xi_JNNFM2018} (see \cref{fig:largebox}).
In particular, \citet{Li_Khomami_JNNFM2006} reported that the streamwise length scale of coherent structures can increase by over ten-fold between LDR and HDR.

For vortices, DR is accompanied by the weakening of their strength and reduction of their density
~\citep{Dubief_Lele_FTC2005,Li_Khomami_JNNFM2006,Xi_Graham_JFM2010,Xi_PhD2009,Zhu_Xi_JNNFM2018}.
The attenuated vortices dilate in their transverse scales, raising \RevisedText{the wall-normal positions of vortex axis lines, where} turbulent fluctuations \RevisedText{are strongest,} away from the wall\RevisedText{. This} is reflected in the outward shift of the peaks in $\tau_{\text{R},xy}$, velocity fluctuation\RevisedText{s}, and vorticity profiles~\citep{Warholic_Hanratty_EXPFL1999,Warholic_Hanratty_EXPFL2001,Min_Choi_JFM2003b,Xi_Graham_JFM2010,Dallas_Vassilicos_PRE2010,Mohammadtabar_Ghaemi_POF2017,Zhu_Xi_JNNFM2018,Shaban_Ghaemi_POF2018}.
Numerical investigation of viscoelastic ECS solutions rendered direct evidence that polymers weaken \RevisedText{those} coherent structures and eventually cause their extinction~\citep{Stone_Graham_PRL2002,Stone_Graham_POF2004,Li_Graham_JFM2006,Li_Graham_POF2007}. Not surprisingly, DR is also observed in those solutions and the observed flow statistics (mean and fluctuating velocities) are consistent with the LDR behaviors discussed above.

\RevisedText{The picture described so far targets the explanation of turbulence suppression and enlargement of the buffer layer, which is largely guided by the presumption of} the Virk three-layer model.
\RevisedText{The model, despite} its conceptual appeal \RevisedText{(}for its simplicity\RevisedText{)}, \RevisedText{is not} consistent with later findings.
An obvious miss is the HDR behaviors of the $U_\text{m}^+(y^+)$ profiles.
This suggests that
\RevisedText{structural insights summarized above are}
likely only accurate for LDR.
Physical understanding of HDR requires the study of coherent structures at higher $y^+$ which are, foreseeably, much more complex. Not much progress was made until fairly recently when new tools for the extraction and analysis of those structures have emerged\RevisedText{~\citep{Zhu_Xi_JFM2019,Zhu_Xi_POF2019}}. Details are discussed in \cref{sec:largebox}.

\paragraph{Two stages of DR \RevisedText{with two mechanisms}}\label{par:hdr}
\mbox{}\RevisedText{Despite the} common practice among researchers to use $\mathrm{DR}\%=30-40\%$ as an empirical threshold for the inception of HDR, sharp changes in flow statistics between LDR and HDR indicate that it is a qualitative transition between two stages of DR likely \RevisedText{underlain} by different mechanisms.
There is thus no reason for it to be tied with any particular magnitude of $\mathrm{DR}\%$.
In the absence of a presumed mechanism for the transition, the critical \RevisedText{$\mathrm{Wi}_\text{LDR-HDR}$,} and thus corresponding $\mathrm{DR}\%$, \RevisedText{of the transition} can generally depend on $\mathrm{Re}$, polymer solution properties, and flow geometry\RevisedText{. (This} opinion also \RevisedText{seems to be} shared by \citet{White_Dubief_JFM2018}.\RevisedText{)}
Indeed,
\RevisedText{the LDR-HDR transition was} observed in DNS at $\mathrm{DR}\%$ as low as $\approx 15\%$
at \RevisedText{a} low $\mathrm{Re}$ and \RevisedText{in} small flow domain\RevisedText{s}~\citep{Xi_Graham_JFM2010}.

\begin{figure}
	\centering
 	\includegraphics[width=\linewidth, trim=0 0 0 0, clip]{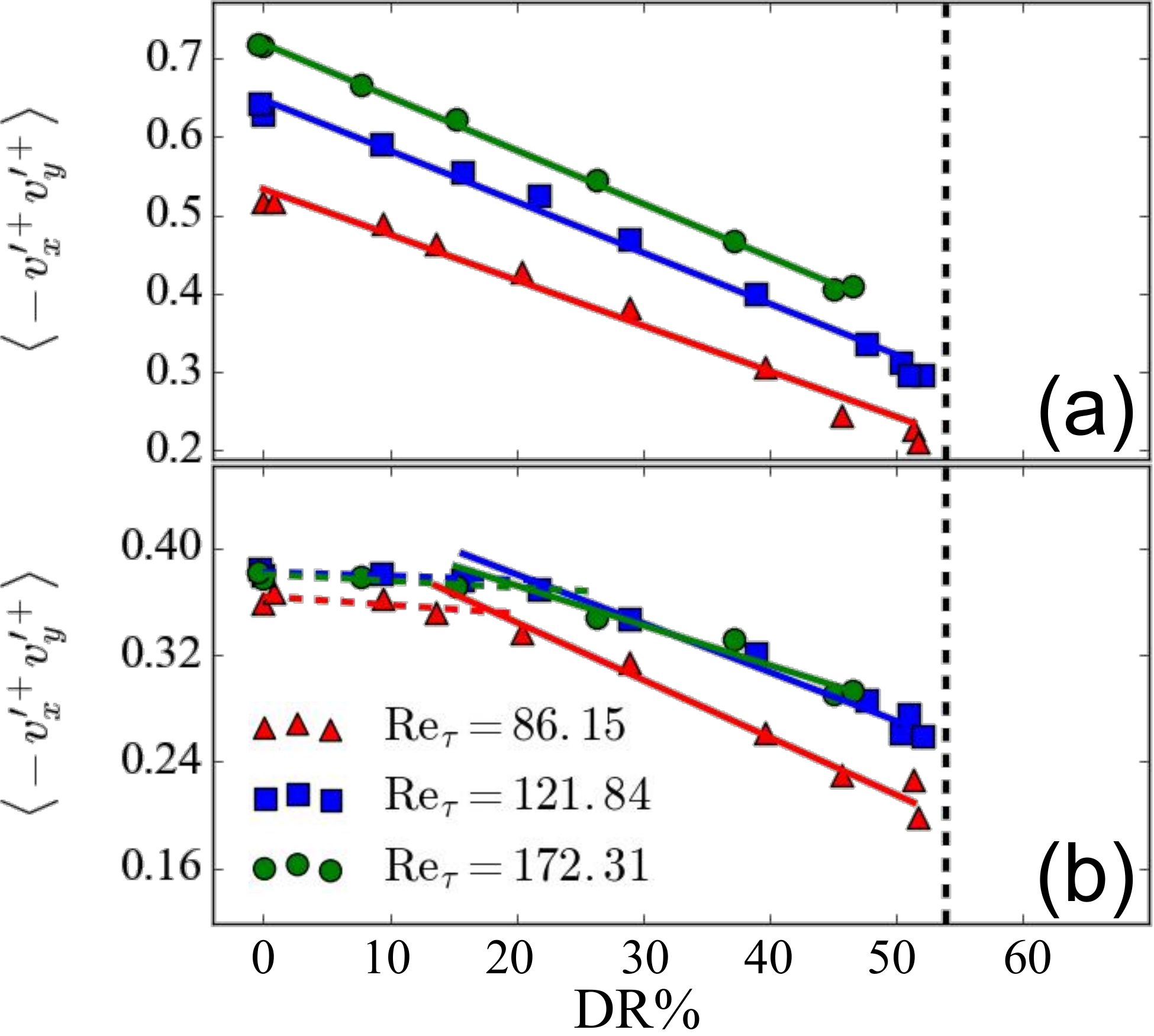}\\
	\caption{Variation of the Reynolds shear stress $\tau_{\text{R},xy}^+$ with increasing $\mathrm{DR}\%$ at (a) $y^+=25$ and (b) y=0.6H in channel flow DNS results with varying $\mathrm{Wi}$ ($\beta=0.97$, $b=5000$).
	(Reprinted from 
	\RevisedText{Journal of Non-Newtonian Fluid Mechanics, 262, 115,
	\citeauthor*{Zhu_Xi_JNNFM2018},
	Distinct transition in flow statistics and vortex dynamics between low- and high-extent turbulent drag reduction in polymer fluids,} 
	Copyright (\citeyear{Zhu_Xi_JNNFM2018}), with permission from Elsevier.)
	}
	\label{fig:rsstrans}
\end{figure}

A systematic investigation of the LDR-HDR transition using DNS was recently reported by \citet{Zhu_Xi_JNNFM2018}. It concluded that the primary difference between LDR and HDR is the region\RevisedText{,} i.e., the \RevisedText{span} of \RevisedText{the} wall layer\RevisedText{, influenced} by DR effects.
At LDR, DR effects are mostly contained in the buffer layer, although the layer does thicken with $\mathrm{DR}\%$. This stage of DR is relatively better understood, per discussion above, and the Virk three-layer model still captures some essential elements\RevisedText{. (The model's depiction} of buffer layer dynamics and its $U_\text{m}^+(y^+)$ profile\RevisedText{, however, is} not accurate\RevisedText{---}see \cref{sec:mdr}\RevisedText{.)}
At HDR, DR effects are felt across nearly the entire domain, with the viscous sub-layer being the only exception. This, of course, \RevisedText{includes} the key observation that the $U_\text{m}^+(y^+)$ profile changes its shape in (what used to be) the log-law layer.
\RevisedText{In addition,} DR effects are more directly measured by the RSS. In \cref{fig:rsstrans}, $\tau_{\text{R},xy}^+$ measured at a position within ($y^+=25$) and one outside the buffer layer ($y=0.6H$ or $y^+=0.6\mathrm{Re}_\tau$; recall \cref{eq:Retaulv}) is plotted against $\mathrm{DR}\%$ for three different $\mathrm{Re}$.
The contrast is clear. Within the buffer layer \RevisedText{(\cref{fig:rsstrans}(a))}, DR is continuous \RevisedText{across the whole range of $\mathrm{DR}\%$} and $\tau_{\text{R},xy}^+$ drops consistently \RevisedText{starting} from the onset ($\mathrm{DR}\%=0$)\RevisedText{. O}utside the buffer layer \RevisedText{(\cref{fig:rsstrans}(b))}, $\tau_{\text{R},xy}^+$ nearly fully retains its Newtonian magnitude until $\mathrm{DR}\%\approx 20\%$ where it starts to descend, indicating that turbulent dynamics at larger $y^+$ remains minimally \RevisedText{impacted} until this threshold.
Note that the transition point $\mathrm{DR}\%\approx 20\%$ is, again, much lower than the commonly cited $30\%-40\%$ threshold for all three\RevisedText{, albeit} relatively low\RevisedText{,} $\mathrm{Re}$ tested.
\RevisedText{The conclusion that the RSS at higher $y^+$ is only suppressed at HDR}
can be verified in the $\tau_{\text{R},xy}$ profiles from various previous studies, using different experimental techniques or numerical algorithms, where the LDR profiles are suppressed only in the buffer layer and stay close to the Newtonian profile at higher $y^+$, \RevisedText{but} the HDR profiles are suppressed \RevisedText{nearly} anywhere \RevisedText{except the} $y^+\lesssim 5$ \RevisedText{region}~\citep{Warholic_Hanratty_EXPFL1999,Warholic_Hanratty_EXPFL2001,Dallas_Vassilicos_PRE2010,Thais_Mompean_IJHFF2013}.

In addition to $U_\text{m}^+(y^+)$ and $\tau_{\text{R},xy}^+$, sharp transitions at higher $y^+$ were also found, by the same study~\citep{Zhu_Xi_JNNFM2018}, in polymer shear stress and, more interestingly, energy spectra (which measure the distribution of TKE over different length scales).
Changes in the latter showed that\RevisedText{,} while at LDR, polymers suppress turbulent fluctuations and redistribute energy towards large scales at all $y^+$ positions, at HDR, there is a sharp increase in this effect at higher $y^+$, where the von~K\'arm\'an \RevisedText{law} used to dominate.
Disappearance of the inertia-dominated layer (i.e., the log-law layer\RevisedText{---}see \cref{sec:mdr} \RevisedText{below}) was also regarded by \citet{White_Dubief_JFM2018} as the sign of HDR based on a mean momentum balance analysis.

Comparing the flow structures of Newtonian/LDR (\cref{fig:largebox}(a)) and HDR/MDR (\cref{fig:largebox}(b)) flows, not only are the vortices attenuated in the latter case, vortex distribution is also more localized, as the small number of remaining vortices tend to appear in \RevisedText{conglomerates}.
\citet{Zhu_Xi_JNNFM2018} quantitatively analyzed the degree of localization at different regimes of DR and found that the localization starts at the LDR-HDR transition.

\RevisedText{Sharp transitions in flow statistics and structures suggest the presence of}
two separate mechanisms for DR.
The first sets in at the onset of DR and, by all accounts, is a general weakening of vortices whose effects are \RevisedText{largely concentrated} in the buffer layer.
The second is triggered at the LDR-HDR transition which suppresses turbulent momentum transport at higher $y^+$ and extends DR effect across most of the domain height.
The distinct changes in what used to be the log-law layer suggest an underlying shift in the coherent structure dynamics in that region.
A plausible hypothesis was proposed by \citet{Zhu_Xi_JNNFM2018} and clear supporting evidences became available with a newest vortex tracking technique~\citep{Zhu_Xi_JFM2019,Zhu_Xi_POF2019}, which will be discussed in \RevisedText{\cref{sec:largebox}.}

\subsubsection{Maximum drag reduction (MDR)}\label{sec:mdr}
\paragraph{\RevisedText{Basic observations}}
\Cref{fig:Um} shows $\mathrm{DR}\%$ increases with increasing fluid elasticity (in their case by increasing polymer concentration). This trend is eventually bounded by an upper limit call the MDR asymptote\RevisedText{---}an ``ultimate'' flow \RevisedText{regime} where certain forms of turbulence still persist and the friction drag \RevisedText{has converged to a level} between the magnitudes of Newtonian turbulence and laminar flow.
The phenomenon of MDR was first reported by \citet{Virk_Merrill_JFM1967}. It was \RevisedText{subsequently} found,
\RevisedText{contrary to intuition},
that the \RevisedText{MDR limit} is universal\RevisedText{---}flows \RevisedText{of} different polymer solutions (changing polymer species, molecular weight, or concentration) \RevisedText{or} different geometric size $l$ would converge to the same MDR asymptote~\citep{Virk_JFM1971}.
That is, $\mathrm{DR}\%$ of various MDR flow states depends solely on their $\mathrm{Re}$\RevisedText{, even when they have different} $\mathrm{Wi}$, $\beta$, or $b$ (using \RevisedText{FENE-P} parameters introduced in section~\ref{par:feneppara}).

\RevisedText{The} $\mathrm{Re}$-dependence can be completely wrapped into turbulent inner scales (note from, e.g., \cref{eq:Retaulv}, that inner scales depend on $\mathrm{Re}$) and the rescaled mean velocity profile \RevisedText{in ``+'' units} appears to be universal for MDR under different conditions.
\RevisedText{\citet{Virk_Baher_CES1970} and \citet{Virk_JFM1971} used}
a logarithmic relation, same in form as \cref{eq:loglaw} \RevisedText{but with different constants, to fit various experimental MDR data. The resulting universal profile}
\begin{gather}
	U_\text{m, Virk}^+=11.7\ln y^+ -17.0
	\label{eq:virk}
\end{gather}
\RevisedText{has a markedly higher slope and seems to well approximate experimental profiles}
for most of the flow domain (i.e., no longer confined to a near-wall layer)\RevisedText{.}
Note that the two highest DR cases ($\mathrm{DR}\%>60\%$) in \cref{fig:Um} are closely approaching the Virk profile (dot-dashed line).

\begin{figure}
	\centering
 	\includegraphics[width=\linewidth, trim=0 0 0 0, clip]{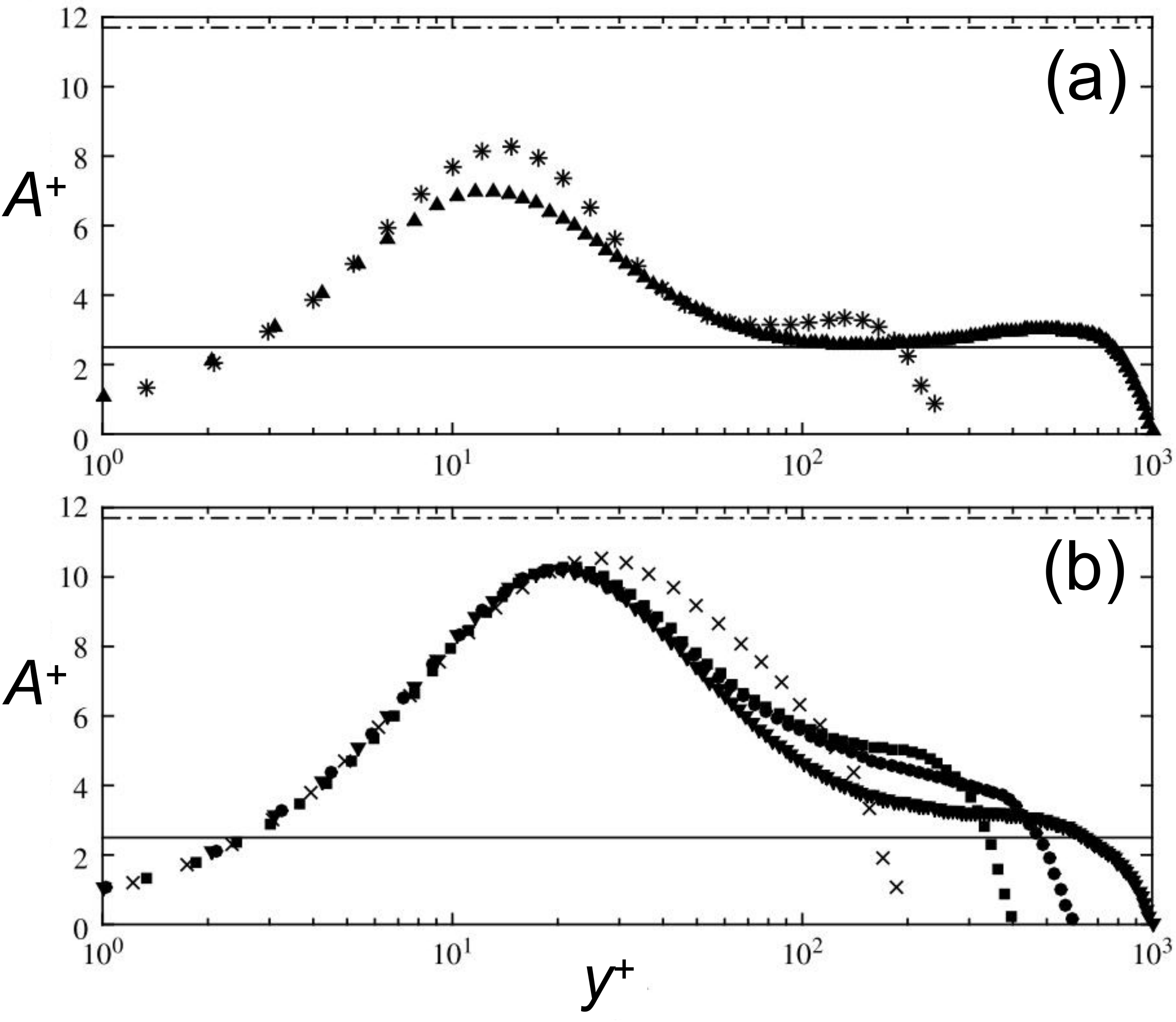}\\
	\caption{\RevisedText{%
	Logarithmic slope function profiles $A^+(y^+)$ (\cref{eq:Aplus}) of (a) LDR ($\mathrm{DR}\%=30\sim35\%$) and (b) HDR ($\mathrm{DR}\%=59\sim62\%$) cases with $\mathrm{Re}_\tau$ ranging from $186$ to $1000$, all from DNS of channel flow.
	Flat lines show the the logarithmic slopes of $A^+=2.5$ (solid) for the Newtonian von~K\'arm\'an law and $11.7$ (dot-dashed) for Virk MDR.
	A clear plateau of $A^+\approx 2.5$ is found in the log-law layer at LDR, whereas at HDR the plateau lifts to higher magnitude and is not discernible at the lowest $\mathrm{Re}$ ($\times$).
	A plateau at $A^+=11.7$, expected per \cref{eq:virk} for MDR, is not seen in any case.
	(\citeauthor*{White_Dubief_JFM2018},
	Properties of the mean momentum balance in polymer drag-reduced channel flow,
	Journal of Fluid Mechanics,
	834, 409,
	(\citeyear{White_Dubief_JFM2018}), reproduced with permission.)
	}}
	\label{fig:Aplus}
\end{figure}

\paragraph{\RevisedText{Validity of the logarithmic relation}}
\Cref{eq:virk} has long \RevisedText{been} seen as a gold standard for MDR, but its validity was recently challenged.
\RevisedText{Taking} derivatives of both sides \RevisedText{of} \cref{eq:loglaw},
\RevisedText{one obtains} 
\begin{gather}
	A^+=y^+\frac{dU_\text{m}^+}{dy^+}
	\label{eq:Aplus}
\end{gather}
\RevisedText{which is the local slope value in the logarithmic relation. A true log law will show a region of nearly constant $A^+$.}
\Citet{White_Dubief_POF2012} calculated the log-law slopes of $U_\text{m}^+(y^+)$ profiles from several recent experimental and DNS studies
and concluded that, for HDR cases, a well-defined wall layer ($y^+$ range) with logarithmic relation is, in their word, ``eradicated''.
\RevisedText{F}or cases with $\mathrm{DR}\%\geq60\%$ (both experimental and numerical, $\mathrm{Re}_\tau\lesssim200$)\RevisedText{,} where the $U_\text{m}^+(y^+)$ profile appears very close to the \RevisedText{Virk MDR profile (}\cref{eq:virk}\RevisedText{)}, the log-law slope $A^+$ does not show any clear plateau, near $11.7$ (as implied by \cref{eq:virk}) or elsewhere. Rather, its magnitude rises up at low $y^+$, reaches its peak \RevisedText{within} $20\lesssim y^+\lesssim30$, and then drops steeply. The peak magnitude is, nonetheless, comparable to (but not the same as) $11.7$.
Later experimental measurements in boundary layer flow by \citet{Elbing_Perlin_POF2013} led to largely similar results, that for their MDR-like case ($\mathrm{DR}\%=64.8\%$), the $U_\text{m}^+$ profile does not show a logarithmic region anywhere near $A^+=11.7$.
For HDR (their $\mathrm{DR}\%=53\%$ case), a roughly constant $A^+$ region is found at $y^+\gtrsim 200$ with a magnitude higher than $2.5$ (Newtonian level) but much lower than $11.7$ (Virk level), which is consistent with \citet{Warholic_Hanratty_EXPFL1999}'s depiction of HDR (i.e., still logarithmic but higher slope).
The missing logarithmic region in \citet{White_Dubief_POF2012} at HDR was likely caused by the relatively low $\mathrm{Re}$ analyzed there\RevisedText{. I}ndeed, in a more recent study by the same authors~\citep{White_Dubief_JFM2018} \RevisedText{(shown in \cref{fig:Aplus})}, DNS channel flow data of $\mathrm{Re}_\tau$ up to $1000$ were included and, at HDR, a quasi-flat region in the $A^+$ profile can be spotted at $y^+\gtrsim 200$ (where $2.5<A^+\ll 11.7$).
Nevertheless, for MDR, conclusions of those studies are unanimous, that a rigorously logarithmic profile with $A^+$ anywhere close to $11.7$ (\cref{eq:virk}) is not found.

\RevisedText{Recall that for Newtonian turbulence, the von K\'arm\'an law of wall was deduced}
from the following scaling argument~\citep{Pope_2000}:
\begin{CmpctItem}
	\item in the turbulent inner layer, \RevisedText{the bulk flow rate and geometry} are no longer relevant and the mean velocity gradient is determined by inner scales (\cref{sec:inner}) only:
	\begin{gather}
		\frac{dU_\text{m}^+}{dy^+}=\frac{1}{y^+}A^+(y^+)
		\label{eq:loglawderive}
	\end{gather}
	($A^+(y^+)$ is an undetermined non-dimensional function universal to different flow conditions.); and
	\item at sufficiently high $y^+$ (well above the viscous sub-layer but still within the inner layer), viscosity effects vanish and the flow is dominated by inertia\RevisedText{---}it is thus postulated that $A^+$ is a constant (because $y^+$ depends on viscosity through the definition of $l_\text{v}$\RevisedText{---}\cref{eq:lv}).
\end{CmpctItem}
Integration of \cref{eq:loglawderive} with a constant \RevisedText{$A^+$} leads to \cref{eq:loglaw} \RevisedText{and the numerical values of $A^+$ and $B^+$ are determined from fitting with experiments and DNS.}
\RevisedText{Since $U_\text{m}^+(y^+)$ profiles at MDR are nearly the same for different $\mathrm{Re}$---i.e., inner scales are still very much relevant,}
the loss of a logarithmic layer at MDR 
\RevisedText{is likely}
associated with
the elimination of \RevisedText{the} inertia-dominated layer in near\RevisedText{-wall} turbulence.

The finding that a well-defined $A^+=11.7$ logarithmic layer is not found at any $y^+$ position and any stage of DR is in direct contrast to one of the key elements in Virk's three-layer model, which postulates that \RevisedText{the Virk} MDR
\RevisedText{logarithmic profile (}\cref{eq:virk}\RevisedText{) should appear even in the intermediate DR regime but only in a limited region.
According to the model, before MDR, $U_\text{m}^+(y^+)$ would follow the Virk profile}
within the enlarged buffer layer or\RevisedText{, as termed by Virk, the} ``elastic sub-layer''
until its crossover to
\RevisedText{a logarithmic dependence with the von~K\'arm\'an slope~\citep{Virk_JFM1971,Virk_AIChEJ1975}.}
\RevisedText{The model not only misses the increased slope in the log law layer at HDR, it is also clear from \cref{fig:Aplus} that, at both LDR and HDR, the buffer layer profile does not follow the Virk logarithmic relation.}
Indeed, it appears that it is the reduction and elimination of the inertia-dominated log-law layer, rather than the generation \RevisedText{and expansion} of an elastic sub-layer, that drives the transition to HDR and MDR~\citep{White_Dubief_JFM2018}.

\paragraph{\RevisedText{Fundamental attributes of MDR}}\label{para:mdrattributes}
Mechanistic understanding of MDR remains the most coveted target in this area, which is not surprising \RevisedText{considering} its mysterious nature.
It is common practice among researchers to use the Virk profile (\cref{eq:virk}) as an identifying trait of MDR: a flow state would be considered MDR if its $U_\text{m}^+(y^+)$ profile \RevisedText{matches} the Virk profile.
\RevisedText{However, let us be reminded that \cref{eq:virk} is based on empirical fitting to a somewhat arbitrarily chosen functional form: (1) there is no \textit{a priori} physical argument} for a logarithmic relationship \RevisedText{other than a simple analogy to the von~K\'arm\'an law (\cref{eq:loglaw}); (2) coefficients in \cref{eq:virk} are empirically determined from experimental data.
L}atest examination of $A^+$ \RevisedText{reviewed above further shows that (1) a logarithmic relationship is probably inaccurate and (2) comparing $U_\text{m}^+(y^+)$ profiles} in semi-log coordinates (such as \cref{fig:Um}) can be misleading: \RevisedText{non-logarithmic profiles are not sufficiently distinguished from logarithmic ones.}

\RevisedText{Therefore, research}
of MDR must
\RevisedText{go beyond this fixation on the Virk logarithmic profile of \cref{eq:virk} and}
qualitative traits
\RevisedText{must be considered}
~\citep{Xi_Bai_PRE2016,Zhu_Xi_JNNFM2018}.
In particular, any complete theory for MDR must consistently explains its three key attributes.
\begin{CmpctDescr}
	\item[Existence] Why does there have to be an upper bound for DR at all? If polymers cause \RevisedText{DR} by suppressing turbulence, why can they not take the flow all the way to the laminar state? This question requires fundamental insights into the turbulent self-sustaining mechanism at MDR.
	\item[Universality] Why is the upper bound universal for different polymer solution properties and for different flow geometric \RevisedText{dimensions}? MDR occurs at the limit of high DR where polymer effects are strongest\RevisedText{---}it is counter-intuitive that this state, or at least the mean flow statistics thereat, is not affected by changing polymer properties.
	\item[Magnitude] Although \cref{eq:virk} may \RevisedText{not have} used the most appropriate functional form, it is still undeniable that $U_\text{m}^+$ magnitudes at MDR are quantitatively close to the Virk profile. Why would DR converge to that particular magnitude? This question must be included because, as discussed below, there are possibly multiple flow states that address the previous two points \RevisedText{and} they cannot all be MDR.
\end{CmpctDescr}
A complete answer remains elusive. However, substantial progress has been made in the past decade, which will be reviewed in \cref{sec:recent}.

\subsection{Mechanistic understanding}\label{sec:mechanism}
Most theoretical attempts at DR predated the discovery of the LDR-HDR transition~\citep{Warholic_Hanratty_EXPFL1999}. Their focus was thus on explaining why polymers cause DR and how to predict the onset of DR based on the supposed mechanism. Efforts were also made \RevisedText{to} address MDR but success has been limited.
Because the LDR-HDR transition was not considered a qualitative transition involving different DR mechanisms until \citet{Zhu_Xi_JNNFM2018}, earlier theories covered in this section were all directed at the first DR mechanism: \RevisedText{i.e.,} the one \RevisedText{responsible for} the onset \RevisedText{of DR and LDR.}
Recent \RevisedText{developments of} new theoretical frameworks and research methodology have led to a series of \RevisedText{mechanistic insights into} HDR and MDR\RevisedText{, which} will be discussed in \cref{sec:recent}.

\subsubsection{Polymer-turbulence interactions}\label{sec:polymturb}
At first glance, the notion of polymer addition causing reduced friction is counter-intuitive to most, as polymers are typically associated with higher viscosity.
It should be clear \RevisedText{by now} that drag-reducing polymer solutions are often too dilute to have significant viscosity increase and, when they do, the definition of \RevisedText{DR} explicitly corrects for the viscosity change \RevisedText{(}\cref{sec:drdef}\RevisedText{)}.
\RevisedText{The term} DR \RevisedText{here} is \RevisedText{thus} referring \RevisedText{not} to any change in the viscous shear stress but \RevisedText{to the} reduction of the extra stress attributed to turbulent motion\RevisedText{---}i.e., the Reynolds stress\RevisedText{---as a consequence} of turbulence-polymer interactions.

To cause DR, polymer molecules have to, one way or another, suppress \RevisedText{turbulent} motion. Availability of detailed polymer conformation (and, consequently, stress) field information, thanks to extensive DNS efforts in the past two decades, \RevisedText{has} provided direct evidence in this regard.
Earlier experiments have shown, by injecting polymers to different wall positions, that DR becomes substantial when polymers reach the near\RevisedText{-}wall region covering the buffer layer and lower log-law layer~\citep{McComb_Rabie_AIChEJ1982a,McComb_Rabie_AIChEJ1982b}.
The buffer layer is also where turbulence production peaks in Newtonian flow and where significant reduction in RSS is observed \RevisedText{when drag-reducing polymers are added}. (For the latter, we now know it only applies to LDR~\citep{Zhu_Xi_JNNFM2018}.)
Investigation of polymer effects on buffer-layer turbulence was thus \RevisedText{the} focus \RevisedText{of earlier DNS studies.}

It is now generally agreed that, within the buffer layer, polymers reduce turbulent intensity by opposing its dominant flow structure\RevisedText{---}streamwise vortices.
Polymers alter fluid dynamics through an additional term in the momentum balance which can be described as a polymer force.
Multiple DNS studies have showed that, in transverse directions (i.e., in the plane of rotation of streamwise vortices), the polymer force counters velocity fluctuations\RevisedText{. D}irect inspection of flow field visualization \RevisedText{images indicated that} the effect \RevisedText{is} strongest immediately next to the vortices~\citep{DeAngelis_Piva_CompFl2002,Sibilla_Baron_POF2002,Dubief_Lele_FTC2005}.
Numerical computation of ECS solutions allowed \RevisedText{those} vortices to be isolated \RevisedText{in a static form} from the complex and dynamical backdrop of turbulent fluctuations\RevisedText{. Investigation of polymer effects on ECS} confirmed \RevisedText{the same} mechanism~\citep{Stone_Graham_POF2004,Li_Graham_POF2007}.
Later study by \citet{Kim_Adrian_JFM2007} constructed statistical representations of \RevisedText{characteristic near-wall} vortices through conditional sampling\RevisedText{---}extracting and averaging local flow structures 
\RevisedText{that satisfy}
certain prescribed conditions~\citep{Adrian_Moin_JFM1988}.
Their results reaffirmed that counter-rotating streamwise vortex pairs are the most significant structure\RevisedText{s} in the buffer layer and the key conclusion of polymer forces counteracting \RevisedText{those} vortices, \RevisedText{deduced} previously from arbitrarily-selected images \RevisedText{in DNS}, is statistically verifiable.

As streamwise vortices lessen in their intensity, they also dilate laterally, which leads to increased spacing between vortices and the \RevisedText{lift} of vortex \RevisedText{axes} away from the wall.
\RevisedText{Since}
velocity streaks are created between counter-rotating vortices%
\RevisedText{, increased vortex separation} is \RevisedText{reflected in the enlargement of} near-wall streak spacings (see section~\ref{par:struct}).
\RevisedText{Meanwhile, upward shift of vortex axes}
is reflected in profile peaks of fluctuating quantities, such as $v_{x,\text{rms}}'$ and $\tau_{\text{R},xy}$, moving away from the wall
\RevisedText{(\cref{fig:vxrms})}.
It also directly accounts for the apparent thickening of the buffer layer.

Suppression of vortex motion reduces the strength of velocity streaks in between\RevisedText{, as measured by their velocity} fluctuation intensity\RevisedText{.}
\RevisedText{This} explains the lowering RSS magnitude.
To see this, note that low-speed streaks have negative $v_x'$ and positive $v_y'$ whereas high-speed streaks have positive $v_x'$ and negative $v_y'$\RevisedText{---}both contribute to \RevisedText{higher} $-\langle v_x'v_y'\rangle$.
Lowering RSS then contributes less to TKE production (see \cref{eq:tkeP}), which leads to an overall flow with less fluctuations and more momentum retained in the mean flow.
Thus far, a convincing depiction of DR mechanism, at least applicable to LDR, has arisen.
The situation of HDR is different, where the dominant structures are more complex and DR is no longer solely attributable to an enlarged buffer layer. 
Study into this second stage of DR has been rather limited and occurred very recently, which will be discussed in~\cref{sec:hdrssp}.

\subsubsection{Classical theories: an introduction}
Distilling a simple quantitative theory from the above micro-mechanical depiction of polymer-turbulence dynamics is not as straightforward.
There has been a long-standing debate between viscous and elastic interpretations of \RevisedText{the} polymer DR mechanism.
Both theories are built on the common foundation of energy cascade, in which turbulence is conceptualized to consist of a hierarchy of eddies of different sizes superimposed on one another \RevisedText{in} the same \RevisedText{flow region}.
TKE is produced at the largest eddies\RevisedText{, whose sizes are} comparable to \RevisedText{that of} the flow domain\RevisedText{,} and ``cascades'' towards successively smaller eddies as larger ones erupt, until it reaches the lower end of the \RevisedText{spectrum---}the Kolmogorov scale\RevisedText{---}where the length scale is small enough for viscous dissipation to dominate.
Experimental and numerical measurements have consistently shown, through energy spectra or Karhunen-Lo\`eve (KL) analysis, that small scale structures are most susceptible to suppression by polymers~\citep{Warholic_Hanratty_EXPFL2001,DeAngelis_Piva_PRE2003,Housiadas_Beris_POF2003,Sibilla_Beretta_FDR2005,Zhu_Xi_JNNFM2018}.
\RevisedText{Indeed}, both viscous and elastic interpretations consider polymers to truncate \RevisedText{or disrupt} the energy cascade at a certain scale larger than the Kolmogorov scale.

\paragraph{Viscous mechanism}
The viscous theory was proposed by Lumley~\citep{Lumley_ARFM1969,Lumley_JPSMacroRev1973}, which postulates that the
\RevisedText{energy cascade is truncated at a larger minimal length scale}
because of the viscosity increment caused by polymer additives.
\RevisedText{The smallest length scale in the energy cascade, i.e., the dissipative scale, of Newtonian turbulence is called the}
Kolmogorov length scale $l_\text{d}$
\begin{gather}
	l_\text{d}\sim\left(\frac{\eta^3}{\rho^3\epsilon}\right)^\frac{1}{4}
	\label{eq:ld}
\end{gather}
\RevisedText{which depends on the fluid viscosity as well as}
density and $\epsilon$\RevisedText{---}the rate of energy dissipation per unit mass of the fluid~\citep{Pope_2000}.
Since drag-reducing polymers often have minimal impact on the shear viscosity of the solution, the higher viscosity responsible for shifting the turbulent dissipation length scale can only be interpreted as its extensional viscosity $\eta^\text{ext}$, which, for viscoelastic dilute polymer solutions, can be much higher than their shear viscosity $\eta$.
Indeed, for drag-reducing polymer solutions, the Trouton ratio
\begin{gather}
	\mathrm{Tr}\equiv\frac{\eta^\text{ext}}{\eta}
\end{gather}
can \RevisedText{be} as high as $O(\num{e4})$~\citep{Metzner_Metzner_RhA1970,Usui_Sano_POF1981}.

\RevisedText{R}elevance of extensional viscosity \RevisedText{is} clear considering the kinematics of turbulence in the near-wall region, where strong but transient extensional motions are constantly being generated near vortices.
Polymer molecules become locally aligned and highly stretched in those extension-intensive spots. Strong retractive forces of \RevisedText{stretched} polymer chains \RevisedText{provide} the resistance to \RevisedText{extensional flow motion, which causes the} suppression of vortex dynamics.

\RevisedText{For dilute solutions of drag-reducing polymers i}n uniaxial extensional flow, as $\mathrm{Wi}^\text{ext}$ \RevisedText{(\cref{eq:Wiext})} increases, \RevisedText{its} $\mathrm{Tr}$ undergoes a sharp upsurge from the Newtonian value of $3$ to \RevisedText{several} orders of magnitude higher, as a result of the abrupt C-S transition~\citep{Arratia_Gollub_NJPhys2009}.
For FENE-P, $\mathrm{Wi}^\text{ext}=1/2$ \RevisedText{is the critical magnitude for this steep transition}~\citep{Bird_Curtis_1987}.
\RevisedText{At lower $\mathrm{Wi}$, there is no appreciable change in $\eta^\text{ext}$ and thus no DR is expected.}
\RevisedText{Therefore, a}n inherent implication \RevisedText{of the viscous mechanism} is that the onset of DR is associated with a critical $\mathrm{Wi}$ independent of polymer concentration, in so far as the solution is still dilute.
The latter is because in a dilute solution, the C-S transition of each individual polymer chain is not affected by the presence and state of other chains.
For a given polymer-solvent pair and given molecular weight, the relaxation time $\lambda_H$ is determined, \RevisedText{which means} that the onset of DR is determined by the flow time scale dropping below a critical value (i.e. $\lambda_H/\tau_\text{flow}$ higher than a critical value)\RevisedText{---}the so-called ``time criterion''~\citep{Hershey_Zakin_CES1967,Lumley_ARFM1969,Lumley_JPSMacroRev1973}.

\RevisedText{After the onset, a}s $\mathrm{Wi}$ or polymer concentration continues to increase, $\eta^\text{ext}$ increases and the energy cascade is further truncated at \RevisedText{larger scales as the dissipative scale increases}. The effect will eventually be bounded\RevisedText{---}i.e., MDR \RevisedText{is} reached\RevisedText{---}when $l_\text{d}$ becomes comparable to the largest scale of the flow \RevisedText{which must be} the geometric length scale $l$.
This idea is in line with Virk's three-layer model~\citep{Virk_JFM1971} which predicts MDR as the limit where the enlarged buffer layer is restricted by the flow geometric size.
\RevisedText{Since this geometric constraint is independent of polymer properties, it offers a simple explanation for}
the universality of MDR, but the mechanism sustaining turbulence thereat remains unspecified.

\paragraph{Elastic mechanism}
The conceptual framework of the elastic theory for DR was constructed by de~Gennes and co-worker~\citep{deGennes_Tabor_EPL1986,deGennes_1990} through their scaling theory for polymer dynamics within the turbulent energy cascade.
\citet{Sreenivasan_White_JFM2000} further elucidated the theory and incorporated the effects of flow heterogeneity for DR prediction in pipe flow.
Differences between the viscous and elastic interpretations of the DR mechanism originate from their underlying molecular assumptions. While the viscous mechanism assumes polymer chains to be highly stretched (past the C-S transition) to display substantial extensional viscosity increase, de~Gennes considered this scenario unlikely, especially immediately after the onset.
He thus postulated that polymer chains are only partially stretched and the deformation is transient in nature. The level of deformation is not sufficient to cause significant increase in extensional viscosity, but the elastic energy stored in \RevisedText{individual stretched} chains can add up. DR is expected when the elastic stress becomes comparable to or higher than Reynolds stress  $-\rho\langle\mbf v'\mbf v'\rangle$\RevisedText{. (E}quivalently, since \RevisedText{this} is a scaling argument, it could also be a comparison between elastic energy and TKE\RevisedText{.)}

As TKE flows towards smaller scales in the energy cascade, the corresponding time scale also decreases.
There is thus a critical scale $r^*$ \RevisedText{above} which polymers \RevisedText{remain undeformed} (i.e., $\lambda_H/\tau_{r*}$ equals a critical $\mathrm{Wi}$ or\RevisedText{,} rigorously\RevisedText{,} $\mathrm{De}$ magnitude).
For $r$ immediately below $r^*$, \RevisedText{polymer chains are stretched but} the elastic stress is small in compassion with the Reynolds stress. Polymers are progressively more stretched at smaller scales.
At another critical scale $r^{**}$, \RevisedText{the} elastic stress starts to exceed the Reynolds stress and \RevisedText{the} energy cascade is presumed to be disrupted.
The entire spectrum is divided into three regimes: at $r>r^*$, polymer chains are unaffected by turbulence; at $r^*>r>r^{**}$, polymers are stretched but not strongly enough to impact the flow; and at $r<r^{**}$, polymers disrupt the energy cascade.

Although $r^*$ is determined by \RevisedText{a} time\RevisedText{-}scale \RevisedText{criterion} and thus independent of concentration, $r^{**}$ is determined by stress or energy\RevisedText{---}quantities proportional to concentration.
\RevisedText{In the energy cascade, smaller length scales are associated with lower magnitudes of TKE or Reynolds stress.
When either the polymer concentration is lower or polymer chains are less stretched, the accumulated elastic energy or stress from all chains are lower.
It would thus intercept the turbulent cascade at a smaller $r^{**}$.}
If $r^{**}<l_\text{d}$\RevisedText{---}the smallest scale in the cascade\RevisedText{---}turbulence is unaffected at all scales.
\RevisedText{The o}nset of DR happens through increasing concentration or increasing
\RevisedText{$\mathrm{Wi}$}
(which determines the level of stretching) until $r^{**}$ is raised above $l_\text{d}$.
\RevisedText{It is thus obvious that a higher $\mathrm{Wi}_\text{onset}$ would be required when the polymer concentration $C_\text{p}$ is lower.}
Compared with the viscous mechanism, \RevisedText{this} concentration-dependence of the onset is a distinguishing feature of the elastic mechanism.

Prediction of MDR from the elastic theory is not as obvious. \citet{Sreenivasan_White_JFM2000} fitted various experimental data points at the margin of MDR (where the flow just reaches MDR) to the theory and \RevisedText{found} that $r^*$ and $r^{**}$ are both in the order of magnitude of the flow geometric size $l$\RevisedText{---}i.e., even the largest eddies, which are least effective at stretching polymers, can generate sufficient elastic stress \RevisedText{for themselves to be disrupted}.
\RevisedText{The study, however, did not address the asymptotic convergence of the flow with further increasing elasticity.}

\subsubsection{Discussion: viscous vs. elastic theories}
The Lumley vs. de~Gennes debate remains one of the most important outstanding \RevisedText{problems} in this field. The gap between the fairly detailed knowledge, at least for LDR, of polymer-turbulence interactions (\cref{sec:polymturb}) and a conclusive predictive theory underscores the complexity of the DR phenomenon.

\paragraph{Experimental and numerical evidences}
Both theories have found supporting evidences but none seems sufficient for an unequivocal conclusion.
\citet{Virk_AIChEJ1975} noted that for PEO solutions \RevisedText{of} different concentration, the onset occurs at the same well-defined $\mathrm{Re_\tau}$ and concluded that the wall-shear rate $\dot\gamma_\text{w}$ of the onset is \RevisedText{independent of concentration} for a given polymer (given $\lambda_H$), which is in line with the time criterion.
Concentration independence was also reported by \citet{Berman_POF1977} \RevisedText{for} both PEO and PAM, who further adjusted polymer relaxation time $\lambda_H$ by adding glycerol to change the solvent viscosity (see \cref{eq:lambdaH,eq:zetascaling}) and found that the onset wall shear rate \RevisedText{$\dot\gamma_\text{w}$} is inversely proportional to $\lambda_H$: i.e., $\mathrm{Wi}_\text{onset}$ is constant. 
On the other hand, data showing sensitivity of the onset to concentration were also reported, most notably by \citet{Nadolink_PhD1987}.

DNS seems naturally suited to test the concentration dependence of the onset as it avoids experimental complexities such as polymer polydispersity, concentration inhomogeneity, and degradation~\citep{Graham_RheologyReviews2004}.
So far, the only DNS study with direct comparison between \RevisedText{$\mathrm{Wi}_\text{onset}$ of} different concentrations was \citet{Xi_Graham_JFM2010} where $\beta=0.97$ and $\beta=0.99$ (a roughly three-fold difference in concentration) have nearly identical $\mathrm{Wi}_\text{onset}$. The study used highly-constrained domain size (motivation will be discussed in \cref{sec:mfu}) and low $\mathrm{Re}$ ($\mathrm{Re}_\tau=84.85$).
\RevisedText{Concentration dependence of $\mathrm{Wi}_\text{onset}$}
at more realistic flow conditions \RevisedText{has} not been \RevisedText{tested}. \RevisedText{Indeed,} accurate determination of $\mathrm{Wi}_\text{onset}$ requires a number of \RevisedText{extended} simulation runs near the onset point.
\RevisedText{Long simulation time is required in each run for lower} uncertainty \RevisedText{in time-averaged flow quantities in order} to discern potential changes in $\mathrm{DR}\%$\RevisedText{. The overall requirement for} computational resources \RevisedText{is non-trivial}.

As to the \RevisedText{underlying} molecular assumptions, DNS has consistently shown that at the onset, average $\mathrm{tr}(\alpha)\lesssim O(0.1)b$\RevisedText{, which is} nowhere close to full extension\RevisedText{~\citep{Housiadas_Beris_POF2003,Yu_Kawaguchi_JNNFM2004,Li_Khomami_JNNFM2006,Xi_Graham_JFM2010,Dallas_Vassilicos_PRE2010,Zhu_Xi_JNNFM2018}.}
\RevisedText{This is often quoted as evidence in support of de~Gennes's argument for an elastic mechanism.
However, one must also consider}
the spatial variation and transient nature of turbulent flows\RevisedText{---}the extent of stretching can vary vastly between different regions.
Indeed, Brownian dynamics simulation of FENE dumbbells in turbulent channel flow have shown that even at low $\mathrm{Wi}$ ($O(1)$ based on $\dot\gamma_\text{w}$), a small portion of the chains are highly stretched in regions with strong local biaxial extension~\citep{Terrapon_Lele_JFM2004}.
\RevisedText{Therefore, d}espite the \RevisedText{low} overall stretching near the onset, high extensional viscosity is still possible at certain spots as a result of the transient occurrence of strong local extension.
In this sense, the extensional viscosity in the Lumley theory should be not interpreted as a domain average.

\paragraph{Further DNS evaluation}\label{par:viselasdns}
DNS results have been interpreted in both theoretic frameworks. \citet{Sureshkumar_Beris_POF1997} tested a very low finite extensibility parameter $b$, which suppresses extensional viscosity, and found no DR for $\mathrm{Wi}$ up to $50$. This seemingly provides direct evidence in support of the viscous mechanism.
However, it is unclear how a low $b$ magnitude may interfere with a elastic mechanism\RevisedText{. I}ndeed, the specific influence of polymer elasticity on turbulence is not \RevisedText{even specified} in the elastic theory.

A common practice in DNS is to inspect the TKE balance (\cref{eq:tkebal}) and infer a mechanism
\RevisedText{based on profiles of different terms in the TKE budget.}
For the elastic conversion rate $\epsilon_\text{p}^k$ profiles, \citet{Min_Choi_JFM2003a,Min_Choi_JFM2003b}\RevisedText{,} and later \citet{Dallas_Vassilicos_PRE2010} have \RevisedText{all} observed a three-layer distribution at LDR:  $\epsilon_\text{p}^k$ is positive (i.e., polymers suppress turbulence) in the viscous sub-layer, becomes negative (i.e., elastic energy feeds turbulence) in a small \RevisedText{region} around $y^+\sim 10-15$, and then turns positive again at higher $y^+$.
The initial interpretation by those authors was that the polymer chains are stretched by the mean shear immediately near the wall and release the elastic energy back to the flow in the buffer layer.
(Recent developments indicated that the second layer with negative $\epsilon_\text{p}^k$ is likely caused by a new state of turbulence; see \cref{sec:edt}.)
At HDR and MDR, the third layer disappears and the second layer expands across the domain.

\RevisedText{Treating the dynamical DR mechanism as a problem of 1D ($y$-direction) transport of averaged TKE budget terms is relatively simplistic. It}
neglects, once again, \RevisedText{the spatial variations of flow patterns and} polymer stretching.
\RevisedText{There is also no direct information on the dynamical sequence of events: e.g., the notion of polymers moving to the buffer layer after getting stretched at the wall is mostly a speculation.
In addition, between}
negative \RevisedText{and} positive $\epsilon_\text{p}^k$\RevisedText{, it is unclear which one} is more relevant to DR.
\RevisedText{Therefore, the overall proposed scenario for an elastic mechanism based on TKE budget analysis is more of a presumption than a conclusion.}

\paragraph{Relationship and distinctions}
\RevisedText{The viscous and elastic} theories are also not as diametrical as their names might suggest. It is not a debate of whether DR is caused by viscosity or elasticity. Rather, both mechanisms require a certain level of elasticity\RevisedText{---}a purely viscous Newtonian fluid would not cause DR under the viscous mechanism.
One may even argue that they are partially reconcilable.
Under the viscous mechanism, polymer feedback to turbulence is almost instant: polymer molecules are extensively stretched in extensional flows as they resist such deformation at the same time. During this process, TKE is converted into elastic energy stored in the stretched chains.
This could as well be one \RevisedText{possible scenario} for the generation of elastic energy or stress in the elastic theory, since the latter does not specify the detailed mechanism for polymer-turbulence interaction.
Of course, the lack of details also allows the \RevisedText{elastic} theory to be more plastic. For instance, \RevisedText{it} may also be interpreted in terms of the ``memory effect'' of viscoelastic fluids\RevisedText{---}polymers store elastic energy at one place and release \RevisedText{it} at \RevisedText{another} (as \RevisedText{in} the case of the \RevisedText{TKE budget analysis discussed above)}.

The fundamental division \RevisedText{between these two theories}
is not on how polymers interact with turbulence\RevisedText{, which is not explicitly specified in the elastic mechanism,} but on how such interactions are felt by turbulence and cause DR.
In the viscous mechanism, polymer effects are solitary and instant. As long as one local extensional flow spot is suppressed, DR\RevisedText{, no matter how small,} will \RevisedText{occur}. DR increases as more \RevisedText{local} flow regions \RevisedText{become suppressed}. \RevisedText{Overall, this mechanism} only requires localized C-S transition and thus depends solely on $\mathrm{Wi}$ (i.e., the time criterion).
In the elastic mechanism, polymer effects are collective. DR is only observed when the total elastic energy exceeds a certain threshold. The onset thus depends on both $\mathrm{Wi}$ and concentration.

\paragraph{Explanation of MDR}
Neither theory offers a complete account for MDR. Both fall short of portraying a turbulence self-sustaining mechanism at MDR, which is essential for the \RevisedText{question of MDR} ``existence'' \RevisedText{(section~\ref{para:mdrattributes})}. 
The viscous theory considers turbulence to be suppressed \RevisedText{beyond} the increased dissipative scale $l_\text{d}$, which leads to an enlarged viscous sub-layer. At the limit of MDR, $l_\text{d}$ becomes comparable to $l$, which would predict near \RevisedText{complete} laminarization.
The elastic theory only compares the magnitudes of turbulent and elastic energies (or, equivalently, stresses) without specifying the specific dynamics. At $r<r^{**}$, it can be interpreted as either elastic energy quenching turbulence or replacing it as the new driving mechanism for instability.
The latter possibility is raised here because of a series of recent studies of turbulent states driven by elasticity, which will be discussed in \cref{sec:edt}.

\citet{Warholic_Hanratty_EXPFL1999} observed in their channel flow experiments that $\tau_{\text{R},xy}$ effectively vanishes at MDR. The deficit left behind in the shear stress balance must have been replaced by polymer shear stress. Since RSS is related to TKE production, it was inferred that at MDR, \RevisedText{the conventional mechanism for turbulence generation is suppressed and}  turbulence \RevisedText{must be} sustained by elasticity.
This finding is not supported by later experimental and numerical observations, all of which showed that although $\tau_{\text{R},xy}$ at MDR is typically significantly lowered compared with its Newtonian magnitude, it remains finitely non-zero~\citep{Ptasinski_Nieuwstadt_JFM2003,Min_Choi_JFM2003b,Dallas_Vassilicos_PRE2010}.
Nevertheless, the magnitude of polymer shear stress exceeding that of RSS is at least consistent with the presumption of the elastic theory.

For ``universality'', an explanation is readily available from the viscous mechanism in which MDR is considered \RevisedText{a flow state where turbulence is suppressed and} polymer effects are \RevisedText{no longer} important.
It poses more challenge for the elastic theory since elastic energy inherently depends on polymer properties.

Finally, neither theory provides quantitative prediction for the velocity ``magnitude'' at MDR.
In a related development, L'vov, Procaccia, and coworkers have proposed a model for the mean velocity profile incorporating a number of empirical approximations~\citep{Lvov_Procaccia_PRL2004,Procaccia_Lvov_RMP2008}.
One notable approximation that allows the coupling of polymer effects into the mean flow prediction involves the definition of an ``effective viscosity'' $\eta^\text{eff}_\text{p}(y)$ (which\RevisedText{,} inevitably\RevisedText{, must} vary with $y$) relating the polymer shear stress to the mean shear rate
\begin{gather}
	\tau_{\text{p},xy}\equiv\eta^\text{eff}_\text{p}(y)\frac{dU_\text{m}(y)}{dy}
\end{gather}
and the assumption that it \RevisedText{grows linearly with} wall distance \RevisedText{$|y-y_\text{w}|$}
outside the viscous sub-layer.
The model predicts DR which increases with \RevisedText{the} $\eta^\text{eff}_\text{p}$ magnitude.
Its solution at $\tau_{\text{R},xy}=0$ gives a $U_\text{m}^+(y^+)$ profile agreeing with the Virk \RevisedText{logarithmic profile} (\cref{eq:virk}).

This development is sometimes characterized \RevisedText{in the literature} as a viscous theory for DR, which is not entirely accurate. The effective viscosity $\eta^\text{eff}_\text{p}(y)$ is not a \RevisedText{molecular} viscosity in Lumley sense \RevisedText{as it includes effects beyond} molecular momentum transport.
\RevisedText{Indeed}, it is an empirical model function
lumping together the effects of polymer-turbulence dynamics, without specifying the nature of those interactions.
It is thus more appropriate to be categorized as a phenomenological model than as a theory.
\RevisedText{Its} MDR prediction builds on the assumption that turbulence is completely quenched (zero $\tau_{\text{R},xy}$), which is what the viscous theory would predict. On the other hand, polymer shear stress $\tau_{\text{p},xy}$ (through the effective viscosity function) is high, which prevents the mean velocity from going back to the laminar value. The source of $\tau_{\text{p},xy}$ is not specified. Thus the model does not necessarily preclude the elastic mechanism which also predicts the polymer stress to exceed the Reynolds stress at $r<r^{**}$.

\section{Recent advances}\label{sec:recent}
The previous section is intended to provide an overview of the major phenomenology and fundamental understanding\RevisedText{s} of DR \RevisedText{that have been more established} over the decades.
There are evidently many outstanding gaps in \RevisedText{those} understanding\RevisedText{s}, most notably in the HDR and MDR regimes, where progress has \RevisedText{historically} been \RevisedText{limited.}
Several new developments and breakthroughs \RevisedText{on these fronts have taken} place over the past ten years.
\RevisedText{These developments show}
significant breakaway from \RevisedText{the} established approaches in this field and, as such, bring forward new lines of thought in the mechanistic inquiry of DR.
\RevisedText{They will be reviewed in this section.}

\subsection{A dynamical framework for DR research}\label{sec:dynamical}
In the classical approach of wall turbulence research, flow quantities are often averaged over homogeneous (streamwise and spanwise) directions and over time.
\RevisedText{R}esults are typically presented \RevisedText{using} profiles of mean velocity, RMS velocity fluctuations, and Reynolds stress\RevisedText{.  B}alances of shear stress and TKE \RevisedText{are also commonly studied}.
Theoretical analysis would thus focus on turbulent transport between different wall layers and neglect both \RevisedText{spatial and} temporal intermittency.
Earlier discussion \RevisedText{of} the viscous mechanism for DR has already shown some limitation of this perspective. In particular, it seems that the Lumley theory can better explain DNS results if \RevisedText{the characteristic} extensional viscosity is interpreted \RevisedText{as the} transient local magnitudes \RevisedText{of $\eta^\text{ext}$} at regions of strong deformation, rather than its ensemble average.
\RevisedText{A number of} studies \RevisedText{from the past decade} have revealed significant new insights hidden in the dynamical intermittency, which prompted a wave of developments in DR research.
Recent review by \citet{Graham_POF2014} specifically focused on intermittency, dynamical systems, and MDR, which covered many of the\RevisedText{,} of course\RevisedText{,} earlier developments in this wave.

\subsubsection{Minimal flow unit (MFU)}\label{sec:mfu}
Coherent structures are spatially and temporally recurrent structural patterns in a turbulent flow field.
Spatial patches showing \RevisedText{such} patterns, such as well-defined vortices and streaks, can be \RevisedText{clearly} identified \RevisedText{in DNS and experimental flow visualization results}. \RevisedText{Those} patterns do not strictly repeat themselves but share many similarities.
To a first approximation, the flow field of near-wall turbulence can be viewed as an ensemble of individual structural units, each evolving through its own life cycle. Although no pair of such units are identical at any given moment, all units are statistically equivalent.

These so-called minimal flow units (MFUs) can be numerically isolated by \RevisedText{constraining} the periodic simulation domain to the smallest size that \RevisedText{still} sustain\RevisedText{s} turbulence. For Newtonian flow, \citet{Jimenez_Moin_JFM1991} pioneered this idea and found that \RevisedText{a} MFU reflects the correct length scales of coherent structures\RevisedText{:} e.g., its spanwise domain size $L_z^+\approx 100$ is comparable to the well-established characteristic near-wall streak spacing\RevisedText{~\citep{Smith_Metzler_JFM1983,Robinson_ARFM1991}}.
At their relatively low $\mathrm{Re}_\tau$ ($100\sim200$), flow structures captured in \RevisedText{a} MFU resemble that of the ECS (\cref{fig:ecs})\RevisedText{---}for a significant portion of time, \RevisedText{the MFU} contains one low-speed streak straddled by staggered streamwise vortices. Unlike ECS which is stationary, the MFU approach captures the dynamical life cycles of such structures.
Remarkably, time-averaged flow statistics of \RevisedText{a} MFU also quantitatively agree with those of large-scale turbulence.

The approach was extended to viscoelastic turbulence more recently by \citet{Xi_Graham_JFM2010}. It was found that the domain size of \RevisedText{a} MFU increases with increasing $\mathrm{Wi}$ and $\mathrm{DR}\%$, which resonates with the observation of increas\RevisedText{ed} streak spacings accompanying DR.
At a low $\mathrm{Re}_\tau=84.85$ and for $\mathrm{Wi}$ up to $29$ tested \RevisedText{(under constant $\beta=0.97$ and $b=5000$)}, viscoelastic MFUs are still dominated by streamwise vortices and streaks.
\RevisedText{However, t}he study \RevisedText{reported} strong intermittency in instantaneous wall shear rate \RevisedText{$\dot\gamma_\text{w}$} magnitudes which also \RevisedText{seems} to correlate with variations in turbulence strength.
Especially at higher levels of $\mathrm{DR}\%$, instants with significantly lower wall shear rate are frequently observed, during which vortex strength, \RevisedText{as} measured by \RevisedText{the} $Q$ quantity (\cref{eq:q}), experiences multifold decrease\RevisedText{s}.

\begin{figure}
	\centering
 	\includegraphics[width=\linewidth, trim=0 0 0 0, clip]{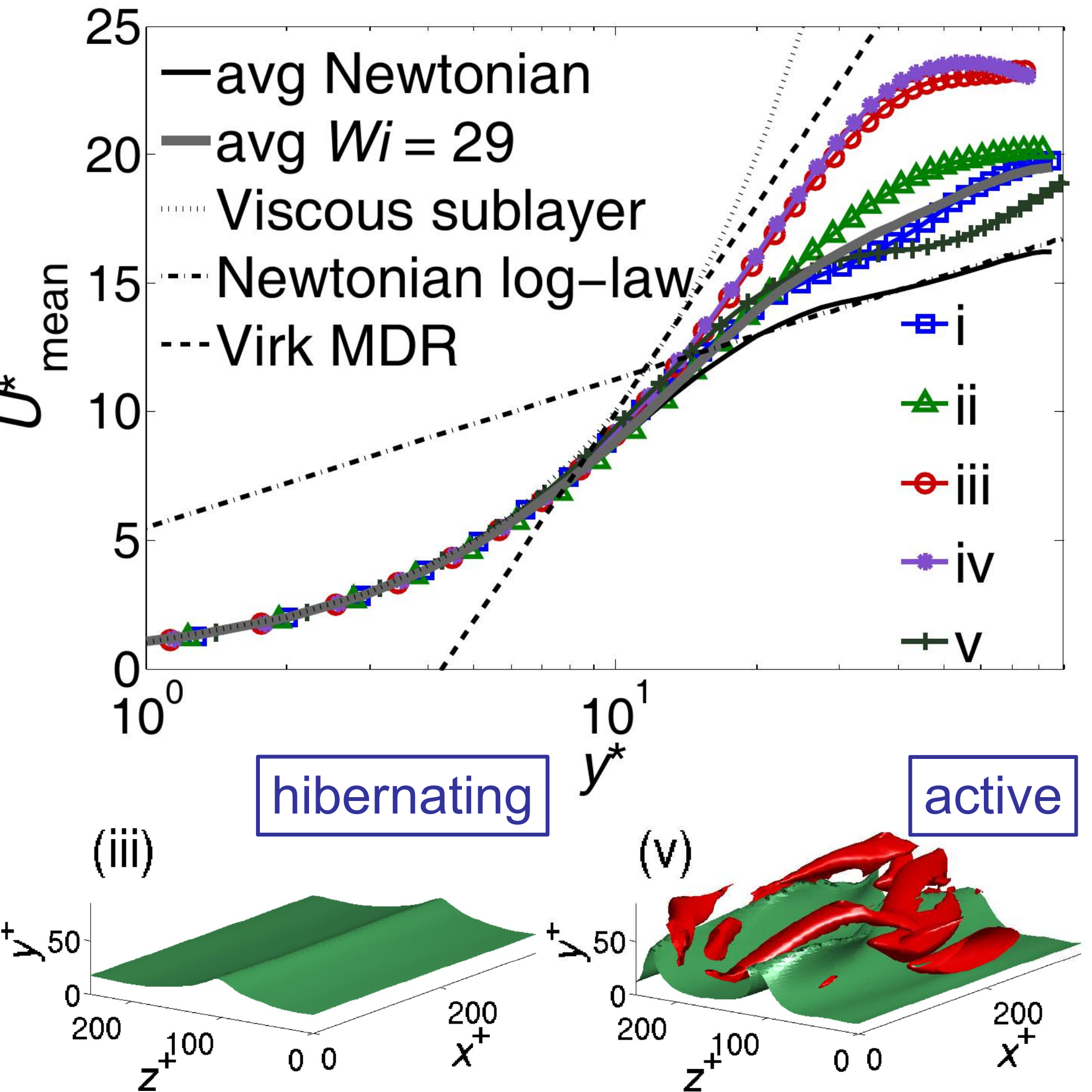}\\
	\caption{Flow statistics and structures around a typical hibernating \RevisedText{interval} in a minimal flow unit (MFU) ($\mathrm{Re}_\tau=84.85$, $\mathrm{Wi}=29$, $\beta=0.97$, $b=5000$): (top) instantaneous mean velocity profiles; (bottom) flow structures during representative moments.
	Moments (i) to (v) are chronologically ordered and cover the transition from active (i), to hibernating (ii, iii, iv), and finally back to active (v) phases.
	Instantaneous wall shear stress at the nearest wall is used to define inner scales and quantities so-nondimensionalized are marked with an asterisk. Time average Newtonian and $\mathrm{Wi}=29$ profiles are shown in solid lines.
	Isosurfaces show constant $v_x$ (light/green) and $Q$ (dark/red, \cref{eq:q}); the same isosurface levels are used for both images.
	(Reprinted figure with permission from
	\RevisedText{\citeauthor*{Xi_Graham_PRL2010},
	Physical Review Letters, 104, 218301,
	\citeyear{Xi_Graham_PRL2010}.}
	Copyright (\citeyear{Xi_Graham_PRL2010}) by the American Physical Society.)
	}
	\label{fig:hibernation}
\end{figure}

\subsubsection{Phenomenology: active and hibernating turbulence}\label{sec:hibernation}
\paragraph{The concept}
These low-shear events are later known as ``hibernating turbulence'', a term coined by \citet{Xi_Graham_PRL2010} in their immediate follow-up study.
They discovered that dynamics in \RevisedText{a} MFU undergoes clear transitions between distinct phases with strong (active turbulence) and weak (hibernating turbulence) turbulent activities (\cref{fig:hibernation}).
By following the time series of common indicators of turbulent activities such as wall shear stress and $\tau_{\text{R},xy}$, hibernating turbulence can be identified as distinct \RevisedText{intervals} during which those quantities drop to uncharacteristically low magnitudes.

Although such events are more often observed at higher $\mathrm{Wi}$ and higher $\mathrm{DR}\%$, viscoelasticity is not a prerequisite. In Newtonian turbulence, the instantaneous wall shear stress or $\tau_{\text{R},xy}$ can be seen fluctuating around its time-average value for most of the time, but occasionally (every $O(1000)l/U$ on average at $\mathrm{Re}_\tau=84.85$), the signal drops significantly as the flow enters the hibernating phase. It would bounce back to the normal level (active phase) after spending $O(100)l/U$ in hibernation.
Hibernating events are rather rare at the Newtonian limit where the time-average flow quantities almost entirely reflect characteristics of active turbulence. Its frequency stays low until well after the onset of DR but rises steadily \RevisedText{after a higher critical $\mathrm{Wi}_\text{c}$}.
For the only case \RevisedText{so far} where direct comparison \RevisedText{has been} made, \RevisedText{$\mathrm{Wi}_\text{c}$ for the} increase of hibernating frequency coincides with \RevisedText{that of} the LDR-HDR transition \RevisedText{$\mathrm{Wi}_\text{LDR-HDR}$, as} observed in \citet{Xi_Graham_JFM2010}.
At the highest $\mathrm{Wi}=29$ reported, the average duration of active \RevisedText{intervals} 
drops by one order of magnitude to $O(100)l/U$\RevisedText{, while, interestingly, that of hibernating \RevisedText{intervals} remain nearly unchanged from the Newtonian limit}.

The term ``hibernating'' turbulence was thus intended as a metaphoric \RevisedText{reference to its} most notable dynamical \RevisedText{trait}, that the flow enters a state of weak activities for an extended time \RevisedText{interval} but will eventually ``revive'' by itself to normal activity.
Indeed, flow fields during hibernation show drastically weaker vortex structures. At moment \RevisedText{(}iii\RevisedText{)} of \cref{fig:hibernation}, $Q$ values are so low across the domain that no isosurface can be found at the prescribed level\RevisedText{,} whereas clear large vortices are found at active turbulence\RevisedText{---}moment \RevisedText{(}v\RevisedText{)}\RevisedText{---}using the same $Q$ level.
A velocity streak is still observed but its strength is markedly lower. In comparison, streaks in active turbulence (moment \RevisedText{(}v\RevisedText{)}) induce stronger distortion on the velocity isosurface and they are also clearly wavier in the flow ($x$) direction.

\paragraph{Proposed link to MDR}
\Cref{fig:hibernation} shows the instantaneous mean velocity profiles for moments before (i), during (ii, iii, iv), and after (v) a representative hibernating \RevisedText{interval}.
(For instantaneous profiles, inner units\RevisedText{---}see \cref{eq:utau,eq:lv}\RevisedText{---}are defined using the wall shear stress of that moment at the nearest wall, with no average between walls or over time. They are thus marked with ``*'' instead of ``+''.)
Time average profiles for Newtonian and viscoelastic ($\mathrm{Wi}=29$) cases are also shown\RevisedText{:} the former follows closely the von~K\'arm\'an log law at $y^+>30$ and the latter is lifted with a raised slope.
Profiles of active turbulence (\RevisedText{(}i\RevisedText{)} and \RevisedText{(}v\RevisedText{)}) imitate Newtonian and LDR profiles\RevisedText{---}they appear roughly parallel to the von-K\'arm\'an \RevisedText{asymptote in the log-law layer,} although with \RevisedText{larger} fluctuations \RevisedText{than time-average profiles}.
Profiles \RevisedText{deep} in hibernation (\RevisedText{(}iii\RevisedText{)} and \RevisedText{(}iv\RevisedText{)}), strikingly, appear close to MDR. Moment \RevisedText{(}ii\RevisedText{)} is transitional and thus less raised.

\RevisedText{Q}uantitative agreement with the Virk asymptote should not be generalized considering the intermittent nature of the dynamics\RevisedText{---}no two hibernating events are identical and the $U_\text{m}^*(y^*)$ profile rises to different levels at different instances of hibernation.
Nevertheless, qualitative similarity to MDR is hard to \RevisedText{ignore}.
Regardless of the magnitude, the profile shape \RevisedText{is} unequivocally different \RevisedText{between active and hibernating phases} with the latter always showing a more lifted silhouette characteristic of HDR and MDR.
In a later study, \citet{Xi_Graham_JFM2012} showed that flow statistics between active and hibernating phases are statistically differentiable. Other than mean velocity, Reynolds stress ($\tau_{\text{R},xy}$ and RMS velocity fluctuations) profiles are significantly more suppressed at hibernation across nearly the entire domain, which is, again, similar to MDR.
Finally, the distinct patterns of turbulent structures observed during hibernation, including weak vortices and straightened streaks, are also consistent with flow visualization at HDR and MDR~\citep{White_Mungal_EXPFL2004,Yu_Kawaguchi_JNNFM2004,Housiadas_Beris_POF2005,Li_Khomami_JNNFM2006,Xi_PhD2009,Zhu_Xi_JNNFM2018}.

Above all, the clearest connection \RevisedText{is the} insensitivity \RevisedText{of hibernating turbulence} to polymer properties, which obviously is reminiscent of the universality of MDR. 
As mentioned above, the average duration of an active \RevisedText{interval} drops for one order of magnitude from the Newtonian limit to $\mathrm{Wi}=29$. Within the same range, the average duration of a hibernating \RevisedText{interval} is nearly invariant with increasing $\mathrm{Wi}$.
\citet{Xi_Graham_PRL2010} performed a numerical experiment in which polymer stress is suddenly turned off (\RevisedText{i.e.}, switching to Newtonian dynamics) at different points during the cycle in \cref{fig:hibernation}.
It was found that at moment \RevisedText{(}i\RevisedText{)}, which is immediately before the system pivots towards hibernation, removing polymer stress prevents upcoming hibernation altogether.
However, once hibernation has started, turning off polymer stress does not stop \RevisedText{the flow} from slipping deeper into hibernation.
Transition between these two behaviors \RevisedText{is} rather sharp\RevisedText{---}there appears to be a turning point after which the flow is on its path to hibernation and polymer stress is no longer relevant.
Therefore, polymers can suppress active turbulence, shorten its duration, and allow hibernation to occur more frequently, but hibernating turbulence itself is largely unaffected by polymers.

This is explained by the substantially reduced polymer extension during hibernation\RevisedText{, as a result of its} weaker turbulent \RevisedText{intensity}. Indeed, during hibernating \RevisedText{intervals}, nearly all polymer stretching can be attributed to the mean shear of the flow and polymer stretching in transverse directions, which can only come from turbulence, becomes negligible~\citep{Xi_Graham_JFM2012}.
\RevisedText{Overall, the level of polymer extension anti-correlates with the level of DR: moments with higher DR (lower drag) show less polymer stretching.
This would be counter-intuitive from the classical ensemble-average perspective of turbulence, where higher DR is associated with higher polymer stress (required for suppressing turbulence), and can only be interpreted in the dynamical framework of active-hibernating intermittency.}

\paragraph{Further evidences}
Intermittency between active and hibernating turbulence has been reported in a number of later studies from different researchers.
\citet{Pereira_Mompean_PRFluids2017} reported a similar investigation of Couette flow where
\RevisedText{the key observations}
are all well consistent with the channel flow findings \RevisedText{of \citet{Xi_Graham_PRL2010,Xi_Graham_JFM2012}}.
Similar findings were \RevisedText{also} reported in pipe flow by \citet{Lopez_Hof_JFM2019} for their low to moderate $\mathrm{Wi}$ regime (their highest $\mathrm{Wi}$ results are in a different regime to be discussed in \cref{sec:edt}).
Existence of hibernating turbulence in Newtonian flow was repeatedly confirmed~\citep{Kushwaha_Graham_PRFluids2017,Whalley_Poole_PRFluids2017,Whalley_Poole_EXPFL2019,Pereira_Mompean_JFM2019}, mostly at $\mathrm{Re}$ close to the L-T transition where these states are more frequently visited.
Conditionally-averaged flow statistics and structures in those hibernating \RevisedText{intervals} again clearly resemble MDR observations.

\citet{Tamano_Graham_JFM2011} reported a DNS study of viscoelastic boundary layer flow, where the streamwise direction is no longer spatially homogeneous and temporal intermittency is reflected in the downstream spatial variation of flow quantities.
They observed that
regions with higher DR show lower polymer stretching \RevisedText{and vice versa,
which is consistent with the same anti-correlation in the temporal dynamics of \citet{Xi_Graham_JFM2012}:
in particular, note the strong similarity between fig.~13 of \citet{Tamano_Graham_JFM2011} and fig.~30 of \citet{Xi_Graham_JFM2012}.}
They also performed a numerical test \RevisedText{similar to that in \citet{Xi_Graham_PRL2012} by turning} off polymer stress after a certain downstream position with low drag and found that the low drag state can survive a substantial distance further downstream without polymer stress\RevisedText{, which reaffirms the conclusion that polymer stress is not important within the hibernating phase.}

\paragraph{Connecting MFU dynamics \RevisedText{with large-scale} turbulence}
\RevisedText{Following the discussion in} \cref{sec:mfu}, MFU can be mapped to realistic large\RevisedText{-}scale flows \RevisedText{by invoking the ergodicity hypothesis of turbulence~\citep{McComb_1990}, which postulates that both} temporal and spatial statistics are \RevisedText{equivalent to} ensemble statistics.
In this sense, temporal intermittency in \RevisedText{a} MFU would map to spatial intermittency in a larger domain.
A single structural unit can experience active and hibernating phase\RevisedText{s} at different times. If a larger flow domain is viewed as a \RevisedText{jigsaw assembled by} such units, at any moment, some units will be caught at the active phase and some in hibernation. \RevisedText{Their p}roportions should reflect the \RevisedText{time share of} each phase \RevisedText{in a single unit}.
Of course, this would be neglecting any correlation between \RevisedText{individual} units, which is a non-trivial assumption (see \cref{sec:largebox}).

This view was adopted by \citet{Kushwaha_Graham_PRFluids2017} and \citet{Wang_Graham_JNNFM2017} for Newtonian and viscoelastic flows, respectively.
They divided DNS flow fields of an extended domain into active and hibernating patches based on magnitudes of local velocity gradient in the buffer layer. Conditionally-averaged flow statistics of these two categories compare well with MFU results.
Conditional average was also applied to large-scale experimental measurements based on local wall shear stress~\citep{Whalley_Poole_PRFluids2017,Whalley_Poole_EXPFL2019}, which again confirmed the same picture.
\citet{Kushwaha_Graham_PRFluids2017} showed that \RevisedText{conditional averages} of active and hibernating turbulence based on temporal intermittency and spatial intermittency analysis are consistently alike.

\subsubsection{DR from a dynamical systems perspective}\label{sec:drdynsys}
\begin{figure}
	\centering
 	\includegraphics[width=\linewidth, trim=0 0 0 0, clip]{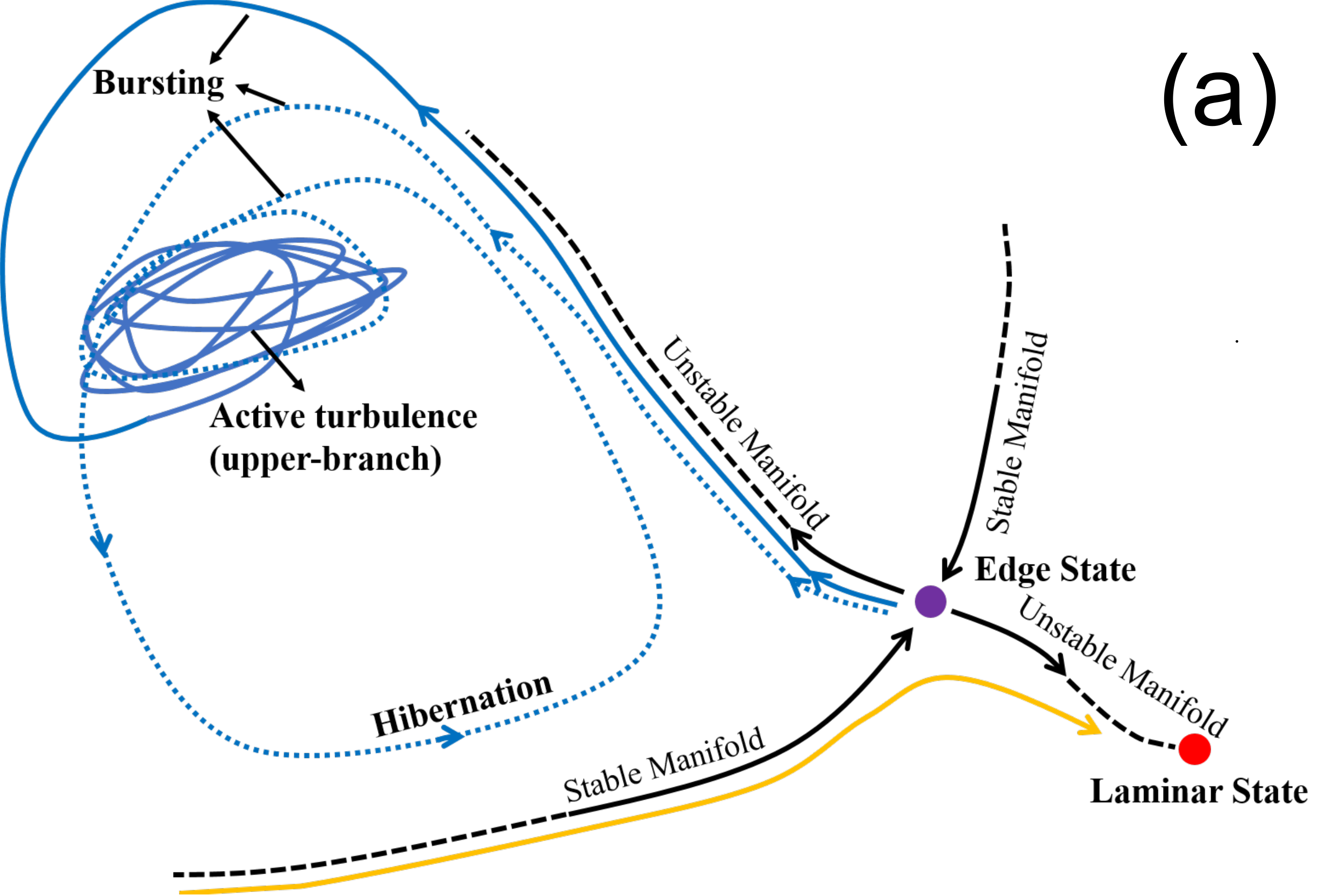}\\
 	\includegraphics[width=\linewidth, trim=0 0 0 0, clip]{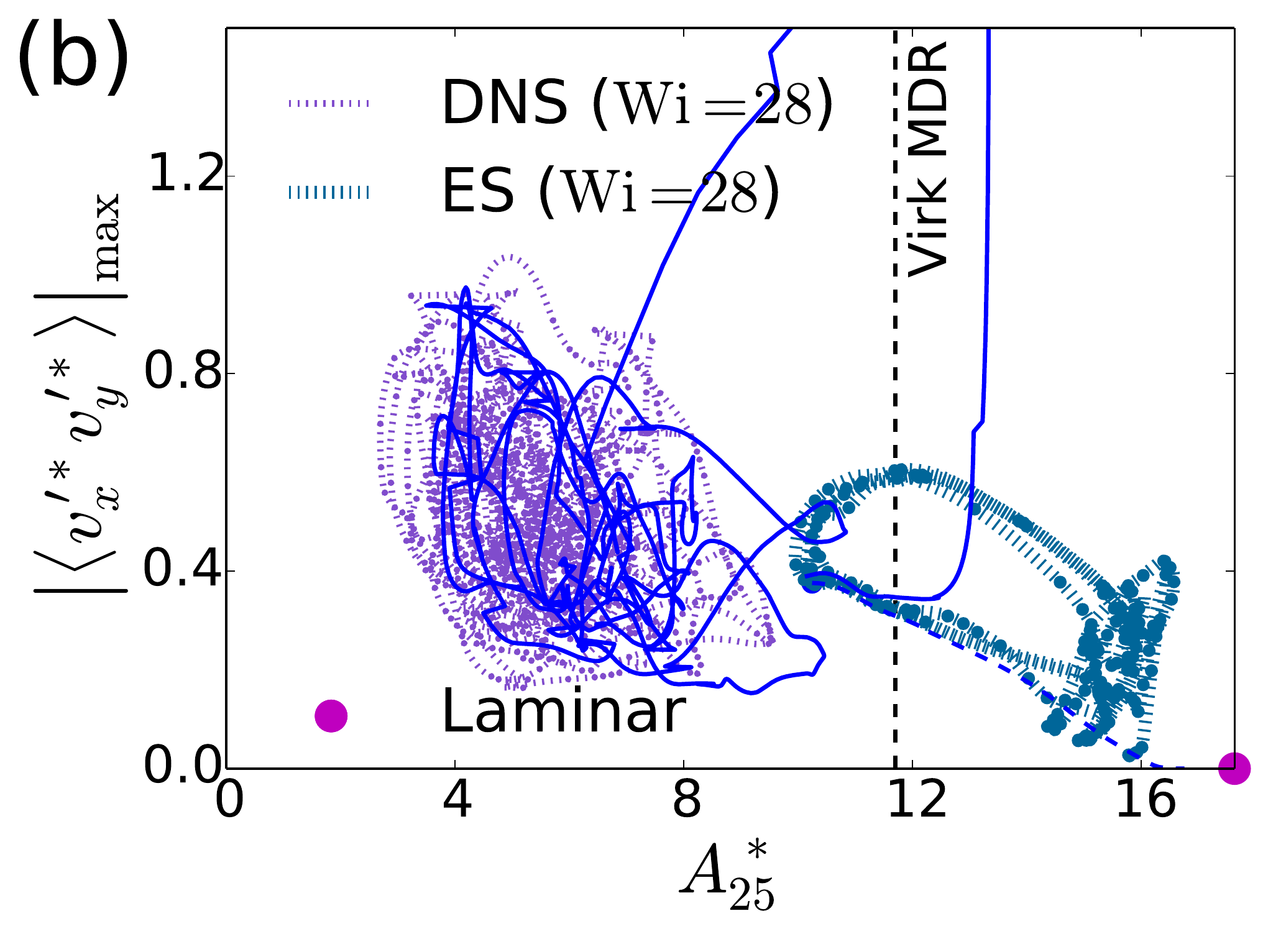}\\
	\caption{State-space dynamics of active-hibernating-bursting (AHB) cycles and relevant solution objects:
	(a) schematic representation
	(\RevisedText{r}eprinted from
	\RevisedText{Journal of Non-Newtonian Fluid Mechanics,
	266, 127,
	\citeauthor*{Zhu_Xi_JNNFM2019},
	Transient dynamics of turbulence growth and bursting: effects of drag-reducing polymers}%
	\RevisedText{; c}opyright (\citeyear{Zhu_Xi_JNNFM2019}), with permission from Elsevier.)
	and
	(b) numerical results of viscoelastic flow projected on a $\tau_{\text{R},xy}^*$ profile peak vs. buffer-layer $U_\text{m}^*(y^*)$ logarithmic slope 2D plane (\RevisedText{r}eprinted figure with permission from 
	\RevisedText{\citeauthor*{Xi_Bai_PRE2016},
	Physical Review E,
	93, 043118,
	\citeyear{Xi_Bai_PRE2016}}
	\RevisedText{; c}opyright (\citeyear{Xi_Bai_PRE2016}) by the American Physical Society.).
	\RevisedText{In (b), s}olid blue line is a trajectory initiated from the edge state and ending at the turbulence kernel\RevisedText{; v}ertical dashed line marks $A^+_\text{Virk}=11.7$.
	The boundary between laminar and turbulent basins, i.e., the edge of chaos (EoS), is formed by the stable manifold of the edge state (ES).
	%
	}
	\label{fig:statespace}
\end{figure}
\paragraph{Solutions in the state space}
From a dynamical systems perspective, turbulent dynamics obtained from \RevisedText{a} MFU can be viewed as a chaotic dynamical trajectory in a state space governed by numerous invariant solutions such as ECS (see \cref{fig:statespace}(a) \RevisedText{for a simple case})~\citep{Ashwin_Timme_Nature2005,Holmes_Lumley_2012,Kawahara_ARFM2012}.
Other than the laminar state, none of \RevisedText{those} solutions are linearly stable. All known ECS solutions are saddle points in the state space, having both stable and unstable directions (stable and unstable manifolds).
(An illustrative visualization of how \RevisedText{invariant} solutions and their unstable manifolds guide the dynamics of turbulence in \RevisedText{a} MFU was shown in fig.~9 of \citet{Gibson_Cvitanovic_JFM2008}.)
For Newtonian turbulence close to the L-T transition (low $\mathrm{Re}$), dynamics \RevisedText{in the ``kernel'' of strong turbulence} is dominated by so-called upper-branch (UB) ECS solutions~\citep{Jimenez_Kawahara_POF2005,Gibson_Cvitanovic_JFM2008,Park_Graham_JFM2015}\RevisedText{, while h}ibernating \RevisedText{intervals} appear to be intermittent escapes from \RevisedText{the kernel}.

Since hibernating turbulence (1) shows distinctly weak turbulent activities and (2) exists in both Newtonian and viscoelastic flows,
\RevisedText{for its origin, one would first}
look for solutions representing weak turbulent states in Newtonian flow.
In canonical flow types, transition to turbulence follows a nonlinear mechanism and requires finite-amplitude disturbances.
(For channel flow, a linear instability
\RevisedText{of the laminar state does occur at $\mathrm{Re}_\tau\approx 107.4$,}
\RevisedText{which leads}
to a new solution called \RevisedText{the} Tollmien-Schlichting \RevisedText{or} T-S wave~\citep{Drazin_Reid_1982,Jimenez_POF1987}\RevisedText{. Since} \RevisedText{$\mathrm{Re}_{\tau,\text{crit}}\approx 44.7$} for the L-T transition \RevisedText{is much lower}~\citep{Nishioka_Asai_JFM1985}\RevisedText{, a nonlinear mechanism is the only pathway for the transition at $\mathrm{Re}_\tau\lesssim107.4$.}
Without further complicating the discussion, \RevisedText{the linear instability scenario will not be discussed which does not exist in pipe or Couette flows.)}
Therefore, the state space has \emph{at least} (\RevisedText{additional solution(s) may exist; see} \cref{sec:edt}) two basins of attraction\RevisedText{:} laminar and turbulent regions.
The ridge separating them can thus be considered the most marginal or weakest form of turbulence. This so-called ``edge of chaos'' (EoC) can be numerically computed, noting that trajectories above and below the \RevisedText{ridge} divert to opposite basins, through repeated bisections~\citep{Skufca_Eckhardt_PRL2006,Schneider_Eckhardt_Chaos2006}.
For the relatively low $\mathrm{Re}$ studied so far, dynamics along the EoC converges to an asymptotic solution called the edge state (ES) (\cref{fig:statespace}(a)), which often appears \RevisedText{as} a quasi-periodic (\cref{fig:statespace}(b)) or non-periodic oscillatory (with varying period length) trajectory itself~\citep{Zammert_Eckhardt_FDR2014,Xi_Bai_PRE2016}.
The ES has only one unstable direction~\citep{Skufca_Eckhardt_PRL2006}\RevisedText{, which} means for an $N$-dimensional system ($N$ being the number of degrees of freedom), its stable manifold \RevisedText{spans} an $(N-1)$-dimensional subspace \RevisedText{and} forms the separating surface between the two basins\RevisedText{---}i.e., the EoC.
Lower-branch (LB) ECS solutions with one-dimensional unstable manifolds have been found and are believed to be the key constituents of the ES ~\citep{Wang_Waleffe_PRL2007,Park_Graham_PRFluids2018}.

\citet{Xi_Graham_PRL2012} computed the ES for viscoelastic channel flow whose flow structure and mean velocity profile turn out to be remarkably similar to hibernating turbulence.
Most importantly, mean flow of the ES was found to be insensitive to viscoelasticity\RevisedText{---}from Newtonian flow to $\mathrm{Wi}=28$, the profiles seamlessly overlap with one another.
The collapsed profile at $\mathrm{Re}_\tau=84.85$ happens to align with the Virk profile (\cref{eq:virk}), although this agreement cannot be generalized to higher $\mathrm{Re}$~\citep{Xi_Bai_PRE2016}.
This insensitivity, same as the case of hibernating turbulence, is again caused by the lack of polymer stretching in this weak turbulent state, where the flow kinematics is dominated by shear rather than extension or rotation~\citep{Xi_Bai_PRE2016}.

\paragraph{Active-hibernating-bursting (AHB) cycles}\label{par:ahbcycle}
The proposed dynamical scenario is sketched in \cref{fig:statespace}(a).
For Newtonian flow, a turbulent trajectory would spend most time sampling a well-defined kernel region of strong (active) turbulence formed by UB solutions.
Occasionally, it may hit certain points of exit in the kernel region and embark on excursion towards the laminar state, only to be blocked by the EoC.
The trajectory will then travel along the EoC until it reaches the vicinity of the ES, where it will be pivoted towards the unstable direction of the latter and start its return journey.
During the excursion, the flow spends substantial time in regions with very weak turbulence, which is reflected as a hibernating \RevisedText{interval} in DNS.

Signs of \RevisedText{such} cycle\RevisedText{s} can be spotted in earlier DNS studies. For instance, \citet{Min_Choi_JFM2003b} obtained their viscoelastic solutions by suddenly switching on viscoelasticity \RevisedText{from} a Newtonian solution. After leaving the Newtonian kernel, all solutions (different $\mathrm{Wi}$) can be seen approaching \RevisedText{the same} low-drag (drag coefficient close to but higher than the laminar magnitude) asymptotic state and stay\RevisedText{ing} in its vicinity for $O(\num{e2})l/U$, before bouncing backwards to \RevisedText{their} corresponding viscoelastic kernel\RevisedText{s}.
\RevisedText{Interestingly, the simulation domain used in that study}
is much larger than MFU\RevisedText{s where dynamical cycles involving active and hibernating states were later more systematically studied~\citep{Xi_Graham_PRL2010,Xi_Graham_JFM2012,Xi_Graham_PRL2012,Xi_Bai_PRE2016}.}

The connection of LB ECSs (which, as discussed, are believed to dominate the ES) to hibernation \RevisedText{is} supported by their direct comparison with conditionally-averaged hibernating events in extended flow domains \RevisedText{from} both experiment\RevisedText{s} and \RevisedText{DNS}~\citep{Whalley_Poole_PRFluids2017,Whalley_Poole_EXPFL2019}\RevisedText{.}
Furthermore, \RevisedText{both} \citet{Pereira_Mompean_JFM2019} and \citet{Kushwaha_Graham_PRFluids2017} found that \RevisedText{the} frequency of hibernation in Newtonian channel flow increases with decreasing $\mathrm{Re}$, which is consistent with the \RevisedText{common belief} that active and hibernating turbulence are organized around UB and LB ECSs, respectively\RevisedText{---}solutions on the two branches \RevisedText{are} closer\RevisedText{, and thus the transition frequency becomes higher,} as $\mathrm{Re}$ decreases towards the bifurcation point.
With increasing $\mathrm{Re}$, hibernation becomes rare, which only becomes unmasked when $\mathrm{Wi}$ is higher~\citep{Wang_Graham_JNNFM2017}.

Returning through the unstable manifold of the ES often leads to a strong overshoot of turbulent intensity\RevisedText{---}a bursting event\RevisedText{---}before it settles back to the active kernel \RevisedText{(\cref{fig:statespace}(a))}.
The term ``bursting'', although prevalent in turbulence literature, is not \RevisedText{unequivocally} defined and interpretation varies depending on the context.
Here, it refers to events marked with strong sharp increase in TKE.
These events are strongly non-equilibrium \RevisedText{in nature} (as opposed to quasi-steady-state solutions like ECS). Their importance in the turbulence self-sustaining cycle was pointed out by \citet{Jimenez_Kawahara_POF2005}.
In time series showing hibernating \RevisedText{intervals}, a strong burst of TKE\RevisedText{, as well as} other quantities \RevisedText{associated with} turbulence intensity\RevisedText{,} usually follows hibernation~\citep{Xi_Graham_PRL2010,Xi_Graham_JFM2012}.
\citet{Pereira_Mompean_JFM2019} showed similar observations. They made the differentiation between ``strong'' and ``moderate'' active states, which, in our terminology, roughly correspond to bursting and active phases.
\citet{Kushwaha_Graham_PRFluids2017} conditionally averaged instances of hibernating \RevisedText{intervals by} aligning them at their end point\RevisedText{s}. The result \RevisedText{clearly} showed that, statistically, hibernating \RevisedText{intervals} are \RevisedText{often} followed by strong overshoots of wall shear stress\RevisedText{---}i.e., ``bursting''.

\citet{Zhu_Xi_JNNFM2019} systematically studied the nature of bursting events and showed that trajectories following the unstable manifold of the ES \RevisedText{\textit{en route} to the turbulent kernel} demonstrate well-defined bursting behaviors.
It starts with quick intensification and lift up of near\RevisedText{-}wall vortices, which results in \RevisedText{a} sharp rise of the RSS (and thus TKE production\RevisedText{---}see \cref{eq:tkeP}).
These primary vortices then undergo swift rupture\RevisedText{, generating} countless much smaller but intense vortex pieces \RevisedText{as their debris}, which is accompanied by TKE quickly reaching its peak.
Viscous dissipation then stabilizes the flow and regulates the fluctuations into larger vortices typical of the kernel \RevisedText{of active turbulence}.
\citet{Park_Graham_PRFluids2018} also showed that strong bursting events in \RevisedText{a} MFU are closely aligned with the unstable manifolds of a class of LB solutions likely sitting on the ES~\citep{Park_Graham_JFM2015}.
The scenario depicted above resonates well with \RevisedText{the} earlier KL analysis of MFU solutions by \citet{Webber_Sirovich_POF1997}. \RevisedText{They} showed that at the early stage of a bursting event (they used the term ``entropy event''), the representational entropy between modes reaches minimum, which indicates that TKE \RevisedText{is} contained in a few large-scale modes\RevisedText{---}i.e., \RevisedText{the} primary vortices. A quick rise in entropy soon follows, which corresponds to the later rupture of \RevisedText{primary} vortices and redistribution of TKE between a wide range of scales.

DNS trajectories of AHB cycles, projected to a two-dimensional (2D) plane, are shown in \cref{fig:statespace}(b)~\citep{Xi_Bai_PRE2016}. A numerically computed ES \RevisedText{solution} is also shown which appears as a quasi-periodic orbit \RevisedText{located} between the laminar state and turbulent kernel.
Starting from the ES and with minimal disturbances \RevisedText{(see the solid blue line in \cref{fig:statespace}(b))}, instabilities would grow and lead to a strong bursting event (overshoot) before the flow decays to the turbulent kernel.
Afterwards, occasional escapes and excursions\RevisedText{, i.e., hibernating events, are} observed, during which the flow visits region\RevisedText{s} close to the ES.
Bursting is \RevisedText{typically} observed
\RevisedText{after those} visits\RevisedText{, which is} reflected in increased $\tau_{\text{R},xy}^*$ but not necessarily in $A^*_{25}$\RevisedText{.
($A^*_{25}$ is the logarithmic slope calculated, according to  \cref{eq:Aplus}, from instantaneous $U_\text{m}^*(y^*)$ profiles at $y^*=25$,}
which is a mean flow quantity \RevisedText{not directly tied to instantaneous turbulent intensity.)}

More detailed state-space visualization can be found in fig.~11 of \citet{Park_Graham_JFM2015} and fig.~9 of \citet{Park_Graham_PRFluids2018}. Although those authors did not directly compute the ES, they showed that hibernating turbulence corresponds to intermittent visits to the vicinity of a class of LB solutions believed to be part of the ES, all having only one unstable dimension.
\citet{Park_Graham_PRFluids2018} also noted that
although hibernation is a necessary precursor to strong bursts, not every hibernating \RevisedText{interval} is followed by clear bursting.
This is not surprising considering the intermittent nature of the dynamics\RevisedText{:} not every hibernation \RevisedText{trip} reaches \RevisedText{close} proximity to the ES. The returning trajectory is only expected to follow the unstable manifold closely\RevisedText{, which leads to strong bursting,} if it gets very close to the ES during hibernation.

\paragraph{Theory for MDR and HDR}
We are now ready to discuss the effects of polymers on this AHB dynamical cycle from which a new theory for explaining MDR will arise.
AHB cycles exist in both Newtonian and viscoelastic flows, but at high $\mathrm{Wi}$, the time scale for active turbulence is severely reduced and the frequency of hibernation, whose drag is substantially lower, is raised.
Although this could as well increase the frequency of bursting, whose drag is even higher than active turbulence, \citet{Zhu_Xi_JNNFM2019} showed that polymers can suppress the strongest form of bursting and reroute the dynamics to avoid vortex rupture and intense fluctuations.
The net \RevisedText{outcome} is that both time-averaged flow quantities and \RevisedText{statistically dominant} flow structures will be increasingly represented by hibernation, which, as discussed in \cref{sec:hibernation}, shows key characteristics of MDR.
It was thus postulated that at sufficiently high $\mathrm{Wi}$, active \RevisedText{intervals} will be so short that hibernation becomes predominant. This is when the overall flow converges to MDR.

A simple yet quantitative mathematical model was derived, by \citet{Xi_Graham_JFM2012}, based on this conceptual picture, assuming that (1) at $\mathrm{Wi}$ higher than a critical value $\mathrm{Wi}_\text{c}$, polymer chains are persistently stretched in active turbulence \RevisedText{where} polymer stress grows exponentially with time, (2) finite extensibility of polymer chains is neglected, (3) polymer stretching rate at active turbulence is independent of $\mathrm{Wi}$, and (4) active turbulence \RevisedText{is} suppressed once the nondimensional polymer stress reaches a threshold value $S_\text{T}$.
It leads to a simple expression for the duration of active \RevisedText{intervals} 
\begin{gather}
	T_\text{act}=\frac{\ln S_\text{T}}{\mathrm{Wi}_\text{c}^{-1}-\mathrm{Wi}^{-1}}
	\RevisedText{\quad(\mathrm{Wi}>\mathrm{Wi}_\text{c})}
	\label{eq:tact}
\end{gather}
which fits perfectly with DNS data at \RevisedText{the} high\RevisedText{-}$\mathrm{Wi}$ \RevisedText{end}.
\RevisedText{(At lower $\mathrm{Wi}$, $T_\text{act}$ is independent of $\mathrm{Wi}$.)}
For $\mathrm{Re}_\tau=84.85$ used in that study, the fitting yields $\mathrm{Wi}_\text{c}=17.36$, which is a stunning agreement with the observed value of $\RevisedText{\mathrm{Wi}_\text{c}}\approx 18$
\RevisedText{(where $T_\text{act}$ starts to decline)}.
At the limit of $\mathrm{Wi}\to\infty$, the model predicts $T_\text{act}\approx 127 l/U$, significantly below the average duration of hibernation $T_\text{hib}\approx 200 l/U$, which affirms the postulation that MDR is a flow state where hibernation becomes \RevisedText{statistically dominant}.

\begin{figure}
	\centering
  	\includegraphics[width=\linewidth, trim=0 0 0 0, clip]{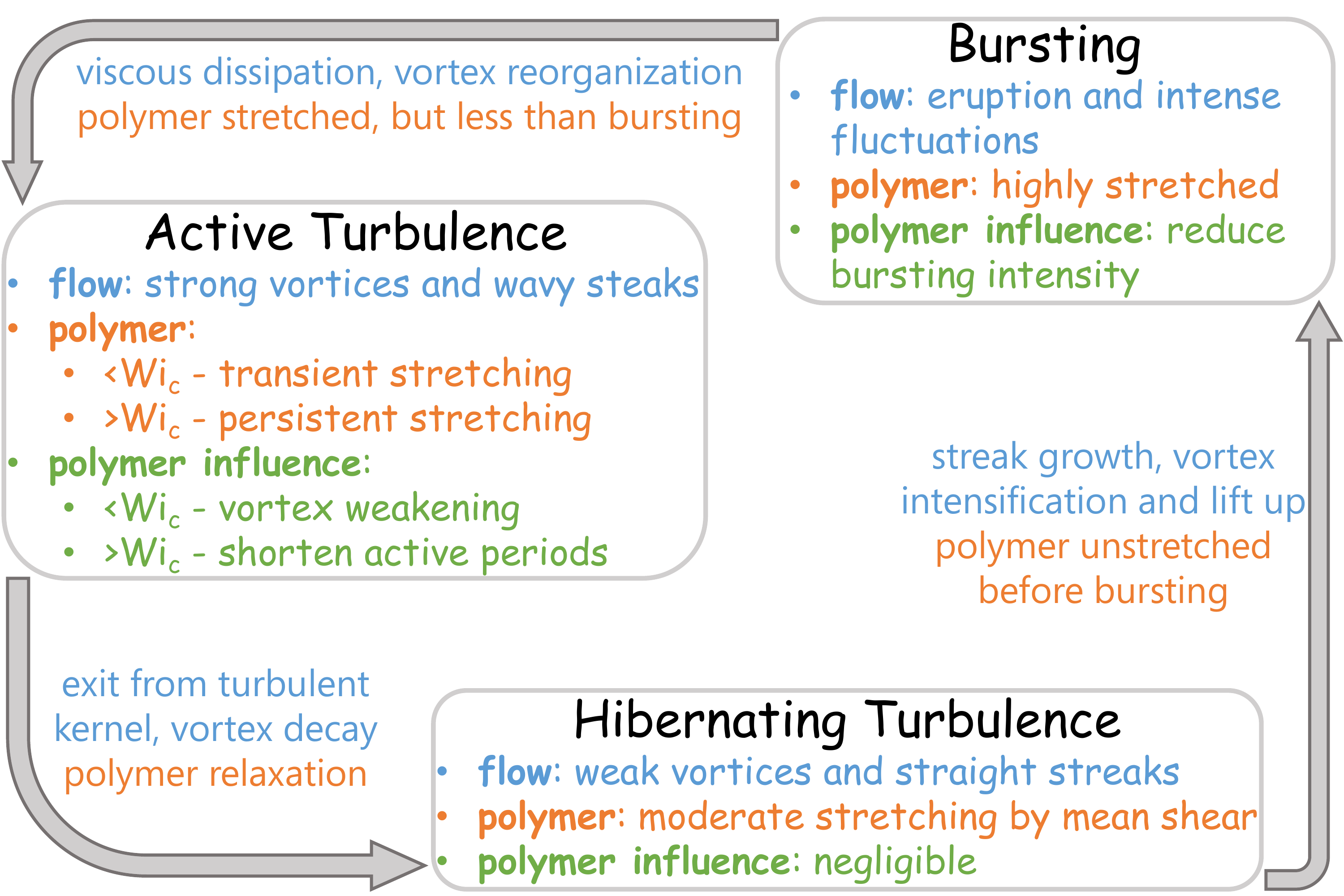}\\
	\caption{Polymer-turbulen\RevisedText{ce} dynamics in active-hibernating-bursting (AHB) cycles\RevisedText{---}a schematic based on findings in \citet{Xi_Graham_PRL2010,Xi_Graham_JFM2010}\RevisedText{,} \citet{Xi_Bai_PRE2016}\RevisedText{, and \citet{Zhu_Xi_JNNFM2019}}.}
	\label{fig:ahbpolymer}
\end{figure}

The overall picture is depicted in \cref{fig:ahbpolymer}. Intermittent AHB transition cycles occur in both Newtonian and viscoelastic flows, only that in the former they are rare and active turbulence lasts for long periods.
After the onset of DR, polymers are stretched at regions of strong flow motion (e.g., local extensional spots), which leads to an overall weakening of vortex structures at active turbulence (and, presumably, during bursting as well), following the mechanism described in \cref{sec:polymturb}.
\RevisedText{The} stretching is likely transient in nature and polymer stress \RevisedText{never grows to $S_\text{T}$, the level required to} disrupt active \RevisedText{intervals}.
\RevisedText{At} $\RevisedText{\mathrm{Wi}>}\mathrm{Wi}_\text{c}$ ($>\mathrm{Wi}_\text{onset}$), polymer stretching \RevisedText{in active turbulence} becomes persistent and polymer stress keeps ramping up until active turbulence is quenched, which triggers early exit to hibernation.
During hibernation, polymer chains recoil and polymer effects are minimal, but, because of the unstable nature of hibernation (both ES and LB solutions have unstable directions), instability will still grow, which \RevisedText{is often followed by} a bursting event \RevisedText{before the start of} a new active \RevisedText{interval}.

Further increasing $\mathrm{Wi}$ continues to reduce $T_\text{act}$ and increase the statistical weight of hibernating turbulence (which remains mostly unaffected).
At MDR, some form of strong turbulence (bursting or active turbulence) must still \RevisedText{exist} and appear intermittently, because polymers do not \RevisedText{stabilize} hibernating turbulence.
Indeed, \citet{Zhu_Xi_JNNFM2019} found that as the flow leaves the ES, polymer effects remain small during the initial growth of \RevisedText{instability} (intensification and lift up \RevisedText{of primary vortices}) and only become significant in later stage\RevisedText{s} of bursting (rupture of primary vortices), which occurs $\approx 50-70 l/U$ \RevisedText{after the initial instability}.
This is the same order of magnitude as the $\approx 127 l/U$ prediction of \cref{eq:tact} at the $\mathrm{Wi}\to\infty$ limit.
\RevisedText{The prediction is higher by a factor of $\sim 2$, which is not unexpected considering the assumptions behind \cref{eq:tact}, especially the neglect of finite extensibility.
It is thus inferred that}
bursting/active phases are terminated almost immediately after polymers are stretched. Thus, the only form of strong turbulence remaining at MDR would be the budding \RevisedText{stage} of instability before polymer stress is significant.

The theory essentially links MDR to a class of fundamentally Newtonian flow \RevisedText{states associated with the L-T transition, in which polymer elasticity is not important}.
This is supported by the observations of \citet{White_Dubief_JFM2018} who compared the mean velocity and RSS \RevisedText{profiles} between MDR and transitional states in Newtonian boundary layer flow (from \citet{Wu_Moin_JFM2009}) and found remarkable similarities\RevisedText{.} 
The main difference, as they noted, is that at MDR, such transient states are stabilized.
\RevisedText{A}ccording to the current theory, this ``stabilization'' is \RevisedText{more accurately described as} an intrinsically intermittent process\RevisedText{---}viscoelasticity does not make hibernation more stable \textit{per se}, but keeps driving the flow back to hibernation to make it statistically more persistent.

Compared with classical attempts, this theory takes a unique dynamical perspective and explains the transition to MDR in terms of \RevisedText{a} shifting dynamical balance between flow states.
\RevisedText{By} offering consistent explanation for two of the three key \RevisedText{attributes of} MDR, as outlined in \cref{sec:mdr}, \RevisedText{it represents} a major step forward \RevisedText{towards solving this mystery}.
For the ``existence'' \RevisedText{attribute}, the theory postulates that MDR is a form of turbulence dominated by weak streaks and vortices (hibernation) with intermittent short-lived eruption\RevisedText{s} of stronger activities (bursting/active turbulence).
The flow does not return to the laminar state because of the insensitivity of the ES and EoS to polymer effects\RevisedText{---}polymers cannot remove the barrier between turbulent and laminar basins.
For the ``universality'' \RevisedText{attribute}, the \RevisedText{explanation} is obvious: hibernating turbulence\RevisedText{, which presumably dominates} MDR\RevisedText{,} is a class of \RevisedText{fundamentally} Newtonian flow states.
It does not rely on viscoelasticity to exist and does not seem to be affected \RevisedText{by} the latter either. Polymers only unmask \RevisedText{those} states by compressing active \RevisedText{intervals}.
The question regarding the ``magnitude'' of MDR remains open, which we \RevisedText{will} discuss shortly below.

\RevisedText{T}he transition from vortex weakening by transiently stretched polymers (first DR mechanism at $\mathrm{Wi}_\text{onset}<\mathrm{Wi}<\mathrm{Wi}_\text{c}$) to the suppression of active turbulence by persistently stretched polymers (second DR mechanism at $\mathrm{Wi}>\mathrm{Wi}_\text{c}$) also offers an explanation for the LDR-HDR transition, where a second mechanism \RevisedText{is} needed to explain the sharp differences between the two regimes.
So far the coincidence between $\mathrm{Wi}_\text{c}$ and the LDR-HDR transition \RevisedText{has} only \RevisedText{been} shown for one parameter setting ($\mathrm{Re}$, $\mathrm{\beta}$, and $b$ combination). More research is \RevisedText{still} required \RevisedText{to establish this link}.

\subsubsection{Limitations and remaining gaps}\label{sec:gaps}
Despite its great promise, this theoretical framework based on ABH cycles
still has a few gaps to fill.
Developments in more recent years (after the \citet{Xi_Graham_PRL2010} 2010 \RevisedText{study}) have \RevisedText{revealed} encouraging new directions, which will be discussed in later sections.
\RevisedText{The answers,}
however, are far from complete.

\paragraph{Quantitative magnitude of MDR}
The theory falls short of offering a mean flow prediction for MDR and cannot \RevisedText{fully} explain the quantitative magnitude of the Virk profile (\cref{eq:virk}). Although Virk-like profiles \RevisedText{are often} observed both during hibernation and at the ES, it \RevisedText{is not always generalizable---}other instances of hibernation \RevisedText{or} ESs at other parameters \RevisedText{may} differ from the asymptote.
The intrinsic difficulty \RevisedText{again} stems from the dynamical nature of the theory, in which MDR is determined by not necessarily one state but the statistical ensemble of many.
Since the ES is not altered by \RevisedText{polymer elasticity}~\citep{Xi_Graham_PRL2012,Xi_Bai_PRE2016}, it is reasonable to anticipate that, near the ES, there is a band of the state space within which \RevisedText{turbulence is} too weak to be influenced by polymers.
Mean flow at MDR, based on this theory, would be determined by the average of those states weighted by their sampling frequency during hibernation.

\paragraph{\RevisedText{A}symptotic convergence}
Direct numerical evidence for the dynamical theory \RevisedText{was} only obtained for limited $\mathrm{Wi}$. For
\citet{Xi_Graham_PRL2010,Xi_Graham_JFM2012} \RevisedText{with their parameter settings}, it is up to $\mathrm{Wi}=29$.
At higher $\mathrm{Wi}$, their DNS trajectories return to the laminar state \RevisedText{after finite time periods}, which is inconsistent with \RevisedText{the expectation of} MDR \RevisedText{as} an asymptotic state of self-sustaining turbulence.
\RevisedText{The theory builds on the immutability of the EoC and ES---these solutions are required to block any laminarization attempt as they are}
asymptotically and dynamically
\RevisedText{approached by the flow trajectory. This immutability is}
postulated based on the insensitivity of ES mean flow quantities to viscoelasticity.
The observed laminarization at higher $\mathrm{Wi}$ apparently suggests that the L-T boundary (EoC and ES) becomes somehow penetrable.
\RevisedText{Although this} does not necessarily contradict the
\RevisedText{immutability of the ES mean velocity, it is at odds with the predicted asymptotic dynamics at MDR.}

\RevisedText{There are two very different, but possibly coexisting, explanations} for the premature laminarization.
One is the \RevisedText{small} domain size used in those MFU \RevisedText{studies}. \citet{Wang_Graham_AIChEJ2014} later reported that the maximum $\mathrm{Wi}$ for sustained turbulence is limited by the domain \RevisedText{size---in particular,} as $L_x^+$ increases from $360$ to $1000$, the highest turbulent $\mathrm{Wi}$ increases from $29$ to $90$ \RevisedText{(same $\mathrm{Re}_\tau, \beta$ and $b$ as \citet{Xi_Graham_PRL2010})}.
\RevisedText{Intermittency between active and hibernating states, similar to that in MFU, was also observed in their larger domains.}
Nevertheless, the asymptotic behavior of MDR was still not truly captured\RevisedText{---}although \RevisedText{their} $\mathrm{Wi}=90$ case has $\mathrm{DR}\%$ close to $60\%$ and its $U_\text{m}^+(y^+)$ profile is comparable to the Virk profile of \cref{eq:virk}, further increasing $\mathrm{Wi}$ would still \RevisedText{cause} laminarization.
Indeed, in other studies with larger domain sizes ($L_x^+$ up to $4000$), laminarization was also observed at sufficiently high $\mathrm{Wi}$~\citep{Housiadas_Beris_POF2003,Li_Khomami_JNNFM2006,Zhu_Xi_JNNFM2018}.
The other \RevisedText{possible explanation emerged} very recently. It was found that under similar flow parameter settings, a new \RevisedText{class of turbulence-like} instabilities, driven at least in part by polymer elasticity, \RevisedText{exist} at high $\mathrm{Wi}$~\citep{Samanta_Hof_PNAS2013,Sid_Terrapon_PRFluids2018}.
As discussed in section~\ref{par:ad}, turbulence \RevisedText{driven by elasticity} is very difficult to capture in DNS when AD is used~\citep{Sid_Terrapon_PRFluids2018,Gupta_Vincenzi_JFM2019}, which was the case in all above-quoted studies.
In this explanation, the \RevisedText{eventual} extinction of lasting turbulence\RevisedText{, }in the conventional sense\RevisedText{, }at high $\mathrm{Wi}$
is real, but \RevisedText{the newly-emerged ``turbulent'' states would keep}
the flow from laminarization.
Further discussion is deferred to \cref{sec:edt}, after \RevisedText{those} states are introduced.

\paragraph{Dynamics at higher $\mathrm{Re}$}
The state-space depiction illustrated in \cref{fig:statespace}(a) is based on numerical invariant solutions \RevisedText{and} DNS \RevisedText{at} fairly low $\mathrm{Re}$, typically $\mathrm{Re}\sim O(\num{e3})$ or $\mathrm{Re}_\tau \lesssim O(\num{e2})$ and \RevisedText{not very far above $\mathrm{Re}_\text{crit}$ for the L-T transition}.
Extension of this scenario to higher $\mathrm{Re}$\RevisedText{,} where experiments \RevisedText{are typically} conducted ($\mathrm{Re}_\tau\gtrsim\num{e3}$)\RevisedText{,} is non-trivial.
The complexity of the state space is expected to grow explosively with increasing $\mathrm{Re}$ because of the rapid bifurcation of existing solutions and emergence of new ones~\citep{Zammert_Eckhardt_FDR2014,Zammert_Eckhardt_PRE2015,Park_Graham_JFM2015}.
\RevisedText{Also,} most known ECS solutions depict streak-vortex structures (\cref{fig:ecs}) representative of turbulence in the buffer layer, which, at higher $\mathrm{Re}$, \RevisedText{accounts only for} a very small portion of the flow domain.
Structures at higher $y^+$ are vastly more complex~\citep{DelAlamo_Jimenez_JFM2006,Jimenez_JFM2018}, most of which are not included in the current framework of dynamical systems.
(Recent\RevisedText{ly,} \citet{Shekar_Graham_JFM2018} reported the \RevisedText{first} class of ECSs \RevisedText{resembling} hairpin vortices, which are \RevisedText{involved in} log-law layer dynamics.)
Understanding polymer effects on those structures is particularly relevant to \RevisedText{the study of} HDR and MDR, where, as reviewed above, DR reduction effects are felt well beyond the buffer layer. 

Overall, the state-space picture of higher $\mathrm{Re}$ and higher $y^+$ is \RevisedText{rather murky at the moment.
It is unclear}
whether any simple dynamical theory can still formed \RevisedText{in a similar bottom-up fashion---from invariant solutions and MFU trajectories.
An alternative approach is to extract and analyze flow structures \textit{a posteriori} from DNS data of realistic flow conditions. Recent efforts on that front will be discussed in \cref{sec:vatip}.}

\paragraph{Dynamics in \RevisedText{extended flow} domain\RevisedText{s}}
MFU allows direct access to the temporal evolution of isolated structures but neglects correlations between different \RevisedText{structural units}.
\RevisedText{Dynamics in a} MFU \RevisedText{can be} mapped to large-scale turbulence \RevisedText{citing ergodicity and spatial-temporal equivalency}, as did in \citet{Wang_Graham_JNNFM2017}, \RevisedText{which, however, still precludes} interactions between structures.
For turbulence in \RevisedText{larger} flow domain\RevisedText{s}, the generation, growth, spreading, and demise of each \RevisedText{unit structure} are inevitably coupled with the dynamics of other structures nearby and, possibly, far away.
For example, \citet{Lopez_Hof_JFM2019} reported that with increasing length of their simulation domain (pipe \RevisedText{geometry}), although a transition from temporal intermittency to spatial intermittency \RevisedText{was indeed observed}, the latter cannot be accurately depicted as a statistical ensemble of different states in the temporal trajectory.
Rather, \RevisedText{with} sufficient \RevisedText{domain length}, highly localized turbulent patches \RevisedText{would form}, \RevisedText{which is} a clear sign for inter-structural correlation.
\RevisedText{In a separate example, \citet{Zhu_Xi_JNNFM2018} found}
that spatial localization
\RevisedText{of vortex structures is one of the key characteristics of HDR, which again would not occur without interaction between structures.}
Detailed analysis of such dynamics in the complex backdrop of \RevisedText{turbulence appears daunting, for which latest development in vortex analysis methodology, discussed below in \cref{sec:vatip}, can be particularly instrumental.}

\subsection{\RevisedText{Elasticity as a driving force for turbulence}}\label{sec:edt}
Conventional turbulence, e.g., \RevisedText{as} in a Newtonian fluid, is driven by \RevisedText{inertial effects}. Adding drag-reducing polymers damps those turbulent motions and coerces the flow to a state of less disorderness and, consequently, less drag. This has been the storyline dominating DR research for decades \RevisedText{and} has also been the \RevisedText{narrative} of \RevisedText{this} review thus far.
\RevisedText{Other than inertia, in polymer solutions, fluid elasticity provides} an additional source of non-linearity \RevisedText{in} system dynamics, which, in principle, could also drive \RevisedText{(}instead of damping\RevisedText{)} flow instabilities.
\RevisedText{Such elastic instabilities are more likely to occur as $\mathrm{Wi}$ increases.
For this reason, it has been long speculated that MDR is formed}
by flow instabilities that are more elastic in nature.
The report of vanishing RSS (and thus large polymer shear stress) at MDR by \citet{Warholic_Hanratty_EXPFL1999} further fueled such speculations.
\RevisedText{On the other hand}, DR is \RevisedText{still} a high-$\mathrm{Re}$ phenomenon\RevisedText{---it is equally possible that stronger conventional turbulence driven by fluid inertia prevails over any}
potential elastic \RevisedText{instability}.
Indeed, solid evidence for the relevance of \RevisedText{such} instabilities\RevisedText{,} driven \RevisedText{at least in part} by elasticity\RevisedText{,} to DR only emerged very recently\RevisedText{: the first and so far only established case is the so-called elasto-inertial turbulence (EIT) reported by} \citet{Samanta_Hof_PNAS2013}.

\subsubsection{Background}\label{sec:etbackground}
\paragraph{Historical context of elastic \RevisedText{instability research}}
It has been widely known that viscoelastic polymer fluids can display instabilities that \RevisedText{are} driven either purely or in part by elasticity\RevisedText{. P}urely elastic instabilities can occur at the limit of vanishing $\mathrm{Re}$ (i.e., \RevisedText{the} inertia-less limit)~\citep{Larson_RhA1992,Shaqfeh_ARFM1996,Larson_Nature2000,Squires_Quake_RMP2005,Morozov_vanSaarloos_PhysRep2007}.
Many earlier studies focused on polymer melts \RevisedText{for} rheometry and polymer processing \RevisedText{applications}. There has also been substantial research \RevisedText{of} dilute solutions of long-chain (i.e., drag-reducing) polymers in, e.g., microfluidics\RevisedText{,} where\RevisedText{,} because of \RevisedText{their} small \RevisedText{geometric} dimensions\RevisedText{,}
$\mathrm{Re}$ is too low to generate turbulence 
\RevisedText{and elastic instabilities are explored for the purpose of promoting fluid mixing.}
The best understood type of elastic instabilities occur in flows with curved streamlines, such as Taylor-Couette flow\RevisedText{---}shear flow in the annular space between two coaxial cylinders with relative rotation~\citep{Larson_Muller_JFM1990}.
\RevisedText{The} mechanism
\RevisedText{of this so-called ``hoop-stress'' instability} 
involves the coupling between polymer normal stress and streamline curvature~\citep{Pakdel_McKinley_PRL1996,Graham_JFM1998}.
\RevisedText{Flow instabilities typically manifest as secondary and often oscillatory flow patterns, which cause increased drag compared with stable laminar flow. In a rotating parallel-plate}
setup,
\RevisedText{\citet{Groisman_Steinberg_Nature2000} observed a particular type of instability}
showing chaotic \RevisedText{flow} patterns\RevisedText{. This instability was shown to exist}
at arbitrarily low $\mathrm{Re}$\RevisedText{, indicating its purely elastic nature.
The term ``elastic turbulence'' was coined to describe this turbulence-like instability that does not rely on inertia.}

Elastic instabilities also occur in flow around a stagnation point, where the streamlines make sharp turns, such as cross-slot flow~\citep{Arratia_Gollub_PRL2006,Poole_Oliveira_PRL2007}. One mechanism \RevisedText{for those instabilities is} the coupling between the incoming convection of polymer stress and fluctuations in the width of the so-called ``birefringent'' strand\RevisedText{---}a \RevisedText{thin} sheet of \RevisedText{fluids carrying} highly stretched polymers\RevisedText{, which initiates} from the stagnation point \RevisedText{but extends far} downstream~\citep{Xi_Graham_JFM2009}.

\begin{table*}
\caption{\RevisedText{Illustrative layout of possible flow states at different parameter regimes. (The actual parameter-space layout varies between specific flow types.)%
}}
\label{tab:instability}
\newcolumntype{V}[1]{%
>{\vbox to 2ex\bgroup\centering\vfill}%
p{#1}%
<{\vfill\egroup}}  
\RevisedText{%
\begin{tabular}{>{\columncolor{LXGry9}}V{\linewidth-8\tabcolsep-3\fboxrule-105ex}|
	V{35ex}|V{35ex}|V{35ex}}
\toprule
	$\mathrm{Re}>\mathrm{Re}_\text{crit}$
	&inertia-driven turbulence (IDT)	
		&IDT with DR	&elasto-inertial turbulence (EIT)
		\tabularnewline
\hline
	$O(1)\lesssim\mathrm{Re}<\mathrm{Re}_\text{crit}$
		&laminar flow/ inertia-driven inability (IDI)
		&laminar flow/ IDI/ inertio-elastic instability (IEI)
		&IEI/EIT
		\tabularnewline
\hline
	$\mathrm{Re}\ll O(1)$
		&laminar flow	&laminar flow/ elasticity-driven instability (EDI)
		&EDI/ elasticity-driven turbulence (EDT)
		\tabularnewline
\midrule[1pt]
\rowcolor{LXGry9}
		&$\mathrm{Wi}\lesssim O(1)$	&$\mathrm{Wi}\gtrsim O(1)$	
		&$\mathrm{Wi}\gg O(1)$	
		\tabularnewline
\bottomrule
\end{tabular}
}%
\end{table*}

\paragraph{\RevisedText{Dual origin for flow instabilities}}
\RevisedText{Mathematically, flow instabilities stem from the nonlinear relationship between the stress and rate of strain, which is found in both}
fluid inertia (measured by $\mathrm{Re}$) and elasticity (measured by $\mathrm{Wi}$).
\RevisedText{Combining} these two effects
\RevisedText{can also}
give rise to
\RevisedText{new}
types of instabilities.
\RevisedText{Such inertio-elastic instabilities (IEIs) were found in a variety of flow geometries.
In a microfluidic planar contraction-expansion flow, \citet{Rodd_Boger_JNNFM2005} observed that the instability secondary flow pattern varies between different $\mathrm{Re}$-$\mathrm{Wi}$ combinations: instabilities occurring at finite $\mathrm{Re}$ (i.e., IEI) are notably different from those at the purely elastic (vanishing $\mathrm{Re}$) limit (both at high $\mathrm{Wi}$).
In a cross-slot flow, \citet{Burshtein_Shen_PRX2017} showed that instability around the stagnation point morphs from an inertia-dominated form to an elasticity-dominated form by raising $\mathrm{Wi}$ at constant finite $\mathrm{Re}$.
}

\RevisedText{According to their physical origin, flow instabilities in viscoelastic fluids can be categorized into three major types:
\begin{CmpctDescr}
	\item[elasticity-driven instability (EDI)] often called purely elastic or inertia-less instabilities in the literature, which are driven \emph{entirely} by elasticity and can occur at the $\mathrm{Re}\to 0$ limit;
	\item[inertia-driven instability (IDI)] instabilities that would also occur in a Newtonian flow at sufficiently high $\mathrm{Re}$;
	\item[inertio-elastic instability (IEI)] instabilities where both inertia and elasticity are essential and both $\mathrm{Re}$ and $\mathrm{Wi}$ need to reach critical threshold values.
\end{CmpctDescr}
}%
These different instabilities were systematically explored in the $\mathrm{Re}$-$\mathrm{Wi}$ parameter space by \citet{Rodd_Boger_JNNFM2005} (see their fig.~15)
\RevisedText{and \citet{Burshtein_Shen_PRX2017} (their fig.~13).
A generic overview of possible flow states at different $\mathrm{Re}$-$\mathrm{Wi}$ regimes is illustrated} in \cref{tab:instability}.

\paragraph{Elastic instabilities in parallel shear flows}
Canonical turbulent flow types are all parallel shear flows whose mean flow has straight streamlines. Instability mechanisms reviewed above clearly do not apply.
Indeed, common viscoelastic parallel shear flows, such as Couette, channel, and pipe flows, are known to be linearly stable at the inertia-less (or purely elastic, i.e., $\mathrm{Re}\to 0$) limit~\citep{Ho_Denn_JNNFM1977,Morozov_vanSaarloos_PhysRep2007}.
Attention has thus been turned to possible nonlinear mechanisms for instability.

\RevisedText{A} linear instability can be triggered with infinitesimal disturbances, whereas \RevisedText{a} nonlinear instability requires a finite-amplitude disturbance\RevisedText{---one that exceeds a certain threshold magnitude.}
A classical example for the latter is the L-T transition in Newtonian fluids: for pipe flow, it typically occurs at $\mathrm{Re}_\text{crit}\approx 2100$ (defined based on bulk velocity) with finite\RevisedText{-}amplitude disturbances, but can be delayed to much higher $\mathrm{Re}$ if \RevisedText{disturbances} are well controlled and minimized~\citep{Bird_Stewart_2002}.
It \RevisedText{has been} speculated that elastic instability in parallel shear flow could follow a similar scenario along the $\mathrm{Wi}$ axis%
~\citep{Morozov_vanSaarloos_PhysRep2007}.
Evidence for such nonlinear transition to purely elastic ($\mathrm{Re}\lesssim O(\num{e-2})$) instability was shown by \citet{Pan_Arratia_PRL2013} in microfluidic ($\approx\SI{100}{\micro\metre}$) open channel flow.

DR \RevisedText{flow} systems can no longer be considered inertia-less.
\RevisedText{At finite $\mathrm{Re}$, \citet{Zhang_Brandt_JFM2013} performed a non-modal linear stability analysis, which studies the transient growth of specific modes of disturbances via linear mechanisms.
They found that a streak mode disturbance becomes transiently amplified with increasing $\mathrm{Wi}$, although it does still eventually decay over time.
The mechanism, however, is probably unrelated with any of the later found IEIs, as the amplification is caused by stronger turbulence production while conversion to polymer elastic energy still acts as a suppressing force for the growth of TKE (see \cref{eq:tkebal}).}
\RevisedText{Several types of IEIs in viscoelastic}
pipe and channel flows
\RevisedText{were reported in latest studies and at least some of them are}
linear in nature~\RevisedText{\citep{Samanta_Hof_PNAS2013,Garg_Subramanian_PRL2018,Shekar_Graham_JFM2018}}.
\RevisedText{In particular, the so-called \emph{elasto-inertial turbulence (EIT)}, discovered and named by \citet{Samanta_Hof_PNAS2013}, has brought significant developments to the understanding of DR, which is the focus of the current section (\ref{sec:edt})}.

\paragraph{Further notes on the terminology}
\RevisedText{Fundamentally, EIT is a type of flow instability that is driven at least in part by fluid elasticity and likely of inertio-elastic nature.
\RevisedText{The latter statement is because} so far EIT has only been reported for finite $\mathrm{Re}$, but without a clear mechanistic understanding, the possibility of it being purely elastic cannot as yet be ruled out.
\citet{Samanta_Hof_PNAS2013} invoked the word ``turbulence'' to describe this instability likely to reflect its chaotic flow patterns, much like ``turbulence'' as in ``elastic turbulence'' of \citet{Groisman_Steinberg_Nature2000}.
In addition, in the context of DR, EIT is viewed as a new state of turbulence that emerges at high levels of polymer elasticity.
Some may frown upon such a loose reference to the term ``turbulence'', especially if they subscribe to a narrower concept of turbulence which must satisfy a certain set of criteria, such as the seven characteristics put forth by \citet{Tennekes_Lumley_1972}.}
Without getting into the philosophical \RevisedText{debate} about what counts as turbulence,
\RevisedText{the term EIT is used here simply to follow its wide adoption in recent literature.}

\RevisedText{On the other hand, one may also encounter the term ``elasto-inertial turbulence'' begin used for flow states of entirely different natures.
For example, \citet{Gillissen_PRL2019} used it to describe a 2D viscoelastic  decaying homogeneous isotropic turbulence, even though the fluctuations are fueled solely by inertia and elasticity plays a suppressing role.
To clarify, in this review, as well as in much of the DR literature, EIT refers a specific category of instabilities that are (1) driven by both inertia and elasticity, (2) showing turbulence-like chaotic flow patterns, and (3) self-sustaining in time.
In DNS, the first criterion can be unequivocally determined by examining the TKE balance equation \cref{eq:tkebal}: elasticity becomes a driving-force for instabilities if and only if $\epsilon_\text{p}^k<0$ (i.e., $-\epsilon_\text{p}^k>0$).

}

\RevisedText{Likewise, EDIs showing turbulence-like chaotic flow patterns will be termed \emph{elasticity-driven turbulence (EDT)}, whereas}
conventional turbulence
\RevisedText{sustained by inertial instability will be}
referred to as 
\RevisedText{\emph{inertia-driven turbulence (IDT)}.}
Many in the literature called
\RevisedText{the latter}
``Newtonian turbulence'' because the turbulence generation mechanism stays the same\RevisedText{, at least qualitatively, as}
that in Newtonian flow, only that in viscoelastic flows, turbulence is attenuated by DR polymers.
That term is a bit ambiguous and even misleading, since it may be misconstrued as turbulence of Newtonian fluids only.
\RevisedText{In \cref{tab:instability}, EDT and EIT are marked at high-$\mathrm{Wi}$ regimes as a representative scenario. In reality, whether and when do these turbulence-like instabilities occur depends on the specific flow type.}

\subsubsection{Phenomenology}
\begin{figure}
	\centering
 	\includegraphics[width=\linewidth, trim=0 0 0 0, clip]{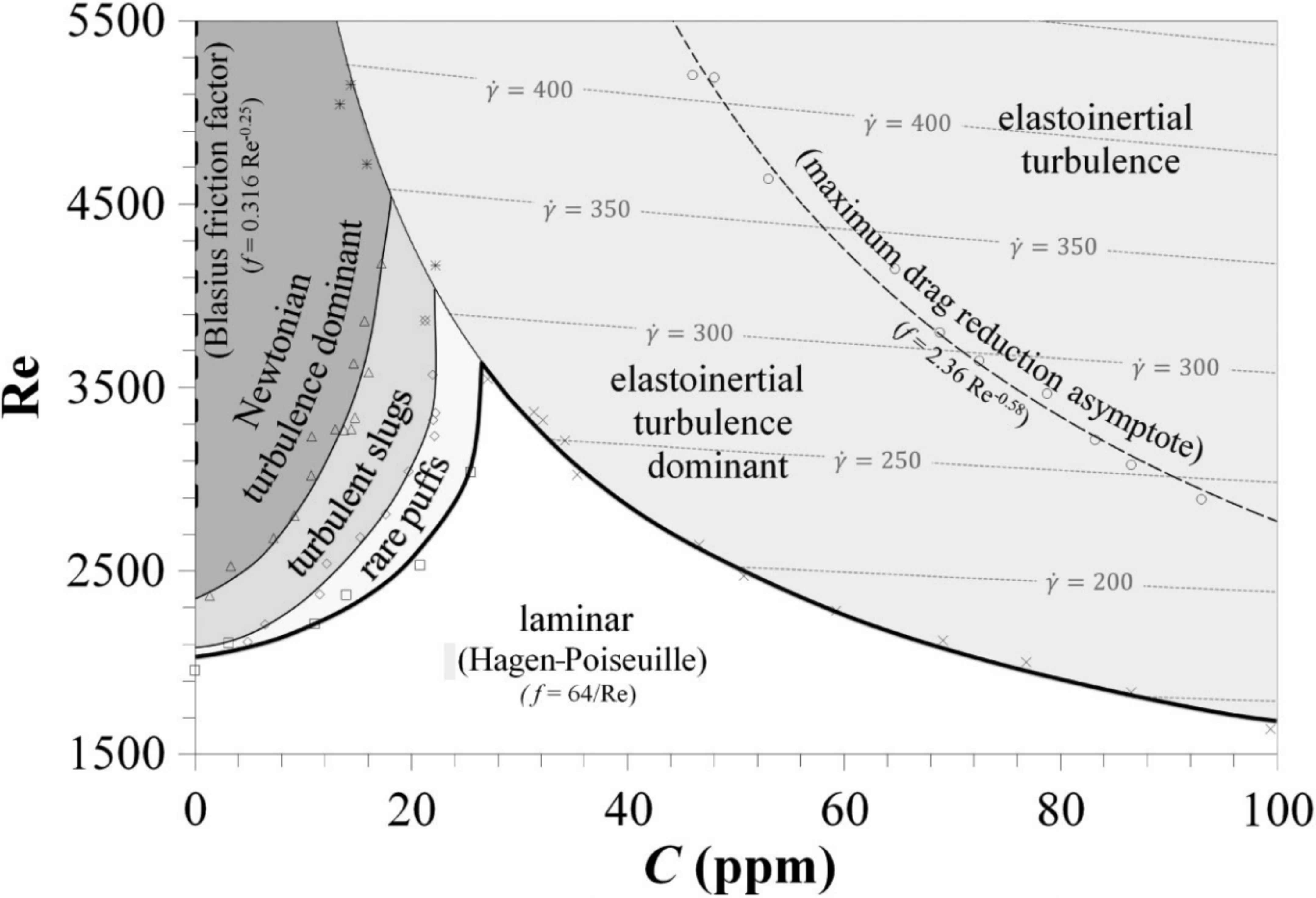}\\
	\caption{Different regimes of turbulent \RevisedText{pipe flow} behaviors in \RevisedText{a} $\mathrm{Re}$-$C_\text{p}$ parameter space, based on the friction factor and spatio-temporal patterns of streamwise velocity measured by PIV.
	Thin lines show constant levels of the characteristic shear rate $\dot\gamma\equiv 8\RevisedText{U_\text{avg}}/D$.
	The term ``Newtonian turbulence'' in the plot corresponds to IDT in our terminology.
	(Reprinted figure with permission from
	\RevisedText{\citeauthor*{Choueiri_Hof_PRL2018}, Physical Review Letters, 120, 124501,
	\citeyear{Choueiri_Hof_PRL2018}.}
	Copyright (\citeyear{Choueiri_Hof_PRL2018}) by the American Physical Society.)
	}
	\label{fig:eitparaspace}
\end{figure}
\paragraph{Transitions between different types of turbulence}
\citet{Samanta_Hof_PNAS2013} measured the $\mathrm{Re}_\text{crit}$ for the L-T transition \RevisedText{in} Newtonian (water) and viscoelastic (aqueous solution of PAM with the same molecular weight
and varying concentration $C_\text{p}$ up to \RevisedText{$\SI{500}{wppm}$}) fluids in pipe flow.
At the same pipe diameter, they found that $\mathrm{Re}_\text{crit}$ initially increases with $C_\text{p}$\RevisedText{---}i.e., drag-reducing polymers delay the transition, which is intuitive considering their role in resisting turbulent motion.
At higher $C_\text{p}$, \RevisedText{however,} $\mathrm{Re}_\text{crit}$ starts to decrease and eventually drops well below the value of Newtonian flow\RevisedText{---the so-called ``}early turbulence\RevisedText{'' phenomenon}.
Both delayed~\citep{Giles_Pettit_Nature1967,Draad_Nieuwstadt_JFM1998,Escudier_Presti_JNNFM1999} and early~\citep{Ram_Tamir_JAPS1964,Hansen_Little_Nature1974} transitions have been \RevisedText{previously} reported in \RevisedText{various} studies, which seemed conflicting at first.
Observing both behaviors in the same system (same polymer-solvent pair and same flow \RevisedText{setup}) with varying $C_\text{p}$ suggests that these are two \RevisedText{coexisting} transition pathways determined by the level of fluid elasticity.

\citet{Choueiri_Hof_PRL2018} \RevisedText{later conducted} a more comprehensive investigation of flow behaviors in a $\mathrm{Re}$-$C_\text{p}$ parameter space\RevisedText{, as} shown in \cref{fig:eitparaspace}.
The boundary between laminar (white) and turbulent (shaded) regions\RevisedText{, i.e., $\mathrm{Re}_\text{crit}$,} clearly \RevisedText{varies} non-monotonic\RevisedText{ally with $C_\text{p}$,} with a peak \RevisedText{at} $C_\text{p,crit}\approx\SI{25}{wppm}$.
For $C_\text{p}< C_\text{p,crit}$, increasing $\mathrm{Re}$ \RevisedText{at} constant $C_\text{p}$ would see the occurrence of localized turbulent patches, the so-called ``puffs'' and ``slugs''~\citep{Mullin_ARFM2011}, before IDT (the authors labeled it as ``Newtonian turbulence'' in the figure) fills the entire domain.
This transition sequence is identical to that in Newtonian flow except that $\mathrm{Re}_\text{crit}$ becomes higher with increasing $C_\text{p}$.
For $C_\text{p}> C_\text{p,crit}$, the transition directly enters, without discernible \RevisedText{structural localization}, a new type of turbulence\RevisedText{, which is presumed to be EIT.}

If we instead explore the parameter space horizontally (i.e., along constant $\mathrm{Re}$ lines), for $\mathrm{Re}$ between $\approx 2300$ (Newtonian $\mathrm{Re}_\text{crit}$) and $\approx 3600$ (highest point of the L-T borderline\RevisedText{, found} at $C_\text{p,crit}$), as $C_\text{p}$ increases, the flow would laminarize first before reentering the turbulence zone.
\citet{Choueiri_Hof_PRL2018} also performed a test where $C_\text{p}$ \RevisedText{was} ramped up at a slow pace with $\mathrm{Re}$ controlled at $3150$.
The Newtonian limit shows well-developed IDT with a friction factor following the Blasius correlation~\citep{Pope_2000}.
Within IDT, the friction factor drops sharply with increasing $C_\text{p}$.
After $C_\text{p}$ reaches the approximate range of \SIrange{15}{20}{wppm}, the flow becomes intermittent and consists of extended quiescent regions separated by occasional bursts of turbulent activities.
For a window around $C_\text{p}\approx\SI{20}{wppm}$, the friction factor matches that of the Virk MDR asymptote. This stage is strongly reminiscent of the high intermittency between hibernating turbulence and active/bursting phases reported by~\citet{Xi_Graham_PRL2010,Xi_Graham_JFM2012} and \citet{Zhu_Xi_JNNFM2019} (see \cref{sec:hibernation,sec:drdynsys}).
\RevisedText{Like} those \RevisedText{earlier} simulations\RevisedText{, t}he flow relaminarizes at $C_\text{p}\gtrsim\SI{20}{wppm}$\RevisedText{. Turbulence, however, returns} at $C_\text{p}\gtrsim\SI{45}{wppm}$. The friction factor \RevisedText{later} converges \RevisedText{again} to the Virk MDR asymptote at \RevisedText{$C_\text{p}\approx\SI{60}{wppm}$ and stays nearly constant with further increasing $C_\text{p}$}.

Although the two regimes of $C_\text{p}\approx\SI{20}{wppm}$ and $C_\text{p}\gtrsim\SI{60}{wppm}$ share nearly the same friction factor, and thus $\mathrm{DR}\%$, and both agree with the Virk asymptote, the fact that they are separated by a laminar window clearly indicates two \RevisedText{distinctly} different stages of turbulence.
Spatio-temporal patterns of streamwise velocity also appear different between \RevisedText{them}. \RevisedText{Although} both show \RevisedText{smooth} elongated streaks\RevisedText{,} the former (IDT with intermittency) \RevisedText{also} shows sporadic bursts of strong turbulence whilst the latter (presumed EIT) does not.
Since the latter stage is the asymptotic limit of high elasticity, the authors proposed\RevisedText{, same as} \citet{Samanta_Hof_PNAS2013}\RevisedText{,} that EIT is the ultimate MDR state.

\RevisedText{\citet{Lopez_Hof_JFM2019} performed DNS of pipe flow to complement the above experiments~\citep{Choueiri_Hof_PRL2018}.
The study focused on one $\mathrm{Re}$ and explored different regimes by adjusting $\mathrm{Wi}$.
Several simulation domain lengths were tested: in all cases, IDT is quenched and the flow laminarizes at sufficiently high $\mathrm{Wi}$.
Consistent with earlier studies~\citep{Xi_Graham_JFM2010,Wang_Graham_AIChEJ2014,Zhu_Xi_JNNFM2018}, the observed laminarization is always}
preceded by a regime of high intermittency in IDT\RevisedText{.}
The intermittency is temporal in shorter domains and resembles the AHB cycles in MFUs~\citep{Xi_Graham_PRL2010,Xi_Graham_JFM2012,Zhu_Xi_JNNFM2019}, but given sufficient domain length, it becomes spatial \RevisedText{and takes} the form of localized turbulence \RevisedText{separated by laminar-like regions.
The localized structures appear}
first \RevisedText{as} slugs and, at higher $\mathrm{Wi}$, \RevisedText{change to} puffs before laminarization.
This
\RevisedText{is strongly reminiscent of}
the L-T transition
\RevisedText{where the same group of structures appear but in an opposite order: with increasing $\mathrm{Re}$, it is}
laminar \textrightarrow puffs \textrightarrow slugs \textrightarrow space-filling IDT.
\RevisedText{Events leading up to laminarization with increasing $\mathrm{Wi}$ were}
thus dubbed ``reverse transition'' by the authors~\citep{Choueiri_Hof_PRL2018,Lopez_Hof_JFM2019}.
\RevisedText{In another agreement with \citet{Xi_Graham_JFM2010}, DR converges to an asymptotic level}
in a small window of $\mathrm{Wi}$
\RevisedText{immediately} before laminarization%
\RevisedText{.
This asymptotic DR level gets higher as the domain length increases and in the longest domain reported ($50D$), where slugs and puffs occur, it agrees with}
the Virk MDR\RevisedText{.}
\RevisedText{At}
a very high finite extensibility parameter $b=40000$\RevisedText{,
the study reported a second stage of turbulence with clearly different flow structures, which was believed to be EIT. }
Same as experiments at comparable $\mathrm{Re}$, it occurs after a window of laminarization
\RevisedText{.
(AD was used in the study which may affect the existence boundary of EIT---see section~\ref{par:ad}.)}

Back to the \RevisedText{parameter space} (\cref{fig:eitparaspace})\RevisedText{, w}ith increasing $\mathrm{Re}$, both the window of \RevisedText{laminarization} and that of \RevisedText{turbulence localization} (puffs and slugs) shrink and eventually vanish, after which the transition from IDT to EIT and the ultimate convergence to MDR become a \RevisedText{continuous} process, \RevisedText{as} typically observed in the literature.
The boundary of the MDR regime, which the authors presumed to be complete EIT, can be determined based on the friction factor magnitude.
However, for the regime immediately preceding MDR, which is labeled ``EIT dominant'' in \cref{fig:eitparaspace}, its underlying turbulent dynamics is harder to determine \RevisedText{using only} experimental \RevisedText{means}.
\citet{Choueiri_Hof_PRL2018} also referred to it as a ``coexistence phase'', which is probable.
Recent DNS by \citet{Zhu_Xi_JNNFM2019} reported the observation of EIT-like structures in a thin near-wall layer while the rest of the channel is populated by typical IDT structures.

\paragraph{Characteristics of flow and polymer conformation fields}
\begin{figure}
	\centering
 	\includegraphics[width=\linewidth, trim=0 0 0 0, clip]{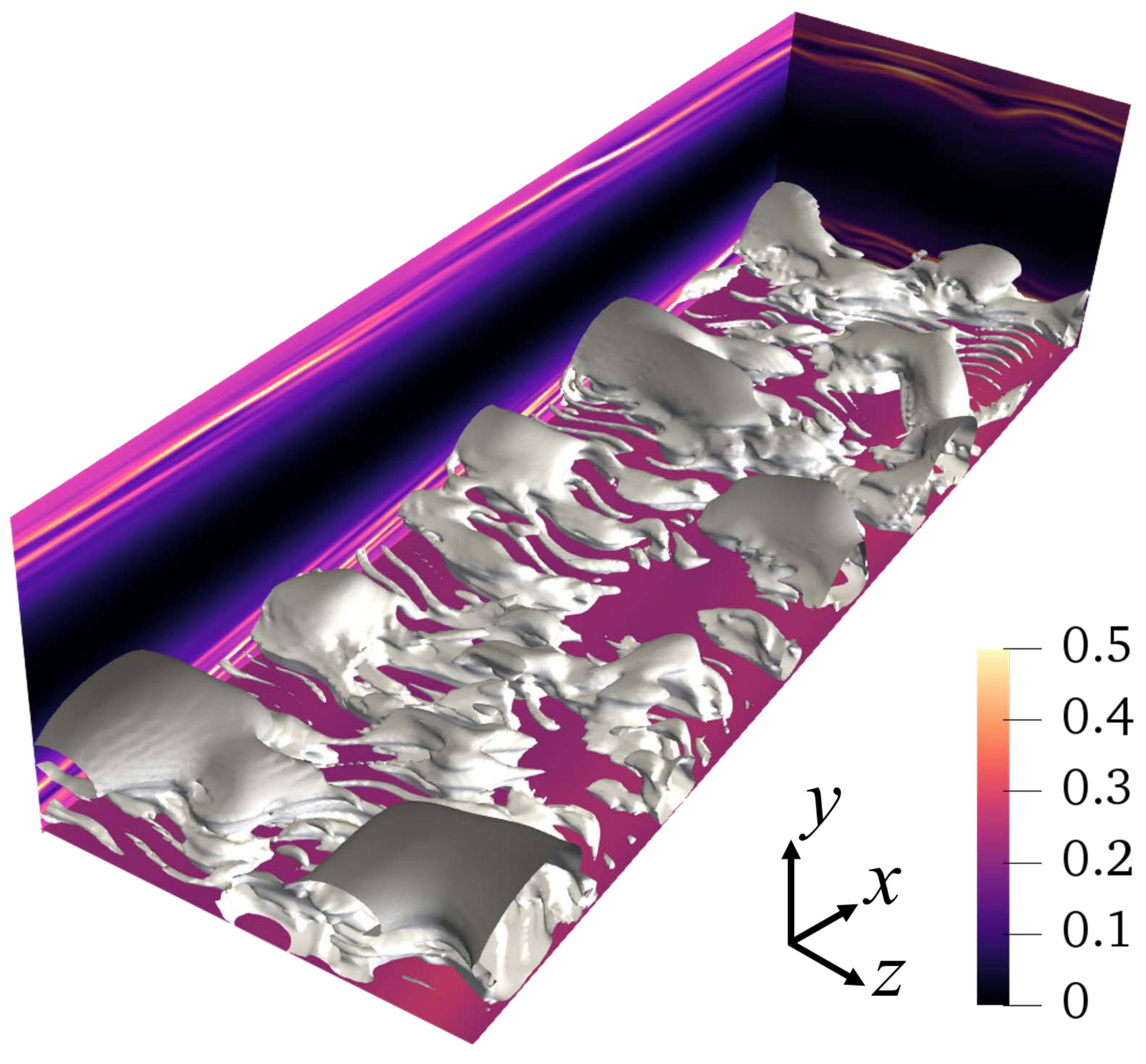}\\
	\caption{Visualization of a typical EIT flow state: isosurfaces show vortices\RevisedText{, in the bottom half of the channel only,} using the $Q$ criterion; color contours on the side and cross-sectional panels show $\mathrm{tr}(\alpha)/b$.
	(Courtesy of Lu Zhu based on data reported in \citet{Zhu_PhD2019}.)
	}
	\label{fig:eitstruct}
\end{figure}
The most compelling evidence for EIT being a different state of turbulence is its distinctive flow structure\RevisedText{s} observed in DNS, which was first reported in \citet{Samanta_Hof_PNAS2013} and \RevisedText{confirmed} by a number of studies~\citep{Dubief_Terrapon_POF2013,Sid_Terrapon_PRFluids2018,Lopez_Hof_JFM2019,Shekar_Graham_PRL2019,Zhu_PhD2019}.
A typical image is shown in \cref{fig:eitstruct}.
In contrast to IDT (see \cref{fig:largebox,fig:ecs,fig:hibernation}), where vortices in the buffer layer align in the streamwise direction (their downstream head may lift up\RevisedText{),}
vortices at EIT align spanwise.
\RevisedText{They} appear in two characteristic sizes with
\RevisedText{larger rolls separated by}
thinner thread\RevisedText{s.}
\RevisedText{The}
polymer conformation \RevisedText{field features distinct} thin tilted sheets of high polymer extension (large $\mathrm{tr}(\alpha)$)\RevisedText{---}bright stripes in the color contours \RevisedText{shown on} the side panel \RevisedText{of \cref{fig:eitstruct}}.
This, as well, differs from IDT where structures of polymer conformation \RevisedText{align closely} with vortices~\citep{Dubief_Lele_FTC2005,Kim_Adrian_JFM2007,Li_Graham_POF2007}.

Recent \RevisedText{DNS} of \citet{Sid_Terrapon_PRFluids2018} \RevisedText{found} flow states with \RevisedText{similar spanwise vortices and tilted} $\mathrm{tr}(\alpha)$ \RevisedText{sheets} in \RevisedText{an} $xy$-\RevisedText{2D channel (i.e., no spanwise dependence), which strongly indicates} that the underlying \RevisedText{instability} for EIT is intrinsically 2D.
EIT in 3D \RevisedText{DNS} appears more \RevisedText{like} the 2D instability as $\mathrm{Wi}$ increases, presumably because 3D structures of IDT, \RevisedText{especially those of} active turbulence, become increasingly suppressed at higher $\mathrm{Wi}$.
Importance of 2D instability in EIT was further confirmed by \citet{Shekar_Graham_PRL2019}
\RevisedText{who found its velocity spectrum to be dominated by spanwise-independent modes.}

\subsubsection{Origin of instability}
\paragraph{Elasticity \RevisedText{for instability}}
\mbox{}\RevisedText{It is clear}
that EIT is a \RevisedText{new} type of turbulence \RevisedText{following a different} self-sustaining mechanism.
\RevisedText{Evidence} for an elastic mechanism \RevisedText{is} abundant.
\RevisedText{B}oth experiments and DNS have shown that EIT can exist at $\mathrm{Re}$ well below the $\mathrm{Re}_\text{crit}$ for \RevisedText{the} Newtonian L-T transition~\citep{Samanta_Hof_PNAS2013,Dubief_Terrapon_POF2013,Choueiri_Hof_PRL2018}\RevisedText{---to see this,} compare the L-T borderline in \cref{fig:eitparaspace} between the Newtonian and high $C_\text{p}$ limits.
Early occurrence of turbulence indicates that an additional \RevisedText{source} for flow instability must at least supplement, and possibly \RevisedText{override}, the classical inertia-driven mechanism, which in viscoelastic fluids can only come from elasticity.
In addition, \citet{Samanta_Hof_PNAS2013} tested the laminar-EIT transition \RevisedText{with} different pipe diameters (\RevisedText{for the same polymer solution and same pipe diameter,} $\mathrm{Re}$ and $\mathrm{Wi}$ \RevisedText{are directly correlated---}see \cref{eq:expath}) and found that the transition occurs at the same critical $\mathrm{Wi}$ \RevisedText{for different} $\mathrm{Re}$.

Direct measure of the \RevisedText{instability} driving mechanism is given in the TKE balance \RevisedText{(}\cref{eq:tkebal}\RevisedText{)}, which can only be \RevisedText{evaluated in} DNS.
\RevisedText{T}urbulence generation from inertial \RevisedText{and elastic} mechanism\RevisedText{s is measured by the production $\mathcal{P}^k$ and elastic conversion $\epsilon_\text{p}^k$ terms, respectively.}
In IDT, polymers suppress TKE and thus $-\epsilon_\text{p}^k<0$. For elasticity to \RevisedText{become} a driving force for turbulence $-\epsilon_\text{p}^k$ must \RevisedText{turn} positive.
In this sense, EIT \RevisedText{was found} in \RevisedText{earlier} DNS \RevisedText{studies by} \citet{Min_Choi_JFM2003a}\RevisedText{,} \citet{Min_Choi_JFM2003b}\RevisedText{,} and \citet{Dallas_Vassilicos_PRE2010}.
As discussed above (section~\ref{par:viselasdns}), positive $-\epsilon_\text{p}^k$ \RevisedText{appears in} a thin slab around $y^+\approx 10-15$ \RevisedText{at LDR}, \RevisedText{which} expands across most of the domain at MDR and at least \RevisedText{some HDR cases}.
\RevisedText{Positive $-\epsilon_\text{p}^k$ was also observed in the transient development of an oblique mode disturbance in the non-modal linear stability analysis of \citet{Zhang_Brandt_JFM2013}.}

The experimental discovery of EIT prompted a more systematic study from \citet{Dubief_Terrapon_POF2013}, where $-\epsilon_\text{p}^k$ at EIT (as identified by its characteristic spanwise \RevisedText{vortices}) \RevisedText{was} found to be positive for all $y^+\gtrsim 8$\RevisedText{.} $\mathcal{P}^k$ is still positive but its magnitude depends on $\mathrm{Wi}$. For $\mathrm{Re}_\tau\approx 130$, $\mathcal{P}^k$ is comparable to $-\epsilon_\text{p}^k$ at $\mathrm{Wi}=96$, but at $\mathrm{Wi}=720$, it becomes \RevisedText{insignificant, suggesting a diminishing role of inertia in EIT with increasing $\mathrm{Wi}$}.
Similarly, at $\mathrm{Re}_\tau=84.85$, \citet{Sid_Terrapon_PRFluids2018} noted that even within in the broadly-defined EIT regime, flow structures in their 3D simulation \RevisedText{evolve} with $\mathrm{Wi}$. At $\mathrm{Wi}=40$, \RevisedText{streamwise vortex} structures typical of IDT appear intermittently and can often been seen blended with the spanwise rolls of EIT. They then claimed that \RevisedText{those} structures vanish \RevisedText{in their} $\mathrm{Wi}=100$ case where the 3D \RevisedText{EIT} solution more closely resembles the 2D one.

This brings up the possibility that the so-called EIT may be a combination of an underlying \RevisedText{2D} instability\RevisedText{, which} \RevisedText{could as well be} driven purely by elasticity (\RevisedText{i.e.,} EDT)\RevisedText{,} and an ensemble of \RevisedText{intermittently occurring} IDT states. The latter become less important at high $\mathrm{Wi}$.
Note that \citet{Samanta_Hof_PNAS2013} proposed the term EIT \RevisedText{prior to any} knowledge of its \RevisedText{instability} mechanism. It was chosen \RevisedText{mainly because} inertial effects \RevisedText{were} presumed to be relevant \RevisedText{at the high $\mathrm{Re}$ where they are found.}
A purely elastic mechanism, which of course awaits further validation, \RevisedText{would} not contradict \RevisedText{that} original study.

\paragraph{Linear vs. nonlinear instability}
The laminar-IDT transition is nonlinear \RevisedText{and} $\text{Re}_\text{crit,IDT}$ would increase when disturbances are reduced.
\RevisedText{In the Newtonian pipe flow of} \citet{Samanta_Hof_PNAS2013}\RevisedText{,} removing the imposed external \RevisedText{perturbation} delay\RevisedText{s} the transition from $\mathrm{Re}_\text{crit,IDT}\approx2000$ to $\approx6500$.
Meanwhile, \RevisedText{for the laminar-EIT transition, which is dominant at $C_\text{p}>C_\text{p,crit}$, they showed that} the unperturbed and perturbed transitions occur at the same $\mathrm{Re}_\text{crit,EIT}$\RevisedText{. This clearly indicates} a linear instability \RevisedText{mechanism whose} $\mathrm{Re}_\text{crit,EIT}$ is unaffected by the disturbance magnitude.
\RevisedText{At $C_\text{p}<C_\text{p,crit}$,} both transition pathways \RevisedText{can} coexist \RevisedText{in} a \RevisedText{certain} range of $C_\text{p}$\RevisedText{. Reducing disturbances would cause the laminar-IDT border to retreat to higher $\mathrm{Re}_\text{crit,IDT}$ and expose more of the laminar-EIT border (see \cref{fig:eitparaspace}).
Indeed, \citet{Samanta_Hof_PNAS2013} did find that at one lower $C_\text{p}$, where the perturbed transition is delayed ($\mathrm{Re}_\text{crit,p}^\text{perturb}>\mathrm{Re}_\text{crit,Newt.}^\text{perturb}$) as expected from the laminar-IDT pathway, removing the external perturbation exposes a new transition point $\mathrm{Re}_\text{crit,p}^\text{unpert.}$ and $\mathrm{Re}_\text{crit,p}^\text{perturb}<\mathrm{Re}_\text{crit,p}^\text{unpert.}\ll\mathrm{Re}_\text{crit,Newt.}^\text{unpert.}$.
This $\mathrm{Re}_\text{crit,p}^\text{unpert.}$ is presumably the laminar-EIT border $\mathrm{Re}_\text{crit,EIT}$ normally hidden behind the laminar-IDT border $\mathrm{Re}_\text{crit,IDT}$ when sufficient disturbances are present.}


In DNS, however, finite-amplitude disturbance is required to trigger EIT from the laminar state~\citep{Dubief_Terrapon_POF2013,Lopez_Hof_JFM2019,Shekar_Graham_PRL2019}. Recent evidence also indicates a dependence on the specific form of disturbance~\citep{Zhu_PhD2019}. Both observations \RevisedText{contradict} the experimental conclusion that EIT is a linear instability.

Two latest stability analysis studies \RevisedText{also} pointed toward opposite directions.
The first, by \citet{Garg_Subramanian_PRL2018}, found a linear instability in viscoelastic pipe flow at finitely large $\mathrm{Re}$ ($\gtrsim O(10^2)$ and depends on polymer properties).
It was a 2D linear stability analysis in the longitudinal plane (spanned by axial and radial directions), the same plane on which 2D EIT solutions are found (spanned by streamwise and wall-normal directions).
Unlike the more commonly studied $\mathrm{Re}\to 0$ limit, which as discussed above (\cref{sec:etbackground}) is linearly stable for viscoelastic parallel shear flow, the study found that at higher $\mathrm{Re}$, a single unstable mode emerges in the laminar state.
The instability consists of a streamwise array of spanwise rolls and neighboring vortices rotate in opposite directions, which resembles the train of spanwise vortices found in EIT.
The structure is, however, localized in a layer adjacent to the pipe axis, whereas EIT structures \RevisedText{are known to} originate from the walls.
A similar instability was also reported in channel flow~\citep{Chaudhary_Shankar_JFM2019}.

The second, by \citet{Shekar_Graham_PRL2019}, performed linear stability analysis in viscoelastic channel flow at the dominant wavenumbers of the EIT solution.
The laminar state \RevisedText{was} found to be linearly stable. However, the least stable eigenmode appears strikingly similar to the corresponding Fourier mode (same wavenumbers) of EIT.
Slow decay of this mode makes it susceptible to nonlinear amplification of finite-amplitude disturbances. \RevisedText{T}he most-amplified mode again closely resembles the same Fourier mode of EIT.
The study \RevisedText{further connected} EIT to \RevisedText{the} Tollmien-Schlichting (T-S) wave, a self-sustaining TW solution to the N-S equation \RevisedText{in channel flow, which, for Newtonian flow, stems} from a linear instability of the laminar state at $\mathrm{Re}_\tau\approx 107.4$~\citep{Drazin_Reid_1982,Jimenez_POF1987,Soibelman_Meiron_JFM1991}.
The bifurcation is subcritical and the solution \RevisedText{extends to} much lower $\mathrm{Re}$ (down to \RevisedText{$\mathrm{Re}_\tau \approx 75$ at the wavenumbers studied in \citet{Shekar_Graham_PRL2019}}) where finite-amplitude disturbance would be required \RevisedText{for it to be triggered} (i.e., nonlinear instability).
\RevisedText{For comparison, however,} EIT has been found at $\mathrm{Re}_\tau$ as low as $40$~\citep{Dubief_Terrapon_POF2013}.

\RevisedText{The study} computed fully nonlinear T-S wave solutions for viscoelastic channel flow\RevisedText{, where} sheets of high polymer extension, a signature structure of EIT (\cref{fig:eitstruct})\RevisedText{, were observed}. 
\RevisedText{They} are generated by near-wall stagnation points in the T-S wave, in a manner similar to the formation of a birefringent strand \RevisedText{in cross-slot flow}~\citep{Xi_Graham_JFM2009}. This offers \RevisedText{an} explanation for an otherwise peculiar structural feature of EIT.
Unlike the linear instability of \citet{Garg_Subramanian_PRL2018}, \RevisedText{polymer sheets in T-S wave} are localized near the walls.
However, \citet{Shekar_Graham_PRL2019} were not able to find direct overlap between T-S wave and EIT in the parameter space\RevisedText{---}the latter exists at much higher $\mathrm{Wi}$ than the former. Therefore\RevisedText{, any} specific dynamical connection between the two\RevisedText{, if existent, likely involves additional} nonlinear mechanisms.
\RevisedText{Generalization to other flow types is also a challenge. For example, T-S wave is not found in pipe flow which is believed to be linearly stable for all $\mathrm{Re}$~\citep{Drazin_Reid_1982,Mullin_ARFM2011}.}

\subsubsection{Discussion: implications for MDR}
Discovery of EIT not only provides direct proof for the existence of instabilities in parallel shear flow that are (fully or partially) elastic in nature, it also reconcile\RevisedText{s} the seemingly conflicting observations of both delayed and early turbulence \RevisedText{in} drag-reducing polymer \RevisedText{solutions}.
\RevisedText{Furthermore}, it brings new perspectives and raises new questions for the MDR problem, especially in light of the major \RevisedText{knowledge} gaps in the dynamical systems framework (\cref{sec:gaps}).
Considering this latest development, three possible scenarios \RevisedText{are discussed here} for the asymptotic limit of high polymer elasticity.

\paragraph{Scenario I: EIT is \RevisedText{the} single form of instability underlying MDR}
This is the most straightforward possibility and also seems to be the opinion held by most researchers. 
\RevisedText{E}xperimental and numerical \RevisedText{findings} reviewed above seem to \RevisedText{demonstrate} that, at least at the relatively moderate $\mathrm{Re}$ examined, EIT is the ultimate state \RevisedText{of the} high polymer elasticity \RevisedText{limit}.
This is \RevisedText{the conclusion of \citet{Choueiri_Hof_PRL2018} as shown} in the parameter space of \cref{fig:eitparaspace} and \RevisedText{also supported by the} diminishing presence of IDT structures with increasing $\mathrm{Wi}$~\citep{Sid_Terrapon_PRFluids2018}.
The dynamical AHB cycle, in its originally proposed form described in \cref{sec:drdynsys}, was clearly \RevisedText{also} observed \RevisedText{in those studies} but only for a limited window of $C_\text{p}$ or $\mathrm{Wi}$, which eventually gives way to EIT\RevisedText{---}a self-sustaining process itself~\citep{Choueiri_Hof_PRL2018,Lopez_Hof_JFM2019}.
This addresses the question of \RevisedText{the} ``existence'' of MDR (see three attributes of MDR in \cref{sec:mdr}).
In addition, mean velocity \RevisedText{of} EIT seems to follow the Virk MDR asymptote~\citep{Samanta_Hof_PNAS2013,Dubief_Terrapon_POF2013,Choueiri_Hof_PRL2018,Lopez_Hof_JFM2019},
\RevisedText{in accordance} with the ``magnitude'' attribute of MDR.

However, it \RevisedText{would appear inexplicable} for a flow instability\RevisedText{, which relies on} elasticity, to be independent of polymer solution properties\RevisedText{---}the attribute of ``universality'' is \RevisedText{a conspicuous} gap to fill.
\RevisedText{The scenario} will also \RevisedText{have difficulty with} the experimental observation that solutions of rigid polymers, which under FENE-P would be described by a very low $b$ parameter (\cref{eq:binterpret}) and presumably cannot \RevisedText{support} EIT, \RevisedText{are bounded by} the same \RevisedText{Virk} MDR asymptote~\citep{Virk_Sherman_AIChEJ1997,Benzi_Ching_PRE2005}.

\paragraph{Scenario II: EIT is \RevisedText{a phenomenological reflection of a new form} of \RevisedText{the} AHB cycle}
This scenario \RevisedText{interprets} the term EIT in existing literature \RevisedText{not as a reference to a single form of flow instability but} as an imprecise umbrella term encompassing many flow states \RevisedText{with varying extent} of elastic effects.
\RevisedText{Indeed,}
DNS results do show the presence of \RevisedText{different amounts of} IDT structures \RevisedText{blended} with spanwise rolls in \RevisedText{3D} flow states generically described as EIT~\citep{Dubief_Terrapon_POF2013,Sid_Terrapon_PRFluids2018,Zhu_PhD2019}.
It is highly possible that the complex appearance of \RevisedText{3D EIT} simply reflects a statistical ensemble containing \RevisedText{both} states dominated by IDT and those belonging to a ``pure'' form of EIT\RevisedText{---}the latter is likely \RevisedText{the corresponding} 2D EIT~\citep{Sid_Terrapon_PRFluids2018,Garg_Subramanian_PRL2018,Shekar_Graham_PRL2019}.
\RevisedText{The 2D form} could also be purely elastic (\RevisedText{i.e.,} EDT) rather than elasto-inertial considering that the magnitude of \RevisedText{average} TKE production $\mathcal{P}^k$ \RevisedText{in 3D EIT} diminishes with \RevisedText{increasing} $\mathrm{Wi}$~\citep{Dubief_Terrapon_POF2013}.
\RevisedText{In this case, t}he observed inertial characteristics \RevisedText{would} be solely attributed to the intermittent occurrences of IDT.

In this scenario, intermittent AHB cycles are \RevisedText{likely} still \RevisedText{involved}, except that \RevisedText{their} specific dynamics has to change at higher $\mathrm{Wi}$.
The original theory\RevisedText{,} discussed in \cref{sec:drdynsys} and depicted in \cref{fig:statespace}\RevisedText{, would still be} valid but only up to a certain $\mathrm{Wi}$. At sufficiently high $\mathrm{Wi}$, the EoC, which is originally in charge of shielding the flow from laminarization, becomes penetrable.
EIT \RevisedText{in its pure form (}or EDT \RevisedText{if it is purely elastic)} takes over to keep turbulence self-sustaining (\RevisedText{i.e.,} prevent laminarization).
\RevisedText{I}n 3D flow, IDT still grows intermittently to initiate active phases of turbulence, which \RevisedText{would be quickly} suppressed by polymers.
The destruction of EoC and emergence of EIT/EDT do not \RevisedText{always} coincide\RevisedText{, which would explain the appearance of a laminar window at lower $\mathrm{Re}$ (\cref{fig:eitparaspace})}.

\RevisedText{The scenario's r}eliance on intermittency does not necessarily contradict the seeming convergence to an ``EIT'' state \RevisedText{observed} by \citet{Choueiri_Hof_PRL2018}.
(Indeed, flow structures observed in both experiments and DNS \RevisedText{do} not strictly \RevisedText{converge} with $C_\text{p}$ or $\mathrm{Wi}$~\citep{Choueiri_Hof_PRL2018,Lopez_Hof_JFM2019}.)
The observed ``converged EIT'' state could be an evolving dynamical cycle between IDT and \RevisedText{EIT/}EDT, only that \RevisedText{variations} between states are difficult to \RevisedText{detect} from streamwise velocity patterns\RevisedText{, which are relied on in experiments} for \RevisedText{flow} state determination.
The lack of apparent bursting in velocity patterns \RevisedText{also} does not preclude the existence of active IDT\RevisedText{, since under polymer stress} the flow can bypass the strongest form of bursting \RevisedText{and energy eruption \textit{en route} to active turbulence~\citep{Zhu_Xi_JNNFM2019}}.

For the attributes of MDR, this scenario again offers explanation for ``existence'' but with a self-sustaining mechanism that is more dynamical in nature.
The \RevisedText{challenge} with ``universality'' still exists, similar to the previous scenario, but there appear to be more avenues for its explanation.
For example, because the intermittent occurrence of active turbulence keeps the flow from fully reaching the pure form of EIT/EDT, its mean flow is not the same as that of the \RevisedText{latter} but determined by \RevisedText{the} dynamical balance \RevisedText{between EIT/EDT and} IDT.
Both EIT/EDT and IDT statistics could change with increasing $\mathrm{Wi}$. How do their dynamical average stays constant at the Virk level would be the key to addressing both the ``universality'' and ``magnitude'' of MDR.

\paragraph{Scenario III: EIT is only important at low $\mathrm{Re}$ and/or \RevisedText{in} small domains}\label{para:scenarioIII}
\mbox{}\RevisedText{This scenario assumes that IDT would not be terminated by polymer effects and replaced by EIT, if higher $\mathrm{Re}$ or larger flow domains are used.
It will only be suppressed to a weaker form in which polymer effects are minimal---i.e., the theory for MDR based on AHB cycles still holds.
Compared with previous scenarios, here EIT is no longer a part of the self-sustaining process of MDR, which thus easily addresses its ``universality''.}

\RevisedText{Existing knowledge of EIT and its related transition scenario was based on studies}
at fairly low $\mathrm{Re}$ (in turbulence standards)
\RevisedText{where the total thickness of the wall-normal layer, from the wall to the channel center, equals}
$\mathrm{Re}_\tau\lesssim O(100)$
\RevisedText{wall units}.
This limits IDT to primarily buffer\RevisedText{-}layer structures\RevisedText{,} such as streamwise vortices\RevisedText{,} and only some lower log-law layer structures\RevisedText{,}
\RevisedText{while leaving out structures of the upper log-law and outer layers.}
\RevisedText{Complexity of t}urbulent structures and dynamics
\RevisedText{increases with}
$y^+$~\citep{Robinson_ARFM1991,Jimenez_JFM2018}.
\RevisedText{It is thus possible that structures at higher $y^+$ are not affected by polymers in the same way as low-$\mathrm{Re}$ IDT and their self-sustaining cycles are not completely disrupted by polymer elasticity.}

\RevisedText{
From a scaling argument, since the log-law layer extends from $\approx 30$ wall units to $\sim O(0.1)l$ (\cref{sec:inner}), it is not fully developed and the inner and outer layers are not completely separate unless $l/l_\text{v}=\mathrm{Re}_\tau\gg O(100)$.
Indeed, in Newtonian flow, the mean velocity profile does not even fully collapse on to the von K\'arm\'an law (\cref{eq:loglaw}) until $\mathrm{Re}_\tau\gtrsim 400$~\citep{Moser_Kim_POF1999,Zhu_Xi_JFM2019}.
At $\mathrm{Re}_\tau>5000$, a secondary peak would arise at higher $y^+$ in the $v_{x,\text{rms}}^{\prime +}$ profile, reflecting the emergence of new structures not captured at lower $\mathrm{Re}_\tau$~\citep{Marusic_Nagib_PoF2010}.
Large-scale coherent structures at high $\mathrm{Re}_\tau$ have attracted much interest among turbulence researchers in recent years.
One example is the so-called very-large-scale motions (VLSMs), multiple packets of strong turbulent structures organized into higher-order patterns, which are typically studied at the $\mathrm{Re}_\tau\sim O(\num{e3})$ regime~\citep{Kim_Adrian_PoF1999}.
Another example is the tall attached structures studied by \citet{LozanoDuran_Jimenez_JFM2012}, which originate from the buffer layer but can extend across the entire channel height at $\mathrm{Re}_\tau\approx 2000$.

Although the TKE production rate peaks in the buffer layer for all $\mathrm{Re}$, at higher $\mathrm{Re_\tau}$, the log-law and outer layers are larger in volume. It was thus estimated that at $\mathrm{Re}_\tau\approx 4200$ their accumulated contribution to TKE production exceeds that of the buffer layer~\citep{Smits_McKeon_ARFM2011}.
Whether EIT can still overtake IDT in high-$\mathrm{Re}$ regimes where vastly different coherent structures are at play is anything but certain.}


Domain size constraints, especially in the streamwise direction, could also be a factor.
\RevisedText{We know} that\RevisedText{, in the streamwise direction,} the velocity correlation length and \RevisedText{minimal} domain size \RevisedText{required} for self-sustaining IDT increase rapidly with $\mathrm{Wi}$ and $\mathrm{DR}\%$~\citep{Li_Khomami_JNNFM2006,Wang_Graham_AIChEJ2014,Lopez_Hof_JFM2019}.
\RevisedText{It is well possible} that in a highly extended domain beyond the current computational power, IDT\RevisedText{, }in a form with strong spatial intermittency (i.e., puffs separated by large laminar-like regions~\citep{Lopez_Hof_JFM2019})\RevisedText{, }can persist \RevisedText{to} higher $\mathrm{Wi}$.
The flow may eventually laminarize if the ``reverse transition'' hypothesis~\citep{Choueiri_Hof_PRL2018,Lopez_Hof_JFM2019} is proven accurate. However, as long as this ultimate $\mathrm{Wi}$ for laminarization is too high to be practically relevant,
\RevisedText{spatially localized turbulence would appear as the asymptotic state.
Obviously, domain size is only meaningful in simulation, which is however the only way the nature of instability (IDT, EIT, or EDT) can be unambiguously determined.
Experimentally, it is inferred from flow patterns, which is how \cref{fig:eitparaspace} was obtained~\citep{Choueiri_Hof_PRL2018}.
For this reason, the above discussion of the domain size factor is of general relevance.}

\subsection{Dynamical analysis of coherent structures in extended domains}\label{sec:largebox}
Recent progress reviewed in the above sections builds on a shift of focus from ensemble average quantities to spatio-temporal intermittency.
Numerical simulation played a pivotal role in \RevisedText{those} developments\RevisedText{.
MFU-based DNS is the most common approach for tracking the temporal evolution of individual structures, which is complemented by
the direct computation of}
invariant solutions (ECS).
\RevisedText{Although many DNS} studies \RevisedText{reviewed in \cref{sec:dynamical,sec:edt}}
were not performed in \RevisedText{domains that are strictly minimized, they}
are still sufficiently constrained
\RevisedText{that dynamical evolution of structures is reflected in the time series of domain-average quantities.
For DNS in more extended domains, patches of different flow states can be analyzed by invoking the}
equivalency between temporal and spatial intermittency, as in \citet{Wang_Graham_JNNFM2017}.
\RevisedText{This naive approach views the dynamical life cycles of turbulent structures as statistical sampling of individual flow states in an ensemble, without regard to any correlation or interaction between structures.}
Such \RevisedText{correlation} become\RevisedText{s} increasingly important at higher $\mathrm{Wi}$, which is evident from the localization of vortices and their organization into slugs and puffs~\citep{Zhu_Xi_JNNFM2018,Lopez_Hof_JFM2019}.
\RevisedText{This} dynamics \RevisedText{is excluded in MFU but} fully captured in DNS \RevisedText{in extended domains. E}ffective extraction of such information from complex turbulent flow fields \RevisedText{is a non-trivial challenge.}

In the area of DR,
\RevisedText{two major approaches were most used in the literature to analyze}
coherent structures \RevisedText{in} large-scale turbulent flow fields.
One is KL analysis or proper-orthogonal decomposition (POD)~\citep{Pope_2000,DeAngelis_Piva_PRE2003,Housiadas_Beris_POF2005,Li_PhD2007}, which decomposes velocity fields into orthogonal spatial modes\RevisedText{---}a basis set for velocity fields.
None of the modes directly represents any real occurring coherent structure, but the leading modes can be viewed as the most important constituents (in terms of the amount of kinetic energy contained) \RevisedText{of} statistically-representative structures. 
\citet{Wang_Graham_AIChEJ2014} extended the analysis to also include the polymer conformation field in the decomposition.
Another approach is conditional sampling, which finds the statistical representation of local flow fields around a predefined event of interest.
The event can be the occurrence of streamwise vortices, in which case patches of flow field\RevisedText{s} around qualifying \RevisedText{vortices} are aligned at \RevisedText{their} center \RevisedText{axes} and statistically averaged~\citep{Sibilla_Beretta_FDR2005,Zhu_Xi_JPhysCS2018}.
It can also be high streak intensity, such as in \citet{Kim_Adrian_JFM2007} where patches are selected based on the detection of strong eruption ($v_x'<0,v_y'>0$) events making largest contributions to the RSS (\cref{eq:rss}).
Compared with KL analysis,
\RevisedText{conditional sampling is a local approach with no direct tie to the imposed simulation domain.
Its results thus connect}
more \RevisedText{closely} with \RevisedText{individual} coherent structures\RevisedText{. However, its outcome is inevitably influenced by the subjectivity in choosing} the selection condition \RevisedText{for the} event.

Those earlier approaches
\RevisedText{generate statistical representations of coherent structures and neglect their individuality}.
Recent attention to spatio-temporal intermittency calls for
\RevisedText{the capability to extract the real instances of individual structures and trace their temporal evolution in complex flow fields.}
\RevisedText{The first step, i.e., extraction of structural instances, is static and recently made possible by a method called VATIP (vortex axis tracking by iterative propagation)~\citep{Zhu_Xi_JFM2019}.
This section}
take\RevisedText{s} the mechanism of LDR-HDR transition as a case study
\RevisedText{for illustrating the application of VATIP in vortex conformation analysis.
Future work of extending VATIP for dynamical analysis (the second step listed above) is also briefly discussed.}

\subsubsection{Case study: vortex regeneration mechanism for HDR}\label{sec:hdrssp}

\begin{figure*}
	\centering
	\begin{minipage}[c][][c]{0.65\linewidth}
 	\includegraphics[width=\linewidth, trim=30 5 50 5, clip]{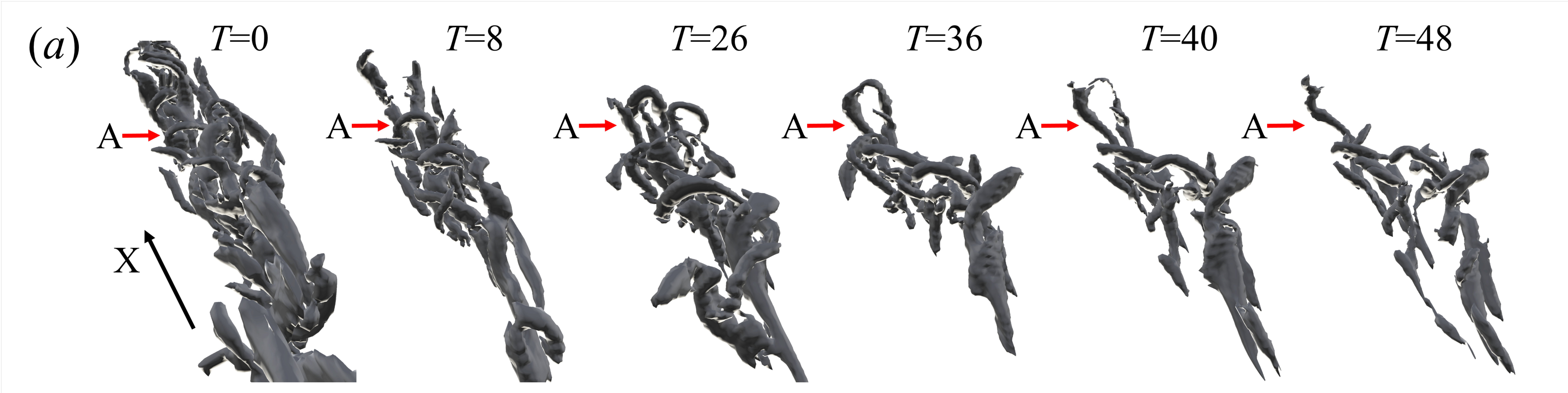}\\
 	\includegraphics[width=\linewidth, trim=30 5 50 5, clip]{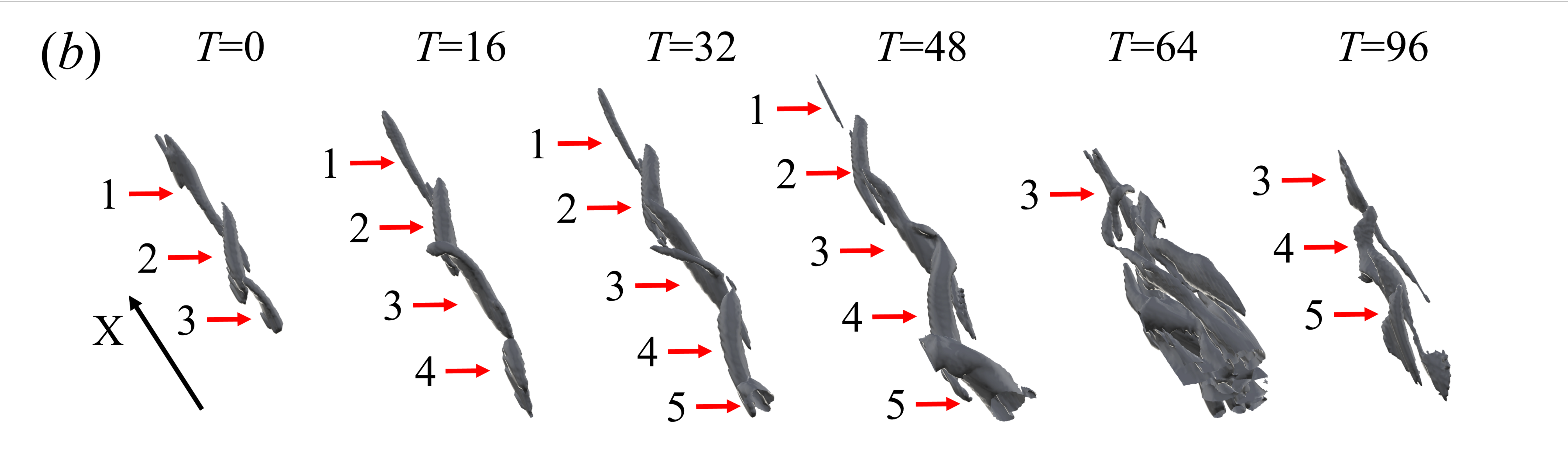}
	\end{minipage}%
	\begin{minipage}[c][][c]{0.35\linewidth}
	\caption{Instances of different vortex regeneration pathways at $\mathrm{Re}_\tau=86.15$: (a) streak-instability mechanism (Newtonian) and (b) parent-offspring mechanism ($\mathrm{Wi}=96,\beta=0.97,b=5000$).
	\RevisedText{Isosurfaces show $Q=0.7Q_\text{rms}$.}
	(Reprinted from
	\RevisedText{Journal of Non-Newtonian Fluid Mechanics,
	262, 115,
	\citeauthor*{Zhu_Xi_JNNFM2018},
	Distinct transition in flow statistics and vortex dynamics between low- and high-extent turbulent drag reduction in polymer fluids,
	Copyright (\citeyear{Zhu_Xi_JNNFM2018}),} with permission from Elsevier)."
	}
	\label{fig:hdrssp}
	\end{minipage}
\end{figure*}

Distinctions between LDR and HDR were thoroughly studied by \citet{Zhu_Xi_JNNFM2018} and discussed in section~\ref{par:hdr}.
The LDR-HDR transition is marked by sharp changes in a variety of flow statistics and believed to have its roots in the localization of vortex structures.
Understanding the origin of localization requires the analysis of individual vortex objects and, when possible, their dynamics and interactions.

\citet{Zhu_Xi_JNNFM2018} proposed a mechanistic explanation citing the conceptual framework of vortex regeneration mechanisms summarized by \citet{Schoppa_Hussain_JFM2002}.
In this framework, new turbulent vortices are \RevisedText{continually} generated from existing ones following two distinct pathways.
The first, the streak-instability pathway, is an extension of the classical conceptual model for \RevisedText{the} turbulent self-sustaining \RevisedText{process by} \citet{Waleffe_POF1997}.
Counter-rotating streamwise rolls (straight vortices) generate low speed streaks in between.
\RevisedText{Adding streamwise sinuous perturbations} can trigger the so-called ``streak breakdown''~\citep{Schlatter_Henningson_POF2008}\RevisedText{---}strong instabilities leading to \RevisedText{streak} intensification and \RevisedText{vortex} growth\RevisedText{. Those} vortices have a strong tendency to lift up\RevisedText{---}a process in which the downstream end of a vortex pivots away from the wall and extends vertically into higher $y^+$ regions.
Lifted vortex heads \RevisedText{are} often \RevisedText{swung} sideways \RevisedText{by} local transverse flows and many form the so-called ``hairpins''\RevisedText{---}$\Omega$-shaped vortices whose downstream head, the arch in $\Omega$, is lifted up while both streamwise legs extend upstream and stay closer to the wall~\citep{Smith_Walker_PTRSLA1991,Wu_Moin_POF2009}.
Rapid lifting and growth of vortices ends with their sudden eruption\RevisedText{---}a bursting phase (section~\ref{par:ahbcycle}), which was studied in detail by \citet{Zhu_Xi_JNNFM2019}. The resulting fragments and intense fluctuations can spread and trigger new streak instability elsewhere to start the next cycles.
An example displaying part of the process is shown in \cref{fig:hdrssp}(a). A vortex tube, marked with ``A'' merges with a nearby vortex to form a crescent at $T=8$, which lifts up and grows into a full hairpin ($T=36$) and then burst\RevisedText{s} ($T=40$) into fragments \RevisedText{(not shown)}, leaving only one leg in its residue ($T=48$). 

\citet{Zhu_Xi_JNNFM2018} proposed that strong polymer effects can block this cycle by suppressing vortex lift up.
\RevisedText{In a separate study}, \citet{Zhu_Xi_JNNFM2019} also \RevisedText{found} that polymers \RevisedText{can subdue} bursting and \RevisedText{suppress the} production of high-intensity fluctuations.
Without lifting and bursting, vortices are stabilized in the streamwise direction, which allows them to be extensively elongated\RevisedText{, as clearly shown in} the conditional eddies at HDR obtained by \citet{Kim_Adrian_JFM2007}.
\RevisedText{It would also disrupt the streak-instability pathway and}
expose the second so-called ``parent-offspring'' pathway to dominate regeneration cycles\RevisedText{.
In the latter, as illustrated in \cref{fig:hdrssp}(b),}
vortices are successively generated in close proximity to form a chain\RevisedText{---}new ``offspring'' vortices are generated at the tip of an existing ``parent'', in the shear layer between the parent and the wall.
Obviously, this mechanism is intrinsically more local than the other.
Both pathways exist in Newtonian flow but at high $\mathrm{Wi}$, the parent-offspring one becomes more prominent, leading to the apparent clustering and localization of \RevisedText{vortices}.
\RevisedText{Applying the VATIP algorithm (discussed below), \citet{Zhu_Xi_POF2019} showed that the suppression of vortex lift-up coincides with the LDR-HDR transition, which presents a compelling depiction of the second DR mechanism required for its explanation (see section~\ref{par:hdr}).}

This explanation is compatible with the earlier theory based on AHB cycles (\cref{sec:drdynsys}) as vortex lift up is part of the bursting and active phases and its suppression leads to higher presence of hibernation.
It, however, goes one step further and explains the vortex localization mechanism, which requires interactions between vortices not considered in the AHB framework.

\subsubsection{Vortex conformation analysis by VATIP}\label{sec:vatip}
The above mechanism was proposed based on direct anecdotal inspection of visualized flow field images (\cref{fig:hdrssp}), which is too often \RevisedText{relied on} in turbulence research.
Reliable conclusions cannot be reached without the objective and quantitative analysis of vortex conformation and dynamics.
Although methods for vortex identification, i.e., derivation of quantitative indicators for vortex regions\RevisedText{---such as $Q$, have} been extensively studied (\cref{sec:vortexid}), they 
\RevisedText{only tell if a region belongs to \emph{any} vortex without identifying which regions constitute the same well-defined vortex, which is thus}
not adequate for vortex conformation analysis.

\begin{figure*}
	\centering
	\begin{minipage}[c][][c]{0.65\linewidth}
 	\includegraphics[width=\linewidth, trim=0 0 0 0, clip]{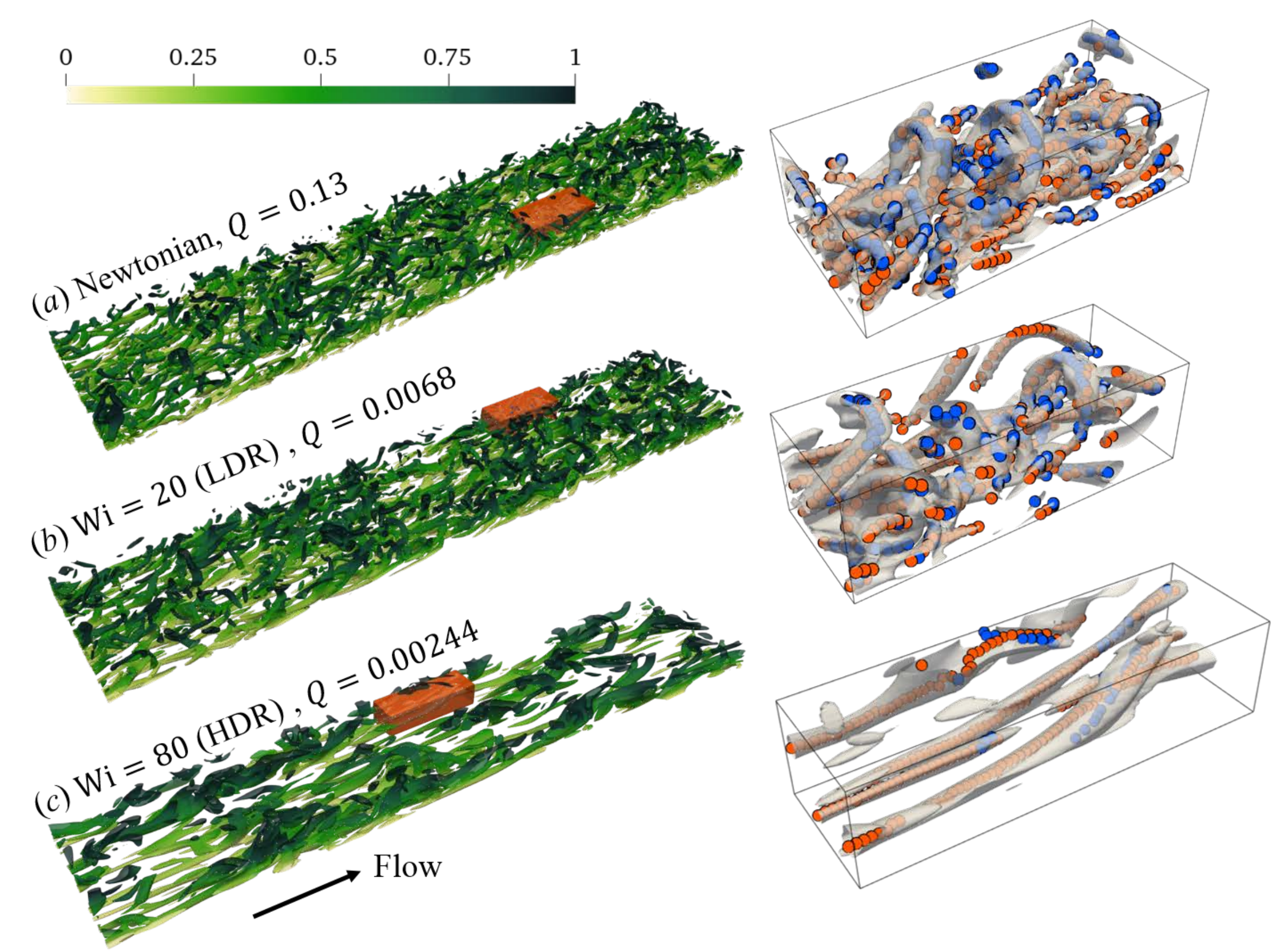}\\
	\end{minipage}%
	\begin{minipage}[c][][c]{0.35\linewidth}
	\caption{
	Instantaneous vortex configurations of (a) Newtonian, (b) $\mathrm{Wi}=20$ (LDR), (c) $\mathrm{Wi}=80$ (HDR) cases at $\mathrm{Re}_\tau=172.31$, $\beta=0.97$, and $b=5000$. 
	Isosurfaces show $Q=0.4Q_\text{rms}$ \RevisedText{for} bottom half of the channel \RevisedText{only}.
	Color shades \RevisedText{map to} wall-normal position\RevisedText{: $0$ and $1$ for the wall and channel center, respectively.}
	A selected box is enlarged in each case\RevisedText{, in which} axis-lines identified by VATIP\RevisedText{~\citep{Zhu_Xi_JFM2019}}\RevisedText{are shown with} circular dots\RevisedText{:} orange/light for axis-points on the $yz$ plane and blue/dark for those on \RevisedText{the} $xy$ \RevisedText{or} $xz$ plane.
	(Reprinted from
	\RevisedText{\citeauthor{Zhu_Xi_POF2019},
	Physics of Fluids,
	31, 095103,
	\citeyear{Zhu_Xi_POF2019},}
	with the permission of AIP Publishing.) }
	\label{fig:vatipve}
	\end{minipage}
\end{figure*}

Each vortex has an axis of rotation which fully describes its topological shape and instantaneous conformation.
An axis-line is formed by connecting the center of rotation\RevisedText{---}the axis-point\RevisedText{---}on each cross-section of the vortex tube, which can be defined as the 2D maximum of $Q$
\RevisedText{(or any other indicator of vortex strength)}
on the plane.
For linear streamwise vortices, grouping axis-points into axis-lines is relatively straightforward, through a so-called ``cone-detective'' procedure by \citet{Jeong_Hussain_JFM1997}. Axis-lines so extracted \RevisedText{provide the reference points} for aligning streamwise vortices\RevisedText{, which} attracted much attention in earlier turbulence research\RevisedText{,} for conditional sampling~\citep{Jeong_Hussain_JFM1997,Sibilla_Beretta_FDR2005,Zhu_Xi_JPhysCS2018}.

\RevisedText{As discussed in \cref{sec:polymturb},} 
existing \RevisedText{studies} on polymer-vortex interaction \RevisedText{were mostly limited to} streamwise vortices with few exceptions~\citep{Kim_Adrian_JFM2007,Kim_Adrian_PRL2008,Agarwal_Zaki_JFM2014}.
Understanding how complex vortices found at higher $y^+$, such hairpins, are affected by polymers is essential for \RevisedText{understanding DR at} higher $\mathrm{Re}$.
As shown in \cref{fig:vatipve}, even at a fair\RevisedText{ly} low $\mathrm{Re}_\tau=172.31$, hairpin vortices are prevalent especially in Newtonian and LDR \RevisedText{cases}.
\citet{Kim_Adrian_PRL2008} \RevisedText{extracted representative hairpin vortices using conditional sampling and studied their transient development with DNS.}
\RevisedText{Their} approach is \RevisedText{akin} to \textit{in vitro} experiments in the sense that model hairpins are studied in isolation from its ``living'' environment.
\RevisedText{The new VATIP algorithm, developed by \citet{Zhu_Xi_JFM2019},}
is capable of extracting axis-lines of \RevisedText{general} \RevisedText{3D} vortices\RevisedText{, including} hairpins\RevisedText{, from near-wall turbulence ``\textit{in vivo}''.}

\RevisedText{VATIP} borrows the cone-detective idea from \citet{Jeong_Hussain_JFM1997} but extends the search
\RevisedText{to all three dimensions} to capture vortices with nonlinear and branched shapes.
Without getting into \RevisedText{the} details, the algorithm \RevisedText{appears} like the video game ``snake''\RevisedText{,} where growing axis-line\RevisedText{s} keep looking for new axis-points to ``swallow'' (i.e., connect \RevisedText{to}). If no more eligible axis-points are found in \RevisedText{their} growing direction, the propagating end\RevisedText{s} turn to the next search direction.
\RevisedText{Iteration between different search directions continues until none of the axis-lines can grow further.}
The method reliably captures all known types of vortices generated by \RevisedText{a} no-slip wall, including both symmetric and asymmetric hairpins.
This is clearly shown in \cref{fig:vatipve} where axis-lines extracted by VATIP (colored dots) faithfully reproduce the contour shape\RevisedText{s} of all vortices (gray tubes), from Newtonian flow to HDR.
\RevisedText{Extracted axis-lines can be classified based on their topologies}
into categories such as streamwise vortices, hairpins, branches, etc.
\RevisedText{(It, however, may not work as well with structures having no direct interactions with the wall, which typically occur at much higher $y^+$ than those studied so far for DR~\citep{DelAlamo_Jimenez_JFM2006,Jimenez_POF2013}---see detailed discussion in \citet{Zhu_Xi_JFM2019}.)}


Application of VATIP to DNS data at different regimes of DR allows a comprehensive investigation of polymer effects on vortex conformation \RevisedText{(\cref{fig:vatipve})}~\citep{Zhu_Xi_POF2019}.
The results show that although streamwise vortices can be found either lying flat or lifted up, hairpins and hairpin-like vortices are almost always lifted, which supports the above \RevisedText{postulation that hairpins are formed} from the lift up of streamwise vortices.
At LDR, vortex intensity is reduced but their distribution pattern and conformation statistics remain \RevisedText{largely unchanged from} Newtonian flow.
At HDR, the number of lifted vortices decline\RevisedText{s} sharply.
Suppression of vortex lift up interrupts the generation of hairpins and hairpin-like vortices. It also reduces the turbulent momentum flux between buffer (streamwise vortices) and log-law (lifted and hairpin vortices) layers, which would explain the changing flow statistics in the log-law layer \RevisedText{between LDR and HDR}.
Overall, the results are consistent with the proposed mechanism in \cref{sec:hdrssp}.

The current analysis is static. It shows, within any frozen instant of turbulence, what is the state and conformation of each vortex, without \RevisedText{establishing the} temporal connection \RevisedText{between vortices found at different} instants.
Expansion of the VATIP framework to cover temporal dynamics is foreseeable\RevisedText{. For instance, a} recent method for temporal \RevisedText{coherent structure} analysis by \citet{LozanoDuran_Jimenez_JFM2014} \RevisedText{may be adapted} to the axis-lines extract\RevisedText{ed} by VATIP\RevisedText{.} 
This would open the door for the detailed, quantitative, and \textit{in situ} analysis of \RevisedText{vortex} life-time dynamics \RevisedText{in realistic extended flow domains.} 

\section{Summary and outlook}\label{sec:summary}
For \RevisedText{over} seventy years, turbulent drag reduction by polymers \RevisedText{has remained} an active area of research, as new approaches and new discoveries continue to emerge.
The past decade (2010s) has witnessed a surge of interesting developments in the \RevisedText{fundamental} understanding of DR\RevisedText{. Many of them} have challenged the established line of thought and brought significant progress in \RevisedText{answering} some of the most \RevisedText{puzzling questions---}%
in particular, \RevisedText{the nature of} MDR.
A central theme among \RevisedText{those latest} advances is to go beyond \RevisedText{the} ensemble statistics of fluctuating turbulent flow fields and study the dynamical evolution and intermittency of coherent flow structures\RevisedText{.} 

The current review covers both classical contributions and recent developments in the fundamental inquiries \RevisedText{into} polymer DR\RevisedText{.} 
\RevisedText{The ultimate target is the mechanistic understanding of transitions between different turbulence regimes and t}he status of current research is summarized briefly as follows.
\begin{CmpctDescr}
	\item[Onset of DR] This problem is equivalent to understanding
	\RevisedText{the mechanism of polymer DR that leads to LDR.}
	\RevisedText{Great strides have been made in deciphering} the interaction\RevisedText{s} between polymers and turbulent flow structures\RevisedText{, thanks to the advancement of numerical simulations over the past two decades.}
	\RevisedText{It is well established now that polymers can counteract and suppress}
	streamwise vortices\RevisedText{---dominant structures in} turbulent buffer layer where the onset is believed to occur (\cref{sec:polymturb}).
	However, the debate of whether such effects should be wrapped theoretically as a viscous (Lumley) or elastic (de~Gennes) mechanism of polymers remains unsettled.
	\item[HDR] \RevisedText{Historically,} HDR has received much less attention
	\RevisedText{because of its later discovery but also because it is often}
	presumed to be \RevisedText{simply} a precursor of \RevisedText{MDR}.
	There is certainly \RevisedText{justification} for this argument. Most notably, HDR is marked by the expansion of DR effects from a limited region (buffer layer) to nearly the entire channel~\citep{Zhu_Xi_JNNFM2018,White_Dubief_JFM2018}. Increasing length scale of drag-reduced turbulence to the scale of flow geometry is also central to several theories for MDR~\citep{Virk_JFM1971,Lumley_ARFM1969,Lumley_JPSMacroRev1973,Sreenivasan_White_JFM2000}.
	On the other hand, if the nature of MDR is indeed \RevisedText{at least partially} elastic (\cref{sec:edt})\RevisedText{, as believed by many},
	\RevisedText{the fact that the LDR-HDR transition can be observed without EIT (or EDT)}%
	~\citep{Zhu_Xi_JNNFM2018} \RevisedText{would indicate the} decoupling between \RevisedText{HDR and MDR}.
	\RevisedText{This} transition \RevisedText{reflects the start of} a second DR mechanism, in \RevisedText{relation} to the first mechanism at the onset of DR.
	This new mechanism \RevisedText{has been} explained in terms of	changing vortex regeneration mechanisms in recent studies~\citep{Zhu_Xi_JNNFM2018,Zhu_Xi_POF2019}. Despite some strong evidence from vortex conformation analysis (\cref{sec:largebox}), further investigation, such as temporal \RevisedText{analysis} of vortex lifetime cycles, is still needed.
	\item[MDR]
	This long-standing mystery finally seems to \RevisedText{have} open\RevisedText{ed} up a few cracks.
	A complete theory must consistently address three attributes of MDR: existence (self-sustaining mechanism), universality (insensitivity to changing polymer solutions and domain size), and magnitude (empirical Virk profile).
	Two different but interconnected streams of developments occurred in the past decade.
	The first is the theoretical framework based on AHB cycles (\cref{sec:drdynsys}). 
	It offers straightforward explanation for ``universality''. As to ``existence'', a plausible self-sustaining mechanism is proposed but so far cannot be fully reconciled with numerical \RevisedText{observations} (\cref{sec:gaps}).
	The second builds on the discovery of new flow states driven by polymer elasticity\RevisedText{---}EIT (or possibility EDT).
	For the ``existence'' attribute, it offers a self-sustaining mechanism backed by experimental evidences.
	However, how to address the ``universality'' \RevisedText{attribute remains an open question.}
	Both streams fall short of a quantitative theory for the ``magnitude'' of MDR.
\end{CmpctDescr}

In addition to the outstanding questions summarized above, many research opportunities lie in areas that have been under-explored so far.
	
\paragraph{High $\mathrm{Re}$}
	Existing understanding and knowledge are mostly based on numerical simulations at fairly low $\mathrm{Re}$.
	There \RevisedText{have} been some limited attempts \RevisedText{at} DNS at higher $\mathrm{Re}$\RevisedText{: }the highest, to my knowledge, \RevisedText{is} $\mathrm{Re}_\tau=1000$ in \citet{Thais_Mompean_IJHFF2013} and \citet{Pereira_Mompean_JFM2017}.
	\RevisedText{However,} in-depth analysis of coherent structures, flow states, and dynamical \RevisedText{intermittency} has been performed mostly at $\mathrm{Re}_\tau\sim O(100)$ where the state space is relatively simple.
\RevisedText{As discussed in section~\ref{para:scenarioIII}, new families of coherent structures would emerge at higher $\mathrm{Re}$, which brings new turbulent self-sustaining dynamics into the picture.
Those}
complex structures \RevisedText{are unlikely to follow the same interaction mechanism with polymers}
as streamwise vortices (described in \cref{sec:polymturb}).
	\RevisedText{B}oth frameworks for MDR, based on AHB cycles and EIT respectively,
\RevisedText{need to be reexamined for higher $\mathrm{Re}$.} 
\RevisedText{T}he newly developed VATIP algorithm\RevisedText{, for its capability of processing complex 3D vortex structures,} could be instrumental (\cref{sec:vatip}).

\paragraph{Connecting numerical models with realistic polymer solutions}
	Numerical investigation has been overly reliant on ``FENE-P polymers'', which, recall section~\ref{par:fenepbg}, is \RevisedText{only} an idealized model with \RevisedText{substantial} simplifications.
	There \RevisedText{has} not been much effort of quantitatively connecting simulation models with experimental drag-reducing solutions.
	Understanding which polymer-solvent combinations are more accurately modeled by FENE-P and how do model parameters map to realistic experimental systems is essential for making chemically specific predictions from DNS\RevisedText{, which is of practical interest for the selection and design of drag reducers.}
	It is also important for interpreting simulation results \RevisedText{through} the practical lens\RevisedText{. For instance,} $\mathrm{Wi}=500$ and $b=\num{e4}$ map to \RevisedText{what} molecular weight and shear rate \RevisedText{for} a given polymer? What parameter ranges are practically relevant? Discussion in section~\ref{par:feneexp} provides the theoretical basis for the determination of materials properties in FENE-P. In practice, parameters are more often obtained by fitting with experimental rheology data~\citep{Anna_McKinley_JRh2001} and parameters from the best fit do not \RevisedText{always} match theoretical expectations~\citep{Purnode_Crochet_JNNFM1998}.

\paragraph{\RevisedText{DR by rigid polymers}}
	Comparison between flexible and rigid polymers is intriguing. Although both seem to be bounded by the same MDR asymptote, the transition patterns before reaching MDR are totally different, based on which Virk dubbed them type~A (flexible) and type~B (rigid) drag reducers~\citep{Virk_Nature1975,Virk_Sherman_AIChEJ1997}.
	\RevisedText{Streamwise v}elocity fluctuation patterns from flexible and rigid polymers also appear \RevisedText{to differ}~\citep{Mohammadtabar_Ghaemi_POF2017,Shaban_Ghaemi_POF2018} \RevisedText{(}section~\ref{par:fluc}\RevisedText{)}.
	Since FENE-P is a model for flexible (type~A) polymers, insight into DR by rigid \RevisedText{(type~B)} polymers is limited.
	Meanwhile, rigid polymers also provide a reference system where significant DR is achievable but elastic instabilities are not expected, which can help us separate the roles of different mechanisms in DR.

\paragraph{\RevisedText{DR by} surfactant\RevisedText{s}}
	Surfactants are also highly potent drag reducers. They assemble into cylindrical ``wormlike'' micelles\RevisedText{---long chain structures held together} not by covalent bonds but through intermolecular (vdW \RevisedText{and} electrostatic) interactions.
	\RevisedText{Wormlike micelles}
	can be viewed as ``living'' polymers\RevisedText{: t}hey \RevisedText{would} break in regions with strong turbulence but can reassemble once \RevisedText{external strain} is removed.
	Experiments have shown that surfactant DR systems can exceed the Virk MDR asymptote~\citep{Bewersdorff_Ohlendorf_CollPolySci1988,Zakin_Myska_AIChEJ1996}, \RevisedText{which indicates} a different mechanism of DR at least in that regime.
	This could be related to the ``living'' nature of the microstructure, or to the formation of higher-order assemblies such as shear-induced structures~\citep{Cates_Candau_JPhysCM1990,Liu_Pine_PRL1996}. Fundamental \RevisedText{understanding is rather limited.}
	
\paragraph{Flow-induced scission of polymer chains}	
	Long-chain polymers are subject to
	\RevisedText{mechanical} degradation in strong turbulence\RevisedText{, which is responsible for the gradual loss of their} drag-reducing capacity~\citep{Vanapalli_Solomon_POF2005}. This effect is not considered in current numerical models.
	Quantitative prediction of \RevisedText{chain} scission is, however, of practical significance. \RevisedText{Any model for chain scission} would be similar to \RevisedText{those} for surfactants as living polymer\RevisedText{s}\RevisedText{,} less the reassembly part\RevisedText{.}
	One idea to combat flow-induced scission is to use supramolecular chemistry~\citep{Wei_Kornfield_Science2015}, where smaller polymeric building blocks are jointed together through non-covalent interactions to form much longer chain\RevisedText{s}. \RevisedText{The} non-covalent bonding sites \RevisedText{would} break in strong flow, but, like surfactants, they can reconnect afterwards.

\begin{acknowledgments}
Much gratitude is due to many coworkers (mentors, colleagues, students) with whom I have collaborated in this area or had enlightening discussions, in particular, Michael~D.~Graham (\RevisedText{U Wisconsin}-Madison), who initially led me into this area with patient guidance and visionary insights, and Lu~Zhu (McMaster U.), whose recent contributions have injected new energy to this this area of research in my group.
Many thanks should also go to John~F.~Gibson (U. New Hampshire), for sharing his Newtonian DNS code \texttt{ChannelFlow}~\citep{Gibson_ChFlowCode} which benefited our research as well as that of many others, to Hecke~Schrobsdorff (Max Planck Inst.), Tobias~Kreilos and Tobias~M.~Schneider (EPFL) for helping with the parallelization of our in-house DNS code, and to Javier~Jim{\'e}nez (Polytechnic U. Madrid) for discussion that inspired our interest in vortex analysis\RevisedText{, which} led to VATIP. 
National Science Foundation (NSF) Grant No. NSF PHY11-25915 partially supported my stay at the Kavli Institute for Theoretical Physics (KITP) at UC Santa Barbara.
Financial support from the Natural Sciences and Engineering Research Council of Canada (NSERC) through its Discovery Grants Program (No.~RGPIN-2014-04903), the facilities of the Shared Hierarchical Academic Research Computing Network (SHARCNET: \texttt{www.sharcnet.ca}), and the computing resources allocation awarded by Compute/Calcul Canada are all acknowledged.
\end{acknowledgments}


%
%

%


\bibliography{FluidDyn,Polymer,General}

\end{document}